\documentclass[journal,twoside,web]{ieeecolor}
\usepackage{jsen}
\usepackage{cite}
\usepackage{amsmath,amssymb,amsfonts}
\usepackage{algorithmic}
\usepackage{textcomp}
\usepackage{url}
\usepackage{verbatim}
\usepackage{graphicx}
\usepackage{wrapfig}
\def\BibTeX{{\rm B\kern-.05em{\sc i\kern-.025em b}\kern-.08em
    T\kern-.1667em\lower.7ex\hbox{E}\kern-.125emX}}
\definecolor{abstractbg}{rgb}{0.89804,0.94510,0.83137}
\usepackage{cite}
\usepackage{multirow}

\usepackage{xcolor}

\newcommand*{\tran}{^{\mkern-1.5mu\mathsf{T}}}

\usepackage[colorlinks, pagebackref=true]{hyperref}
\hypersetup{colorlinks=true,
	linkcolor=blue,   
	urlcolor=cyan}

\begin{document}
\title{Geometric Wide-Angle Camera Calibration: A Review and Comparative Study}
\author{Jianzhu Huai,~\IEEEmembership{Member,~IEEE}, Yuan Zhuang$^\dagger$,~\IEEEmembership{Senior Member, IEEE}, 
	 Yuxin Shao, Grzegorz Jozkow, Binliang Wang, Yijia He, and Alper Yilmaz,~\IEEEmembership{Senior Member, IEEE}
\thanks{Jianzhu Huai, Yuan Zhuang, Yuxin Shao, and Binliang Wang are with 
	the Laboratory of Information Engineering in Surveying, Mapping and Remote Sensing (LIESMARS),
	Wuhan University, 129 Luoyu Road, Wuhan, Hubei, China.
	Y. Zhuang is also with Hubei Luojia Laboratory, Wuhan, Hubei, China, 
	and Wuhan University Shenzhen Research Institute, Shenzhen, Guangdong, China. %
}
\thanks{Grzegorz Jozkow is with the Institute of Geodesy and Geoinformatics, Wroclaw University of Environmental and Life Sciences, Wroclaw, Poland
}%
\thanks{Alper Yilmaz is with the Department of Civil, Environmental, and Geodetic Engineering, The Ohio State University, Columbus, OH, US}
\thanks{$^{\dagger}$Corresponding author: yuan.zhuang@whu.edu.cn}
}

\IEEEtitleabstractindextext{%
\fcolorbox{abstractbg}{abstractbg}{%
\begin{minipage}{\textwidth}%
\begin{wrapfigure}[12]{r}{3in}%
\includegraphics[width=3in]{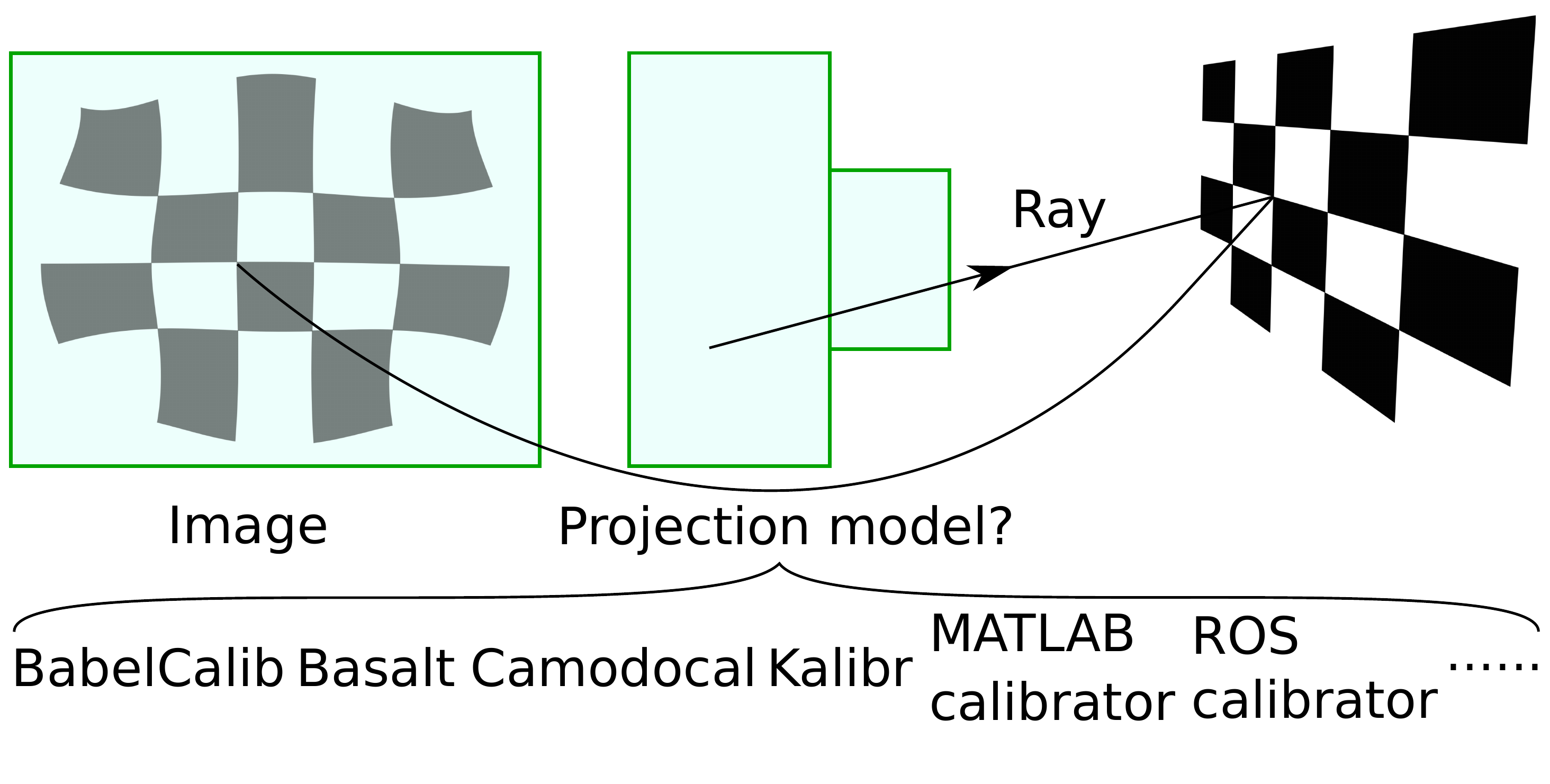}%
\end{wrapfigure}%

\begin{abstract}
Wide-angle cameras are widely used in photogrammetry and autonomous systems which 
rely on the accurate metric measurements derived from images.
To find the geometric relationship between incoming rays and image pixels, geometric camera calibration (GCC) has been actively developed.
Aiming to provide practical calibration guidelines, this work surveys the existing GCC tools and evaluates the representative ones for wide-angle cameras.
The survey covers camera models, calibration targets, and algorithms used in these tools, highlighting their properties and the trends in GCC development.
The evaluation compares six target-based GCC tools, namely, BabelCalib, Basalt, Camodocal, Kalibr, the MATLAB calibrator, and the OpenCV-based ROS calibrator,
with simulated and real data for wide-angle cameras described by four parametric projection models.
These tests reveal the strengths and weaknesses of these camera models, as well as the repeatability of these GCC tools.
In view of the survey and evaluation, future research directions of wide-angle GCC are also discussed.
\end{abstract}

\begin{IEEEkeywords}
geometric camera calibration, calibration tool, camera model, calibration target, calibration algorithm.
\end{IEEEkeywords}
\end{minipage}}}

\maketitle

\section{Introduction}
\label{sec:introduction}
\IEEEPARstart{C}{ameras} have been used in a variety of applications for obtaining metric measurements, such as remote sensing  \cite{tothRemoteSensingPlatforms2016}, 
cultural heritage \cite{kongPreserving2023}, and robotics \cite{miaoJoint3DShape2019}.
These applications need the geometric camera calibration (GCC) information to relate the rays from world points and the image pixels.
The calibration information includes the extrinsic component that transforms points from the object space to the camera coordinate system and 
the intrinsic component that projects rays in the camera frame to the imaging sensor.
For remote sensing applications, the extrinsic and intrinsic components are often unified into a generalized model for computation convenience \cite{taoComprehensive2001}.
For close range applications, the two components are usually treated separately and GCC mainly refers to the intrinsic calibration.

Based on the angle of view (AOV), cameras can be roughly grouped 
\cite{kraus2007photogrammetry,kannalaGenericCameraModel2006}
into conventional cameras,
wide-angle cameras, fisheye cameras, 
and omnidirectional cameras ($\ge 180^\circ$).
The omnidirectional cameras include fisheye cameras with an AOV $\ge 180^\circ$, and catadioptric cameras comprising of lenses and mirrors (``cata'' for mirror reflection and ``dioptric'' for lens refraction).
There are also camera rigs consisting of multiple cameras which achieve a great AOV by stitching images.
Based on whether all incoming rays pass through a single point, cameras can be divided into central cameras of a single effective viewpoint, i.e., the optical center, and non-central cameras.
Central cameras include the conventional cameras, fisheye cameras with an AOV $\le$ 195$^\circ$ \cite{scaramuzzaToolboxEasilyCalibrating2006}, 
and many catadioptric cameras built by combining a pinhole camera and hyperbolic, parabolic, or elliptical mirrors \cite{sturmCameraModelsFundamental2011}.
Analogous to \cite{fanWideangle2022}, we define the term wide-angle cameras to include the wide-angle cameras with AOV $\textless 120 ^\circ$, 
fisheye cameras, and the central catadioptric cameras.
These are the prevalent cameras used in applications requiring metric information.

GCC for these wide-angle cameras have been studied since 1960s \cite{edmundsonRevisitingApolloPhotogrammetric2018} and are still being actively researched today.
This is reflected in the evolution of camera models from the classic parametric models, to the generic models of thousands of parameters and deep neural networks.
The parametric models are realized with a few specific parameters, e.g., focal length, radial distortion.
Since each parameter affect the imaging of all pixels, they are also known as global models \cite{sturmCameraModelsFundamental2011}.
A parametric model is usually designed for a specific type of cameras. 
These models are well supported by the existing calibration tools, and structure from motion (SfM) packages.
The generic model has dense parameters and each parameter subset describes the imaging process in a local area,
e.g., B-spline models \cite{schopsWhyHaving102020} and per-pixel models \cite{ramalingamUnifyingModelCamera2017}.
A generic model is supposed to work with a wide range of cameras, but obviously requires more observations for calibration.
Recently, the deep neural network has been widely used in image rectification as surveyed in \cite{liao2023deep, fanWideangle2022}.
A neural network trained with lots of distorted and undistorted image pairs captured by a variety of cameras, can regress a parametric camera model or 
estimate the distortion vector at every pixel for a new image or video clip.
These deep camera calibration studies usually have attention on distortion removal.
This paper focuses more on the repeatable parametric and generic models which are dominantly used in photogrammetry and robotics for accurate metric measurements.

Numerous tools have been developed for carrying out GCC, each with a unique set of features.
They are often available as proprietary programs, 
such as the camera calibrator in MATLAB \cite{mathworksinc.MATLABComputerVision2021} and Agisoft Metashape \cite{agisoftllcAgisoftMetashape2022}, 
or open-source programs, such as Kalibr \cite{mayeOnlineSelfcalibrationRobotic2016}.
As for similarities, existing tools usually support global camera models and calibration with some planar target.
Notably, many tools are based on the same underlying packages, e.g., OpenCV \cite{bradskiLearningOpenCVComputer2008}, thus,
they tend to have similar limitations.
Moreover, many programs developed independently are very close in functionality, implying a possible duplicate effort.
As for practical differences, these tools usually support different sets of camera models and calibration targets.

The diverse landscape of camera models and calibration tools on one hand offers ready-to-use solutions in a variety of situations,
but on the other hand, it gets overwhelming for practitioners to choose the proper calibration tool.
To address this difficulty, quite a few comparative studies have been conducted.
For instance, three calibration algorithms were compared in \cite{hieronymusComparisonMethodsGeometric2012} for cameras with large focal lengths.
Digital displays and printed targets were compared in \cite{schmalzCameraCalibrationActive2011} for close-range cameras.
Hughes et al. \cite{hughesReviewGeometricDistortion2008} reviewed distortion models for fisheye cameras with AOV \textless 180$^\circ$.
Both \cite{puigCalibrationOmnidirectionalCameras2012} and \cite{zhangSurveyCatadioptricOmnidirectional2013} surveyed calibration methods for omnidirectional cameras.

Considering the prevalence of wide-angle cameras,
there is a lack of qualitative overview and quantitative comparison of existing GCC tools which elucidates choosing the proper camera model and calibration tool.
To fill this gap, we extensively review existing GCC tools for wide-angle cameras from several practical aspects and benchmark several popular tools with simulated and real data.

The contributions of this work are summarized as follows:
First, this review categorizes camera models, calibration targets, and calibration algorithms as used in GCC tools for wide-angle cameras, providing a concise reference for these aspects.
We then qualitatively reveal the strengths and similarities of these calibration tools,
hopefully preventing repetitive development efforts in the future.
Second, an evaluation of six representative calibration tools is conducted for in-house cameras with varying AOV by simulation and real-data tests to show their accuracy and repeatability.
The evaluation clearly shows strengths and weaknesses of four popular global geometric camera models and indicates which calibration tool to use for wide-angle camera applications.
Third, based on the review and evaluation, we highlight future research directions for wide-angle GCC.

\begin{figure}[!tbp]
	\centering
	\includegraphics[width=0.98\columnwidth]{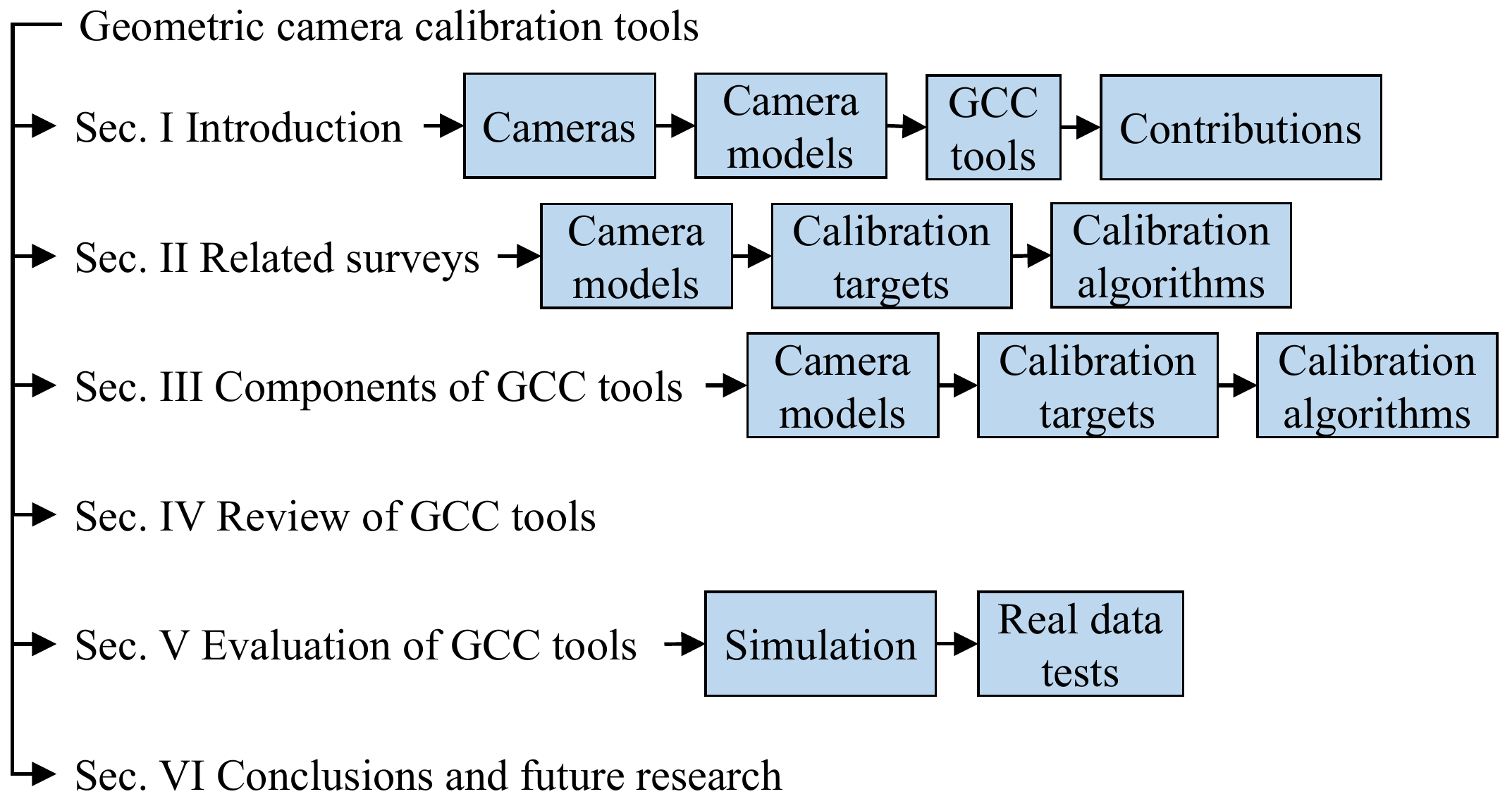}
	\caption{The structure of this survey.}
	\label{fig:structure}
\end{figure}

The following text is organized as shown in Fig.~\ref{fig:structure}.
Next, Section \ref{sec:related-work} briefly reviews related work on comparative studies of GCC.
For the available camera calibration tools, Section \ref{sec:method} sorts out the camera models,
the calibration targets, and the calibration algorithms.
The GCC tools are reviewed in Section \ref{sec:tools}.
Section \ref{sec:results} presents experiments of six calibration tools with a range of cameras and several popular global camera models.
Finally, conclusions and future research trends are given in Section \ref{sec:conclusion}.

\section{Related Work}
\label{sec:related-work}
This section briefly reviews comparative studies and surveys about wide-angle GCC from several aspects including camera models, calibration targets, and calibration methods, 
with an emphasis on approaches based on geometric optimization.
Deep learning-based camera calibration approaches are surveyed in \cite{fanWideangle2022, liao2023deep}.

\subsection{Camera Models}
Comparative studies about camera models are usually conducted in papers proposing new or enhanced models.
For fisheye cameras, in \cite{usenkoDoubleSphereCamera2018}, the double sphere (DS) model was proposed and compared with several global models
including the Kannala-Brandt (KB) model \cite{kannalaGenericCameraModel2006}, the extended unified camera model (EUCM) \cite{khomutenkoEnhancedUnifiedCamera2016}, 
the field of view (FOV) model \cite{devernayStraightLinesHave2001}, validating that its accuracy rivaled that of the KB model with 8 parameters.
In \cite{bergamascoCanFullyUnconstrained2013}, a per-pixel generic model was shown to be more accurate than a pinhole camera model with radial distortion.
The generic B-spline model \cite{beckGeneralizedBsplineCamera2018} was enhanced in \cite{schopsWhyHaving102020} with a denser grid of control points for the cubic B-spline surface, 
and it was shown that well-calibrated generic models led to more accurate results than global models in photogrammetric applications.
Authors of \cite{ramalingamGenericImagingModels2006,sturmCameraModelsFundamental2011} extensively reviewed existing camera models and established a taxonomy based on several criteria.
In this paper, we survey the camera models commonly found in GCC tools and provide their exact formulations for reference (Section \ref{subsec:cammodels}).

\subsection{Calibration Targets}
Though self-calibration has gained momentum over the years, e.g., \cite{kenefick1972analytical,gruenSystem2001,stamatopoulosAutomated2014},
to ensure high accuracy, GCC can be performed with a set of points of known positions, such as a calibration field used in remote sensing, and
calibration targets in close-range applications.
The diversity of calibration targets made necessary comparative analysis of these targets.
Regarding control point detection in camera calibration, circle grids and checkerboards were studied in \cite{mallonWhichPatternBiasing2007} and it was found that 
circles suffered from perspective and distortion biases whereas corner points of checkerboards were invariant to the distortion bias.
Schmalz et al. \cite{schmalzCameraCalibrationActive2011} systematically compared the active targets with digital displays to the printed checkerboard for GCC
with several combinations of displays, cameras, and lenses.
They found that calibration with the active target had much lower reprojection errors, 
but required compensation for the refraction of the glass plate and multiple images per pose and hence a tripod or the like.
In an underwater environment, fiducial markers including the ARToolKit \cite{katoMarkerTrackingHMD1999}, the AprilTag \cite{olsonAprilTagRobustFlexible2011}, and the Aruco \cite{garrido-juradoAutomaticGenerationDetection2014} were compared in \cite{dossantoscesarEvaluationArtificialFiducial2015} 
where the AprilTag showed better detection performance but required higher computation.
In environments with occlusions and rotations, three markers, the ARTag \cite{fialaARTagFiducialMarker2005}, the AprilTag \cite{olsonAprilTagRobustFlexible2011}, 
and the CALTag \cite{atchesonCALTagHighPrecision2010} were compared in \cite{sagitovARTagAprilTagCALTag2017} and the CALTag emprically achieved the best recognition rate.
For pose tracking in surgery, Kunz et al.~\cite{kunzMetricbasedEvaluationFiducial2019} compared the Aruco and AprilTag markers and found that
both could achieve sub-millimeter accuracy at distances up to 1 m.
For localization of unmanned aerial systems, four fiducial markers, the ARTag, the AprilTag, the Aruco, and the STag \cite{benligiraySTagStableFiducial2019}, were compared in \cite{kalaitzakisExperimentalComparisonFiducial2020} in terms of detection rate and localization accuracy.
The AprilTag, the STag, and the Aruco were shown to have close performance whereas the Aruco was the most efficient in computation.
For drone landing, several variants of the AprilTag and the circular WhyCode \cite{lightbodyVersatileHighperformanceVisual2017} were compared in \cite{springerEvaluationOrientationAmbiguity2022} 
on an embedded system and the suitable variants were determined.
Unlikely above comparative studies about targets, our paper briefly surveys the calibration targets (Section \ref{subsec:target}) supported by the available GCC tools.

\subsection{Calibration Algorithms}
The algorithms for GCC are vast, ranging from target-based to self-calibration, from offline to interactive calibration, from sequence-based to single image calibration.
Quite a few papers have reviewed the GCC methods in view of different applications.
For close-range photogrammetry, an overview of developments of camera calibration methods up to 1995 was provided in \cite{clarkeDevelopmentCameraCalibration1998}.
Several calibration techniques up to 1992 for conventional cameras with a pinhole model were reviewed and evaluated in \cite{salviComparativeReviewCamera2002}.
For close-range applications, several target-based and self-calibration methods were compared in \cite{remondinoDigitalCameraCalibration2006} with a 3D target and a checkerboard,
showing that the self-calibration methods based on bundle adjustment often achieved good calibration for consumer-grade cameras.
For time-of-flight range cameras, three intrinsic calibration methods were compared in \cite{lichtiComparisonThreeGeometric2011} for
calibrating camera lens parameters and range error parameters by using a multi-resolution planar target.
For cameras of large focal lengths ($\ge$35 mm), Hieronymus \cite{hieronymusComparisonMethodsGeometric2012} compared three calibration methods, 
one with a test field of a known geometric pattern, and two methods with devices for generating laser beams.
He found that these methods achieved comparable high accuracy for the pinhole model with radial and tangential distortion.
For cameras with lenses of focal lengths $\ge$50 mm in particle tracking velocimetry, Joshi et al.\cite{joshiComparativeStudyCamera2013} 
studied the accuracy of three camera calibration methods,
the direct linear transform (DLT) that ignores the distortion \cite{abdel-azizDirect2015},
a linear least squares method with the rational polynomial coefficient (RPC) model \cite{fraserSensorOrientationRPCs2006} but only using the numerator terms, 
and Tsai's method which determines the intrinsic and extrinsic parameters in two steps \cite{tsaiVersatileCameraCalibration1987}.
They found that errors of the Tsai's method were fluctuant due to the unstable nonlinear optimization.
For omnidirectional cameras, four calibration methods were compared in \cite{puigCalibrationOmnidirectionalCameras2012} and shown to achieved similar calibration accuracy.
For infrared cameras, Usamentiaga et al.\cite{usamentiagaComparisonEvaluationGeometric2018} compared three calibration methods,
a DLT method, an iterative method, and a complete method that considered lens distortion, and unsurprisingly, the last method resulted in best distance measurements.
For roadside cameras, GCC methods based on vanishing points were compared in  \cite{kanhereTaxonomyAnalysisCamera2010}, assuming no lens distortion.
For X-ray cameras ignoring radial distortion, 
the DLT method \cite{hartleyMultipleViewGeometry2003}, Tsai's method \cite{tsaiVersatileCameraCalibration1987}, and Zhang's method \cite{zhangFlexibleNewTechnique2000} were compared in \cite{albiolEvaluationModernCamera2017}, and the DLT showed superiority in accuracy and operation simplicity.
For a camera-projector pair, Tiscareno et al.\cite{tiscarenoAnalysisDifferentCamera2019} 
calibrated the camera with the DLT method, Tsai's method, and Zhang's method, and calibrated the projector with the DLT, through simulation.
They found that Zhang's method gave smaller reprojection errors than the others for camera calibration.
For zoom-lens cameras with varying focal lengths, calibration methods were reviewed in \cite{ayazSurveyZoomlensCalibration2017}.
Different from the preceding surveys and comparisons focusing on calibration methods, this paper reviews and compares GCC tools for wide-angle cameras of fixed intrinsic parameters.

\section{Geometric Camera Calibration Components}
\label{sec:method}
This section reviews geometric camera models, targets, algorithms as available in existing calibration tools, especially for wide-angle cameras.

Before elaborating GCC, some definitions are clarified here.
The focal length is defined to be the distance between the camera's optical center and the sensor as in \cite{ramalingamGenericImagingModels2006}.
The distance's rigorous name is principal distance, but the term focal length is so widely used instead.
Since the optical center is defined only for central cameras, the focal length is not defined for non-central cameras.
Accordingly, the focal length can take a range of values including the one when the camera is focused at infinity.
We define the optical axis as the line passing through the optical center and orthogonal to the sensor chip.
Its foot on the sensor chip is the principal point.
For ease with pinhole cameras, the sensor is often inverted and placed in front of the optical center, forming the image plane \cite{hartleyMultipleViewGeometry2003}.
For a catadioptric camera, the mirror axis refers to the symmetry axis of the mirror.
We define the AOV of a lens to be the maximum angle formed by rays coming into the lens.
Likewise, the AOV of a camera is defined as the maximum angle formed by rays corresponding to the sensor's exposed pixels, along the sensor's horizontal axis, vertical axis, or diagonal,
leading to HAOV, VAOV, or DAOV, respectively.
Thus, the AOV of a camera depends on both the lens and the sensor.

\subsection{Camera Models}
\label{subsec:cammodels}
\begin{figure}[!tbp]
	\centering
	\includegraphics[width=0.98\columnwidth]{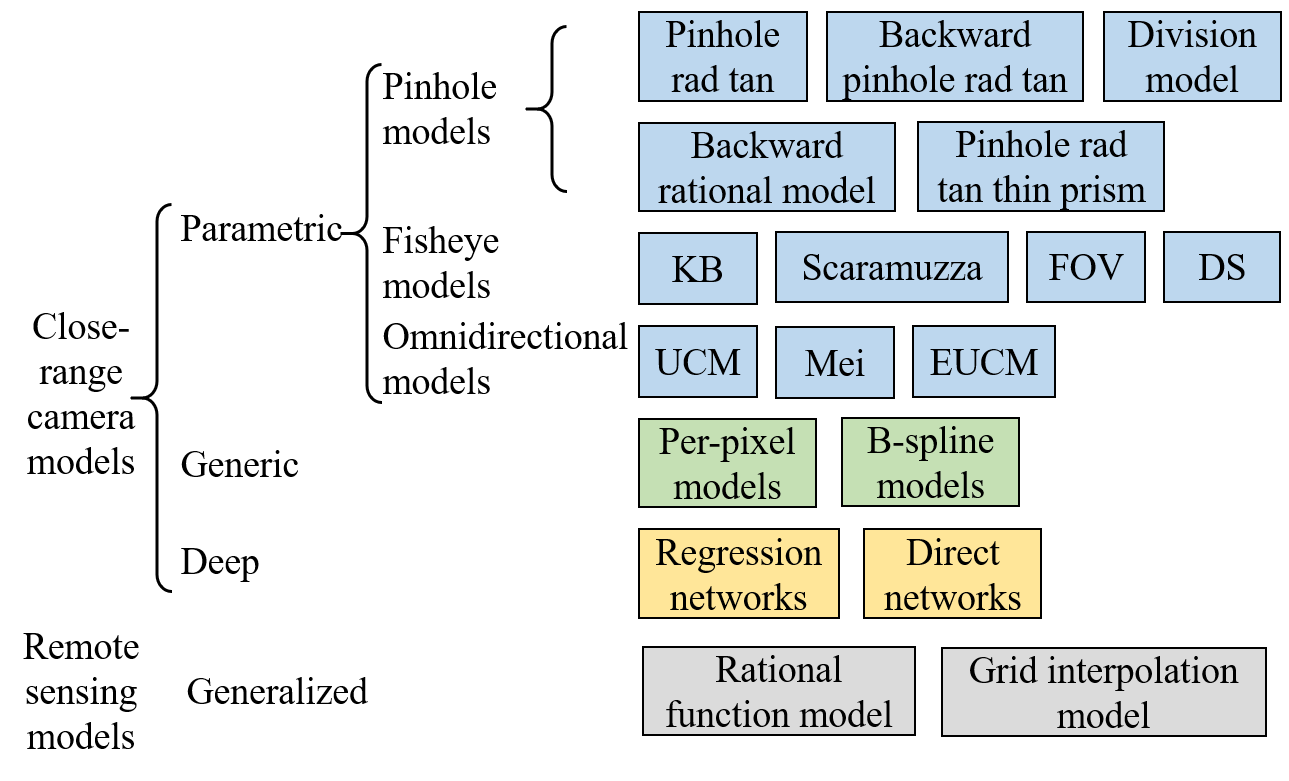}
	\caption{The parametric and generic camera models reviewed in Section~\ref{subsec:cammodels}. 
	For completeness, deep neural network models and the generalized camera models for remote sensing are also shown referring to \cite{fanWideangle2022} and \cite{piRobustCameraDistortion2022}.}
	\label{fig:models}
\end{figure}

The following describes the variety of camera models used in close-range applications, which have been adopted in GCC tools surveyed in this paper.
Camera models used for remote sensing, such as the RPC model \cite{fraserSensorOrientationRPCs2006}, 
the detector directional model \cite{wangOnorbitGeometricCalibration2014}, are referred to \cite{piRobustCameraDistortion2022}.
We begin with global parametric models for central cameras which dominate the GCC tools, and end with local generic models.
These global models are typically defined in a (forward) projection manner where image points are formulated given world points or rays,
although the same formulae may be used the other way round to obtain a ray given an image point, i.e., backward projection / back-projection / unprojection, 
for instance, \eqref{eq:pinhole-radtan} and \eqref{eq:pinhole-radtan-inv}.
For local models, however, the backward projection is usually used to express the camera model as the forward projection can be very complex \cite{schopsWhyHaving102020}.
For the below camera models listed in Fig.~\ref{fig:models}, we describe either the forward or the backward model unless both are closed-form, 
with the understanding that going the other way often requires iterative optimization.

A set of symbols is defined in order here.
We denote a point in the camera frame by $\mathbf{x}_c = [X_c, Y_c, Z_c]$ with Euclidean coordinates $X_c$, $Y_c$, and $Z_c$.
The measured image point is denoted by $\mathbf{u}_m = [u_m, v_m]$ with pixel coordinates $u_m$ and $v_m$.
The world-to-image forward projection is denoted by $\boldsymbol{\pi}(\mathbf{x}_c, \mathbf{i}): \mathbb{R}^3 \rightarrow \mathbb{R}^2$ where $\mathbf{i}$ is the set of intrinsic parameters.
Its inverse, the image-to-world inverse projection model is
$\boldsymbol{\pi}^{-1}(\mathbf{u}_m, \mathbf{i}): \mathbb{R}^2 \rightarrow \mathbb{S}^2$ where $\mathbb{S}^2$ is the set of 3D unit vectors.
We denote by $\theta$ the incidence angle between an incoming ray and the optical axis.
We use the subscripts `m', `d', `n', and `c' to indicate measurement, distortion, normalization, and the camera coordinate frame.

\subsubsection{Global Models for Wide-Angle Cameras}
Conventional and wide-angle cameras usually have little distortion and satisfy well the pinhole model.
The set of parameters in the pinhole projection without distortion are $\mathbf{i} = [f_x, f_y, c_x, c_y]$,
including the focal length and the principal point along the image plane's two axes in units of pixels.
The distortion-free pinhole model is given by
\begin{equation}
	\mathbf{u}_m = \boldsymbol{\pi}(\mathbf{x}_c, \mathbf{i}) = 
	\begin{bmatrix}
f_x X_c/Z_c + c_x \\ f_y Y_c/Z_c + c_y
\end{bmatrix} \label{eq:pinhole},
\end{equation}
with the closed-form inverse model,
\begin{equation}
\boldsymbol{\pi}^{-1}(\mathbf{u_m}, \mathbf{i}) =
\frac{1}{\sqrt{x_n^2 + y_n^2 + 1}}
\begin{bmatrix}
x_n \\ y_n \\ 1
\end{bmatrix},
\end{equation} where
$x_n = (u_m - c_x)/f_x$ and $y_n = (v_m - c_y)/f_y$.

To account for lens distortion, a variety of distortion models for pinhole cameras have been proposed.
A popular one is the radial-tangential polynomial model, i.e., the plumb bob model or the Brown-Conrady model \cite{brownCloserangeCameraCalibration1971}.
Its intrinsic parameters, $\mathbf{i} = [f_x, f_y, c_x, c_y, k_1, k_2, p_1, p_2]$, include the pinhole projection parameters, the radial distortion parameters $k_j, j=1, 2, \cdots, p$ 
(the maximum index $p$ is usually truncated to two in practice), and the tangential / decentering distortion parameters $p_1, p_2$.
The pinhole radial tangential model is given by
\begin{align}
    \begin{bmatrix}
    x_n \\ y_n
    \end{bmatrix} &= \begin{bmatrix} X_c/Z_c \\ Y_c/Z_c \end{bmatrix}, 
    \qquad r_n^2 = x_n^2 + y_n^2, \\
    \begin{bmatrix}
    	x_d \\ y_d
    \end{bmatrix}
    &= \begin{bmatrix}
    x_n(1 + \sum_{j=1}^p k_j r_n^{2j}) + \delta_{ud} \\
    y_n(1 + \sum_{j=1}^p k_j r_n^{2j}) + \delta_{vd} \end{bmatrix},
 \label{eq:pinhole-radtan} \\
\begin{bmatrix} \delta_{ud} \\ \delta_{vd} \end{bmatrix} &=
\begin{bmatrix}
2p_1 x_n y_n + p_2(r_n^2 + 2x_n^2) \\
p_1 (r_n^2 + 2 y_n^2) + 2p_2 x_n y_n
\end{bmatrix} \label{eq:deltauvd}, \\
    \boldsymbol{\pi}(\mathbf{x}_c, \mathbf{i}) &= \begin{bmatrix}
    f_x x_d + c_x \\ f_y y_d + c_y
    \end{bmatrix} \label{eq:affine}.
\end{align}
This model usually suits well lenses with an AOV \textless 120$^\circ$\cite{usenkoDoubleSphereCamera2018}.

The inverse of \eqref{eq:pinhole-radtan} has no closed-form solution, and usually requires an iterative procedure.
Notably, Drap and Lef\`{e}vre \cite{drapExactFormulaCalculating2016} propose an exact formula involving a power series to invert \eqref{eq:pinhole-radtan}.
Alternatively, the pinhole radial tangential model can also be defined in a backward manner, i.e.,
\begin{align}
\begin{bmatrix}
x_d \\ y_d
\end{bmatrix} &= \begin{bmatrix} (u_m - c_x)/f_x \\ (v_m-c_y)/f_y \end{bmatrix}, 
\qquad r_d^2 = x_d^2 + y_d^2, \\
\begin{bmatrix}
	x_n \\ y_n
\end{bmatrix} &=
\begin{bmatrix}
x_d(1 + \sum_{j=1}^p k_j r_n^{2j}) + \delta_{ud} \\
y_d(1 + \sum_{j=1}^p k_j r_n^{2j}) + \delta_{vd} \end{bmatrix},
\label{eq:pinhole-radtan-inv}\\
\begin{bmatrix}
	\delta_{ud} \\ \delta_{vd} \end{bmatrix} &=
\begin{bmatrix}
2p_1 x_d y_d + p_2(r_d^2 + 2x_d^2) \\
p_1 (r_d^2 + 2 y_d^2) + 2p_2 x_d y_d
\end{bmatrix},\\
\boldsymbol{\pi}^{-1}(\mathbf{u}_m,\mathbf{i}) &= 
\frac{1}{\sqrt{x_n^2 + y_n^2 + 1}}
\begin{bmatrix}
x_n \\ y_n \\ 1
\end{bmatrix}.
\end{align}
Obviously, for the same camera, the parameters of the backward model differ from those of the forward model.
This backward model has been used in e.g., the PhotoModeler \cite{photomodelertechnologiesPhotoModeler2022}.

The forward pinhole radial tangential model in \eqref{eq:pinhole-radtan} can be simplified to the division model proposed by \cite{fitzgibbonSimultaneousLinearEstimation2001} 
which is a radial symmetric model with the set of intrinsic parameters $\mathbf{i} =  [f_x, f_y, c_x, c_y, k_1]$,

\begin{align}
\begin{bmatrix}
x_d \\ y_d \end{bmatrix} &= \begin{bmatrix} (u_m - c_x) / f_x \\
(v_m - c_y) / f_y
\end{bmatrix}, \qquad
r_d = \sqrt{x_d^2 + y_d^2}, \\
\begin{bmatrix}
	x_n \\ y_n
\end{bmatrix} &= 
\begin{bmatrix}
	x_d / (1 + k_1 r_d^2) \\ y_d / (1 + k_1 r_d^2)
\end{bmatrix} \label{eq:division}, \\
\boldsymbol{\pi}^{-1}(\mathbf{u}_m, \mathbf{i}) &= 
\frac{1}{\sqrt{x_n^2 + y_n^2 + 1}}
\begin{bmatrix}
x_n \\ y_n  \\ 1
\end{bmatrix}.
\end{align}

A backward rational model is proposed in \cite{liNoniterativeMethodCorrecting2005},
\begin{align}
\begin{bmatrix}
x_n \\ y_n
\end{bmatrix} &=
\begin{bmatrix}
x_d \\ y_d
\end{bmatrix}
\frac{1 + \sum_{j=1}^{p} k_j^1 r_d^{2j}}{1 + \sum_{j=1}^{q} k_j^2 r_d^{2j}} \label{eq:rational}, \\
\boldsymbol{\pi}^{-1}(\mathbf{u_m}, \mathbf{i}) &=
\frac{1}{\sqrt{x_n^2 + y_n^2 + 1}} \begin{bmatrix}
	x_n \\ y_n \\ 1
\end{bmatrix},
\end{align}
with the intrinsic parameters $\mathbf{i} = [f_x, f_y, c_x, c_y] \cup \mathbf{k}^1 \cup \mathbf{k}^2$
where $\mathbf{k}^1 = [k_j^1, j = 1, 2, \cdots, p]$ and 
$\mathbf{k}^2 =  [k_j^2, j=1, 2, \cdots, q]$.
The rational model in OpenCV \cite{bradskiLearningOpenCVComputer2008} supports $p \le 3$ and $q \le 3$.

Furthermore, the thin prism effect is considered in \cite{wengCameraCalibrationDistortion1992} along with radial and tangential distortion, where the model is defined as
\begin{align}
	\begin{bmatrix}
		x_n \\ y_n
	\end{bmatrix} &= \begin{bmatrix} X_c/Z_c \\ Y_c/Z_c \end{bmatrix},
	\qquad r_n^2 = x_n^2 + y_n^2, \\
	\begin{bmatrix}
		x_d \\ y_d
	\end{bmatrix}
	&= \begin{bmatrix}
		x_n(1 + k_1 r_n^2) + \delta_{ud} + \delta_{up} \\
		y_n(1 + k_1 r_n^2) + \delta_{vd} + \delta_{vp} \end{bmatrix}, \\
\begin{bmatrix}
	\delta_{up} \\ \delta_{vp} \end{bmatrix} &=
\begin{bmatrix}
s_1 r_n^2 \\
s_2 r_n^2
\end{bmatrix},\\
	\boldsymbol{\pi}(\mathbf{x}_c, \mathbf{i}) &= \begin{bmatrix}
		f_x x_d + c_x \\ f_y y_d + c_y
	\end{bmatrix} \label{eq:pinhole-radtan-prism},
\end{align}
where the tangential distortion $[\delta_{ud}, \delta_{vd}]$ is given in \eqref{eq:deltauvd}.
Overall, the intrinsic parameter set is
$\mathbf{i} = [f_x, f_y, c_x, c_y, k_1, p_1, p_2, s_1, s_2]$.
The OpenCV considers more terms for the thin prism effect by
$\delta_{up} = s_1 r_n^2 + s_2 r_n^4$ and $\delta_{vp} = s_3 r_n^2 + s_4 r_n^4$.

\subsubsection{Global Fisheye Camera Models}
Fisheye cameras typically have an AOV $\ge 100 ^\circ$, and can reach 280$^\circ$ \footnote{https://www.back-bone.ca/product/entaniya-280/}.
They are quite common but show great distortion, thus, quite a few global 
models have been proposed as listed in \cite{abrahamFisheyestereo2005,schneiderValidation2009}.
The most popular ones are probably the KB model \cite{kannalaGenericCameraModel2006} and the FOV model \cite{devernayStraightLinesHave2001}.

The full KB model proposed in \cite{kannalaGenericCameraModel2006} has 23 parameters where four describe the affine transform \eqref{eq:affine}, 
five describe an equidistant radial symmetric distortion, and the other 14 describe the asymmetric distortion.
The commonly used KB-8 model is radially symmetric and has 8 intrinsic parameters, $\mathbf{i} = [f_x, f_y, c_x, c_y, k_1, k_2, k_3, k_4]$. 
It is defined by
\begin{align}
	\boldsymbol{\pi}(\mathbf{x}_c, \mathbf{i}) &=
	\begin{bmatrix}
	f_x d(\theta) X_c/r_c + c_x \\
	f_y d(\theta) Y_c/r_c + c_y
	\end{bmatrix} \label{eq:kb8}, \\
	r_c &= \sqrt{X_c^2 + Y_c^2} = Z_c tan(\theta), \\
	d(\theta) &= \theta + k_1 \theta^3 + k_2 \theta^5 + k_3 \theta^7 + k_4 \theta^9,
\end{align}
Unlike the KB-9 in \cite{kannalaGenericCameraModel2006}, the KB-8 model sets the coefficient of the term $\theta$ in $d(\theta)$ to be 1.
The KB-8 model can handle an AOV $\ge 180^\circ$, 
but when it is formulated as an equidistant distortion on top of a pinhole projection as in Kalibr \cite{mayeOnlineSelfcalibrationRobotic2016} and OpenCV, 
the projection will fail for points of $Z_c\le0$ \cite{schneiderValidation2009}.

The Scaramuzza model \cite{scaramuzzaToolboxEasilyCalibrating2006} for central catadioptric cameras and fisheye cameras up to a 195$^\circ$ AOV resembles the inverse of the KB-8 model.
It is defined in a backward manner for a measured image point $[u_m, v_m]$ as
\begin{align}
\begin{bmatrix}
u_m \\ v_m
\end{bmatrix} &=
\begin{bmatrix}
c & d \\ e & 1
\end{bmatrix} \begin{bmatrix}
u_h \\ v_h
\end{bmatrix} + \begin{bmatrix}
c_x \\ c_y
\end{bmatrix}, \\
\boldsymbol{\pi}^{-1}(\mathbf{u}_m, \mathbf{i}) &=
\frac{1}{\sqrt{u_h^2 + v_h^2 + w_h^2(\rho_h)}}\begin{bmatrix}
u_h \\ v_h \\ w_h(\rho_h)
\end{bmatrix} \label{eq:scaramuzza}, \\
\rho_h &= \sqrt{u_h^2 + v_h^2}, \\
w_h(\rho_h) &= a_0 + a_2 \rho_h^2 + a_3 \rho_h^3 + a_4 \rho_h^4,
\end{align}
where $u_h$, $v_h$ are the ideal coordinates of the image point on a hypothetical plane orthogonal to the mirror axis.
The parameter vector for the model is $\mathbf{i} = [a_0, a_2, a_3, a_4, c_x, c_y, c, d, e]$.
Since $c$ in the 2$\times$2 stretch matrix is about one, $a_0$ is similar in role to $f_x$ or $f_y$ in \eqref{eq:kb8}.
This model is available in the MATLAB camera calibrator \cite{mathworksinc.MATLABComputerVision2021}.
For projecting a world point to the image, a polynomial approximation of the involved forward projection is adopted in \cite{scaramuzzaToolboxEasilyCalibrating2006} to reduce the computation.

The FOV model \cite{devernayStraightLinesHave2001} has one distortion parameter and a closed-form inversion.
It has been popular for fisheye lenses in consumer products, e.g., Tango phones.
With intrinsic parameters $\mathbf{i} = [f_x, f_y, c_x, c_y, \omega]$, its definition is given by
\begin{align}
\boldsymbol{\pi}(\mathbf{x}_c, \mathbf{i}) &= 
\begin{bmatrix}
f_x X_c \frac{r_d}{r_u} + c_x \\
f_y Y_c \frac{r_d}{r_u} + c_y
\end{bmatrix} \label{eq:fov}, \\
r_u &= \sqrt{X_c^2 + Y_c^2}, \\
r_d &= \frac{1}{\omega} \mathrm{arctan2}(2r_u \tan \frac{\omega}{2}, Z_c).
\end{align}

For backward projection of an image point, the FOV model has a closed-form solution given by
\begin{align}
\boldsymbol{\pi}^{-1}(\mathbf{u}_m, \mathbf{i}) &= 
\begin{bmatrix}
\frac{x_d\sin(r_d \omega)}{2 r_d \tan\frac{\omega}{2}} & \frac{y_d\sin(r_d \omega)}{2 r_d \tan\frac{\omega}{2}} & \cos(r_d \omega)
\end{bmatrix}\tran \\
\begin{bmatrix}
x_d \\ y_d
\end{bmatrix} &= \begin{bmatrix}
(u_m - c_x) / f_x \\ (v_m - c_y) / f_y
\end{bmatrix}, \\
r_d &= \sqrt{x_d^2 + y_d^2}.
\end{align}
Despite only one distortion parameter, the FOV model often requires as much computation as the KB-8 model for forward and backward projections
due to the trigonometric functions.

The DS model\cite{usenkoDoubleSphereCamera2018} fits well large AOV lenses, 
has a closed-form inversion, and does not involve trigonometric functions, thus making it very efficient.
This model contains 6 parameters, $\mathbf{i} = [f_x, f_y, c_x, c_y, \xi, \alpha]$.
In forward projection, a world point is projected consecutively onto two unit spheres of a center offset $\xi$,
and lastly projected onto the image plane using a pinhole model.
The projection model is defined by
\begin{align}
\boldsymbol{\pi}(\mathbf{x}_c, \mathbf{i}) &= \begin{bmatrix}
	f_x \frac{X_c}{\alpha d_2 + (1-\alpha) (\xi d_1 + Z_c)} + c_x \\
	f_y \frac{Y_c}{\alpha d_2 + (1-\alpha) (\xi d_1 + Z_c)} + c_y
\end{bmatrix} \label{eq:ds}, \\
d_1 &= \sqrt{X_c^2 + Y_c^2 + Z_c^2}, \\
d_2 &= \sqrt{X_c^2 + Y_c^2 + (\xi d_1 + Z_c)^2}.
\end{align}
Its closed-form unprojection is given by
\begin{align}
\boldsymbol{\pi}^{-1}(\mathbf{u}_m, \mathbf{i}) &=
\frac{z_d \xi + \sqrt{z_d^2 + (1-\xi^2)r_d^2}}{z_d^2 + r_d^2}
\begin{bmatrix}
	x_d \\ y_d \\ z_d
\end{bmatrix} - \begin{bmatrix}
0 \\ 0 \\ \xi
\end{bmatrix}, \\
z_d &= \frac{1 - \alpha^2 r_d^2}{\alpha \sqrt{1 - (2\alpha - 1) r_d^2} + 1 - \alpha}, \\
r_d^2 &= x_d^2 + y_d^2.
\end{align}
This model has been implemented in Basalt \cite{usenkoDoubleSphereCamera2018} and Kalibr.

\subsubsection{Global Omnidirectional Camera Models}
An omnidirectional camera has an HAOV $\ge 180^\circ$ and a DAOV up to $360^\circ$. %
Several models have been developed for such cameras.

The unified camera model (UCM) in \cite{courbonGenericFisheyeCamera2007} can deal with both fisheye cameras and central catadioptric cameras, defined by
\begin{align}
\boldsymbol{\pi}(\mathbf{x}_c, \mathbf{i}) &= 
\begin{bmatrix}
\gamma_x \frac{X_c}{\xi \rho + Z_c} + c_x &
\gamma_y \frac{Y_c}{\xi \rho + Z_c} + c_y
\end{bmatrix}\tran \label{eq:ucm}, \\
\rho &= \sqrt{X_c^2 + Y_c^2 + Z_c^2},
\end{align}
with intrinsic parameters
$\mathbf{i} = [\gamma_x, \gamma_y, c_x, c_y, \xi]$.
When $\xi$=0, the above model degenerates to a pinhole model.

The unified model is formulated equivalently in \cite{usenkoDoubleSphereCamera2018} for better numeric stability.
The formulation is given by
\begin{equation}
\boldsymbol{\pi}(\mathbf{x}_c, \mathbf{i}) =
\begin{bmatrix}
f_x \frac{X_c}{\alpha \rho + (1-\alpha) Z_c} + c_x \\
f_y \frac{Y_c}{\alpha \rho + (1-\alpha) Z_c} + c_y
\end{bmatrix} \label{eq:ucm-usenko},
\end{equation}
with intrinsic parameters $\mathbf{i} = [f_x, f_y, c_x, c_y, \alpha]$ where
\begin{equation}
	\alpha = \xi /(1 + \xi), \quad f_x = \gamma_x /(1 + \xi), \quad
	f_y = \gamma_y / (1 + \xi) \label{eq:ucm-fxy}.
\end{equation}
The unprojection function for the UCM is given by
\begin{align}
\boldsymbol{\pi}^{-1}(\mathbf{u}, \mathbf{i}) &= 
\frac{\xi + \sqrt{1 + (1-\xi^2) r_d^2}}{1 + r_d^2} \begin{bmatrix}
		x_d \\ y_d \\ 1
	\end{bmatrix} - \begin{bmatrix}
	0 \\ 0 \\ \xi
\end{bmatrix}, \\
x_d &= \frac{u_m - c_x}{f_x (1 + \xi)}, \quad 
y_d = \frac{v_m - c_y}{f_y (1 + \xi)}, \\
r_d^2 &= x_d^2 + y_d^2, \quad \xi = \frac{\alpha}{1 - \alpha}.
\end{align}

For better accuracy with the UCM, Mei and Rives \cite{meiSingleViewPoint2007} also consider the lens distortion, the misalignment and the sensor skew.
The Mei model is defined by
\begin{align}
\begin{bmatrix}
x_n \\ y_n
\end{bmatrix} &=
\begin{bmatrix}
X_c / (Z_c + \xi \rho) \\
Y_c / (Z_c + \xi \rho)
\end{bmatrix}, \qquad r_n = \sqrt{x_n^2 + y_n^2}, \\
\begin{bmatrix}
x_d \\ y_d
\end{bmatrix}
&= \begin{bmatrix} 
x_n d(r_n) + 2p_1 x_n y_n + p_2(r_n^2 + 2x_n^2) \\
y_n d(r_n) + p_1 (r_n^2 + 2 y_n^2) + 2p_2 x_n y_n
\end{bmatrix}, \\
d(r_n) &= 1+k_1 r_n^2 + k_2 r_n^4 + k_3 r_n^6,
\\
\boldsymbol{\pi}(\mathbf{x}_c, \mathbf{i}) &= \begin{bmatrix}
\gamma_x (x_d + s y_d) + c_x \\ \gamma_y y_d + c_y
\end{bmatrix} \label{eq:mei},
\end{align}
with the intrinsic parameters $\mathbf{i} = [\gamma_x, \gamma_y, c_x, c_y, \xi, k_1, k_2, k_3, p_1, p_2, s]$ where $k_1$, $k_2$, and $k_3$ are for radial distortion, 
$p_1$ and $p_2$ for misalignment, and $s$ for skew.
This model is adopted in \cite{liMultiplecameraSystemCalibration2013} and Camodocal \cite{hengCamOdoCalAutomaticIntrinsic2013}.
As pointed out in \cite{khomutenkoEnhancedUnifiedCamera2016}, $k_1$ of the Mei model is redundant with $\xi$.

The extended unified camera model (EUCM) \cite{khomutenkoEnhancedUnifiedCamera2016} enhances the UCM by a parameter $\beta$ to deal with the radial distortion.
Its projection model is given by 
\begin{align}
	\begin{bmatrix}
		u_m \\ v_m
	\end{bmatrix} &= 
\begin{bmatrix}
	f_x \frac{X_c}{\alpha \rho + (1-\alpha) Z_c} + c_x \\
	f_y \frac{Y_c}{\alpha \rho + (1-\alpha) Z_c} + c_y
\end{bmatrix} \label{eq:eucm}, \\
\rho &= \sqrt{\beta(X_c^2 + Y_c^2) + Z_c^2},
\end{align}
with parameters $\mathbf{i} = [f_x, f_y, c_x, c_y, \alpha, \beta]$,
where $\alpha \in [0, 1]$, $\beta > 0$, and 
$\alpha \rho + (1-\alpha) Z_c > 0$.
The unprojection function for the EUCM is given by 
\begin{align}
\boldsymbol{\pi}^{-1}(\mathbf{u}, \mathbf{i}) &=
\frac{1}{\sqrt{x_d^2 + y_d^2 + z_d^2}} \begin{bmatrix}
x_d \\ y_d \\ z_d
\end{bmatrix}, \\
\begin{bmatrix}
x_d \\ y_d \end{bmatrix} &= 
\begin{bmatrix}
(u_m - c_x) / f_x \\
(v_m - c_y) / f_y
\end{bmatrix}, \quad
r_d^2 = x_d^2 + y_d^2, \\
z_d &= \frac{1 - \beta \alpha^2 r_d^2}{\alpha \sqrt{1-(2\alpha - 1) \beta r_d^2} + 1 -\alpha}.
\end{align}

\subsubsection{Local Generic Camera Models}
The preceding global camera models are available in a variety of GCC tools possibly for their simplicity, but they are mostly for a specific type of cameras.
To work with a wide range of cameras, generic models with thousands of parameters have been proposed, such as 
\cite{rosebrockGenericCameraCalibration2012,beckGeneralizedBsplineCamera2018}.
They are still behind the global models in availability among GCC tools and in support by downstream applications.

We briefly describe two generic models implemented in \cite{schopsWhyHaving102020}, a per-pixel model and a B-spline model.
The per-pixel model of \cite{ramalingamUnifyingModelCamera2017} associates a ray direction to every pixel for a central camera and a ray direction and a 3D point on the ray to every pixel for a non-central camera.
Furthermore, interpolation between pixels is used to achieve continuous projection.
A B-spline model adopted in \cite{schopsWhyHaving102020} associates ray parameters to a sparse set of grid points instead of all pixels.
These grid points control the cubic B-spline surface which represents the back projection function.
Notably, this B-spline model is initialized using the relative camera poses computed with the method \cite{ramalingamUnifyingModelCamera2017} developed for the per-pixel model.

\subsection{Calibration Targets}
\label{subsec:target}
\begin{figure}[!tbp]
	\centering
	\includegraphics[width=0.95\columnwidth]{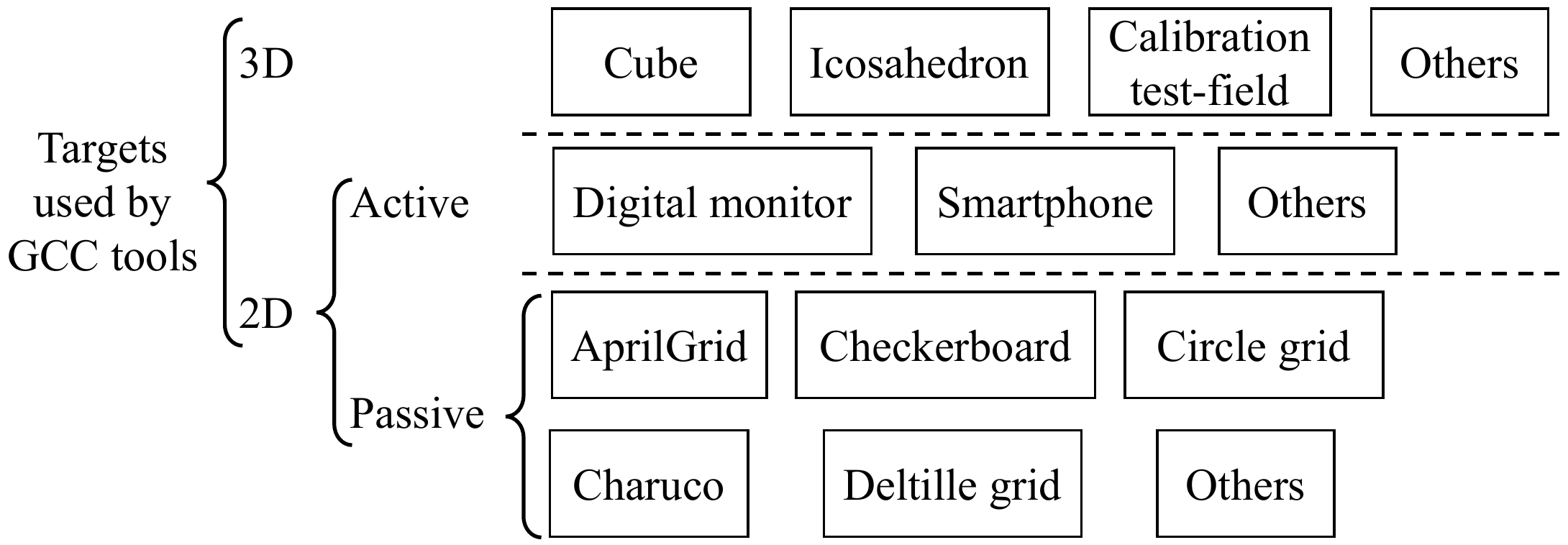}
	\caption{Categories of targets for geometric camera calibration.}
	\label{fig:targetlist}
\end{figure}

GCC usually depends on passive or active man-made objects, e.g., ground control points in remote sensing or planar targets in close-range calibrations.
Recent self/auto-calibration methods, e.g., \cite{keivanConstanttimeMonocularSelfcalibration2014,pix4ds.a.PIX4Dmapper2022}, use opportunistic environmental features, 
whereas infrastructure-based methods \cite{hengCamOdoCalAutomaticIntrinsic2013} use a prior landmark map of the environment.
Since artificial targets are still commonly used for better accuracy control, this section surveys the targets supported by GCC tools, as listed in Fig.~\ref{fig:targetlist}.

There are a few 3D targets, such as cubes \cite{anCharucoBoardbasedOmnidirectional2018} and icosahedrons \cite{haDeltilleGridsGeometric2017},
each of which is usually a composite of multiple planar targets.
The accuracy requirements of length and orthogonality complicate their manufacturing and hamper their accessibility.
The majority of calibration targets are planar, including surveyed markers on flat walls,
and a variety of coded patterns either displayed on digital screens \cite{gaoScreenbasedMethodAutomated2019,schmalzCameraCalibrationActive2011, haAccurateCameraCalibration2015} or printed out.
The targets based on digital displays usually have accurate size and good flatness and can deal with defocusing \cite{haAccurateCameraCalibration2015,bellMethodOutoffocusCamera2016}, 
but such a target usually requires capturing multiple pattern images at each pose and compensating the refraction of the display's glass plate.

\begin{figure}[!tbp]
\centering
	\includegraphics[width=0.45\columnwidth]{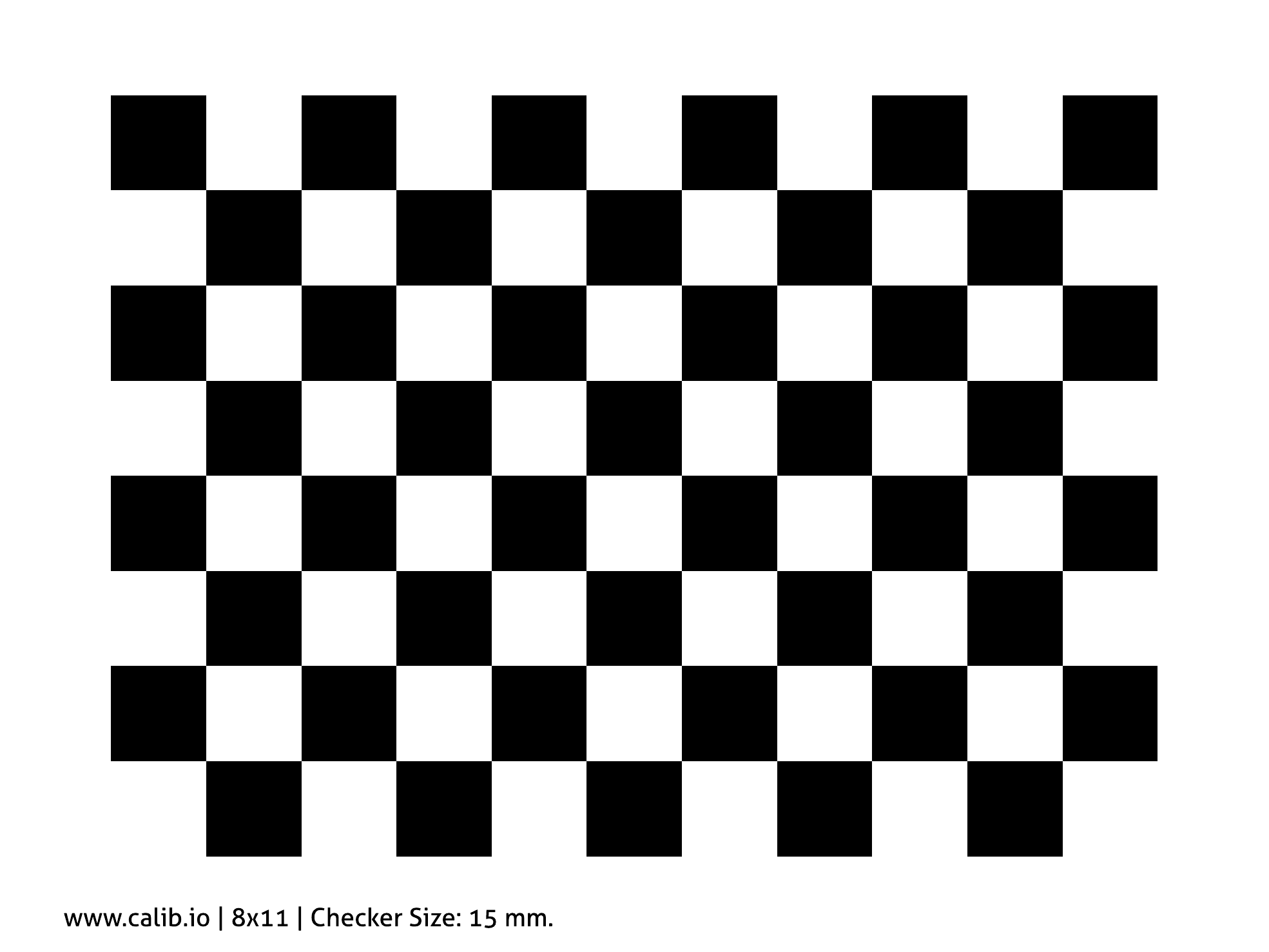} 
	\includegraphics[width=0.45\columnwidth]{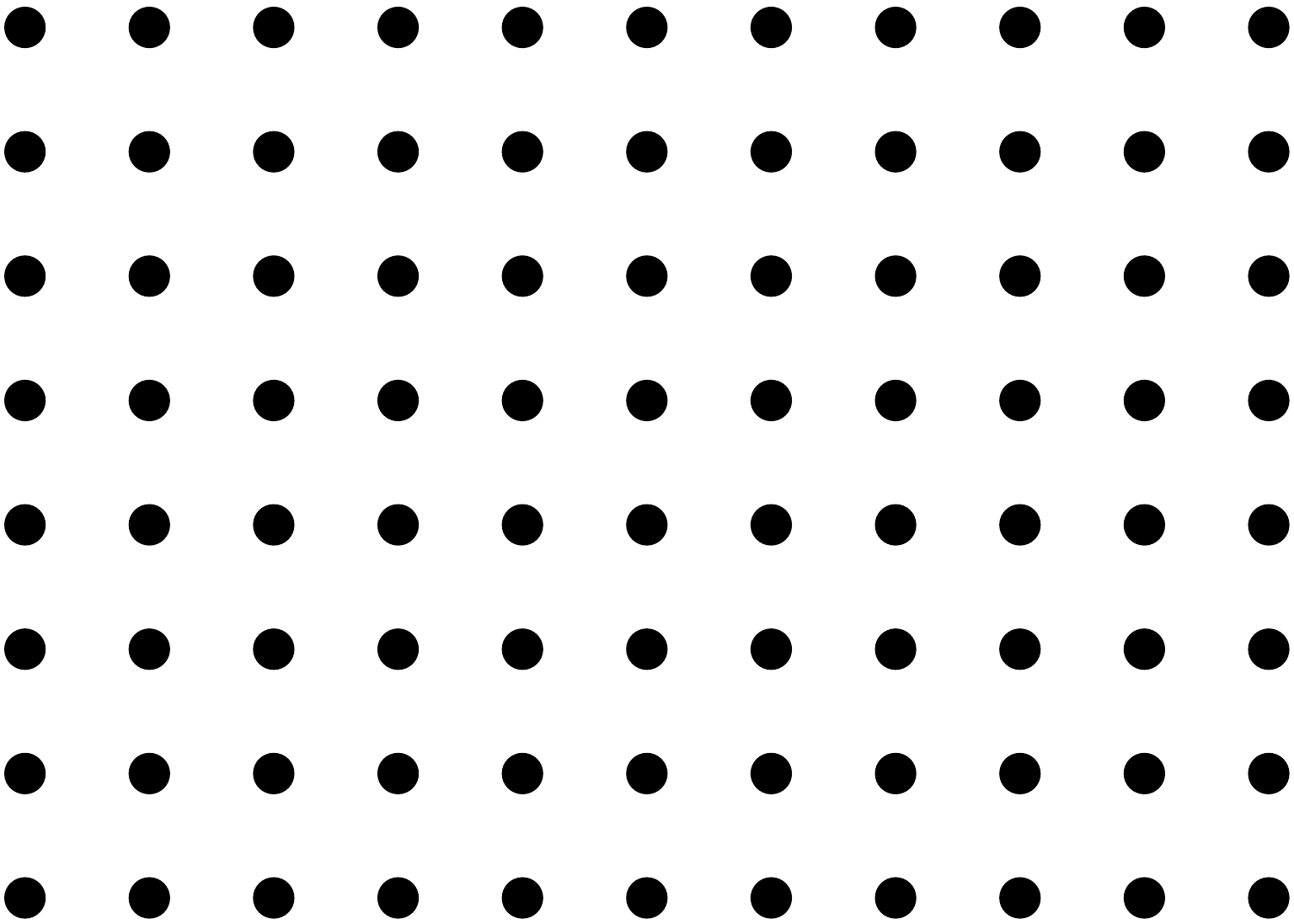}\\
    (a)\hspace{0.45\columnwidth}(b) \\
	\includegraphics[width=0.45\columnwidth]{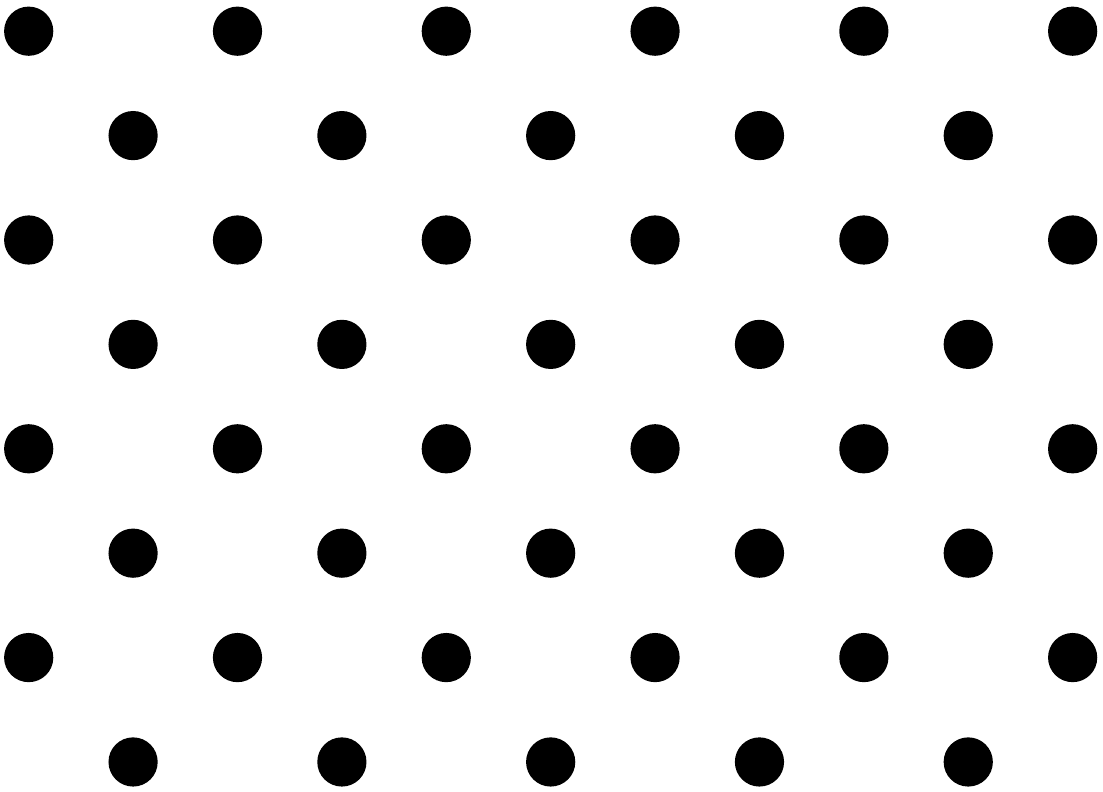}
	\includegraphics[width=0.45\columnwidth]{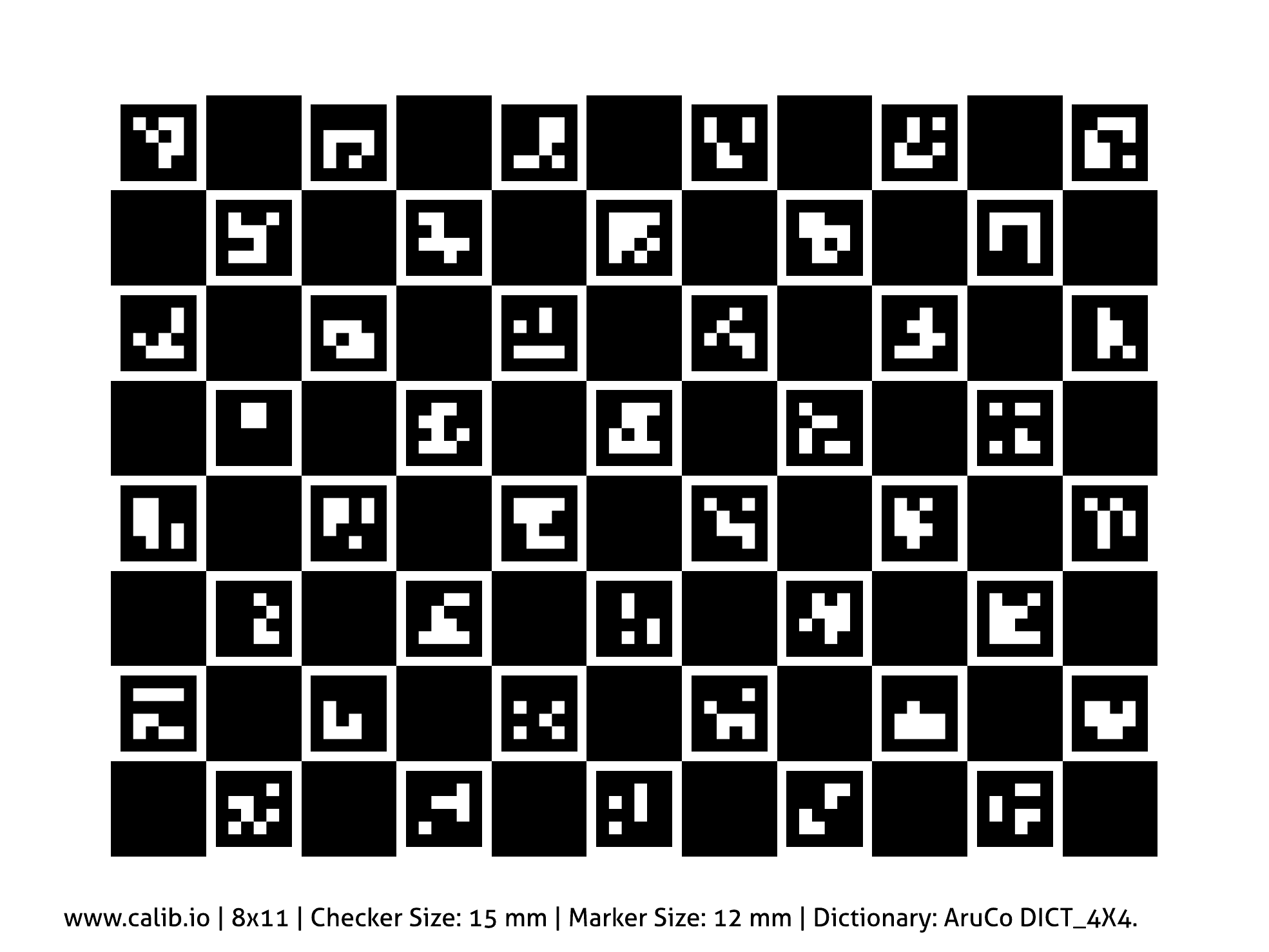}\\
	(c) \hspace{0.45\columnwidth} (d) \\
	\includegraphics[width=0.45\columnwidth]{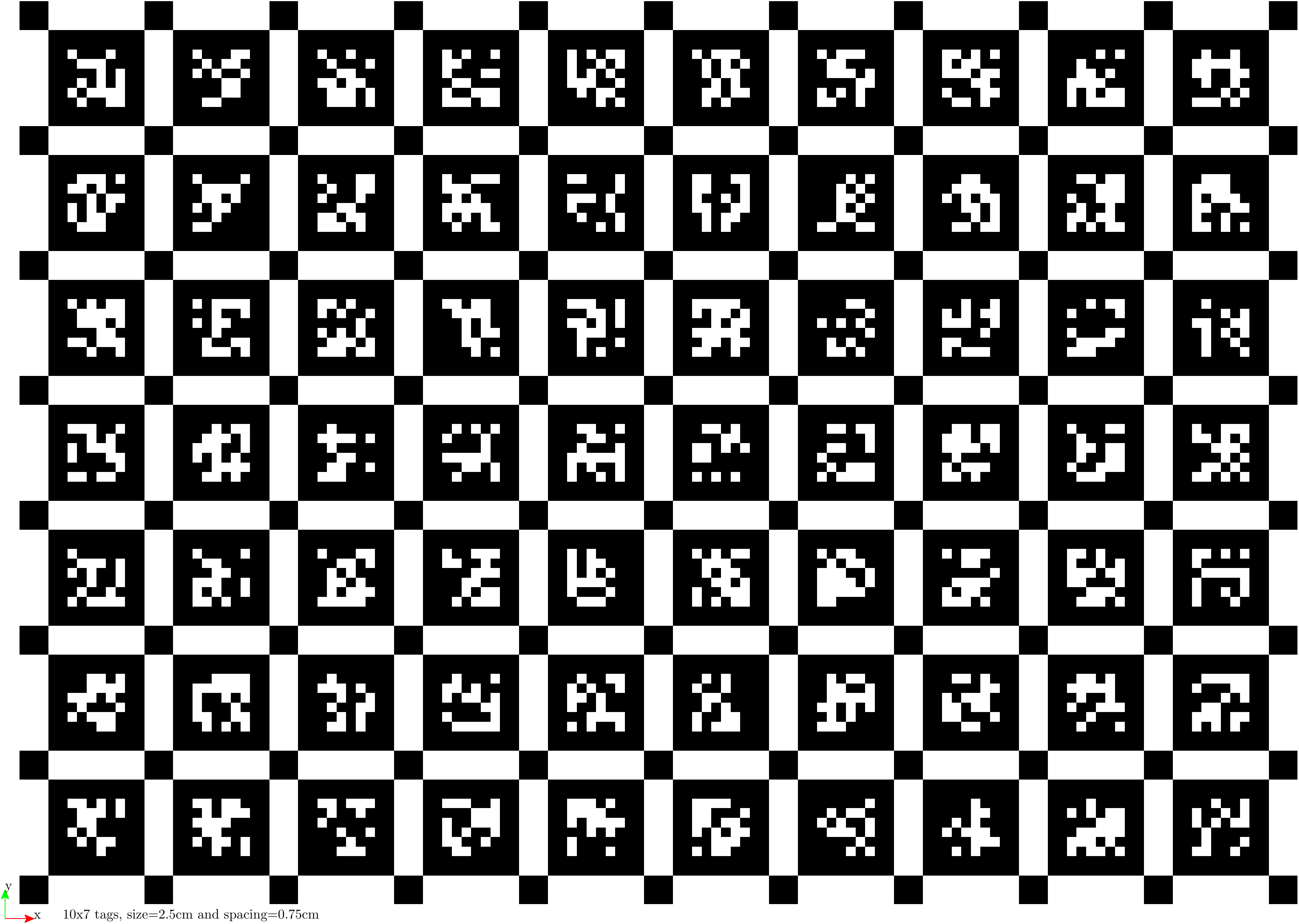}
	\includegraphics[width=0.45\columnwidth]{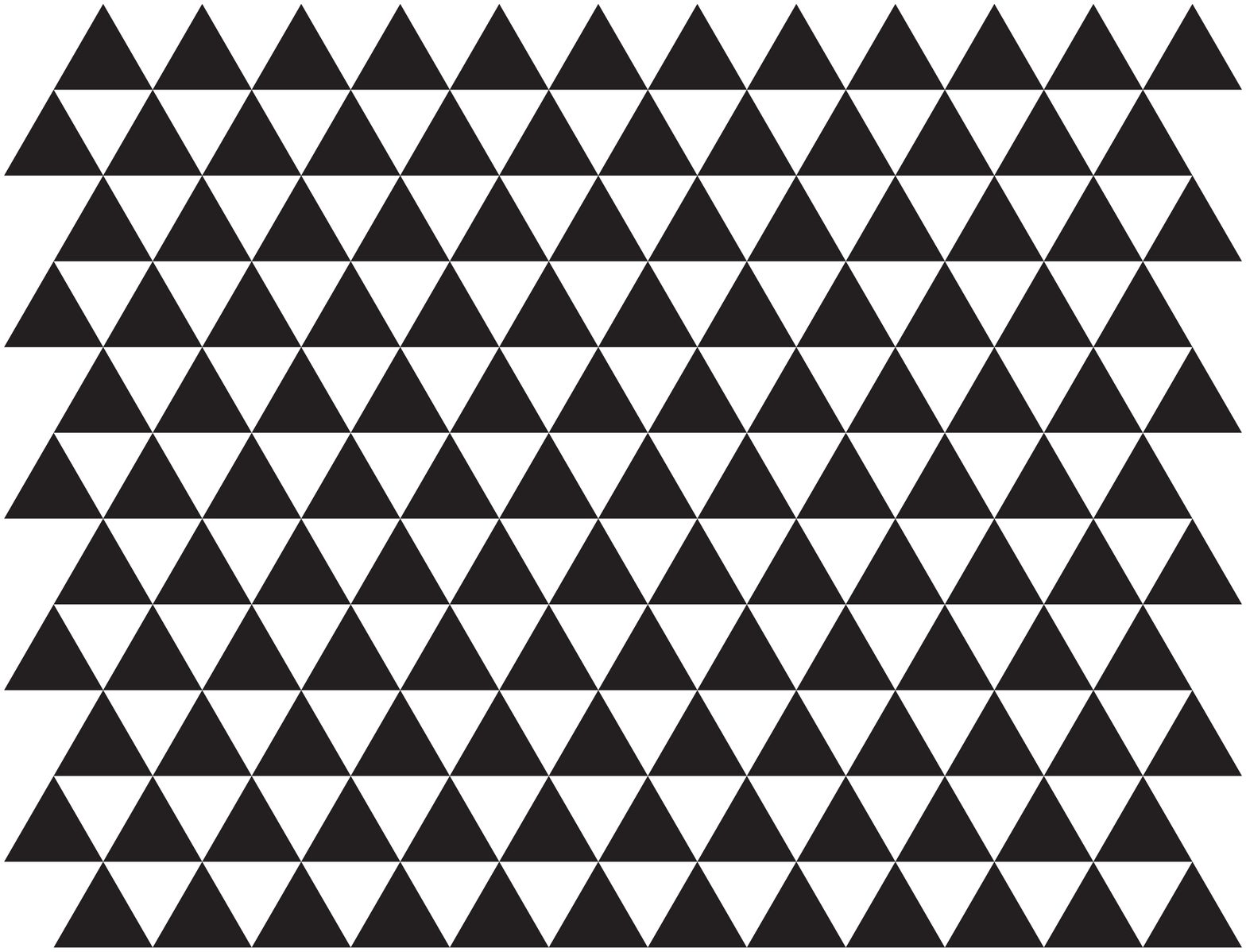} \\
	(e) \hspace{0.45\columnwidth} (f)
	\caption{The passive planar calibration targets: (a) 8$\times$11 checkerboard, (b) 8$\times$11 circle grid, (c) 8$\times$11 asymmetric circle grid, (d) 8$\times$11 Charuco, (e) 7$\times$10 AprilGrid, (f) 10$\times$11 Deltille.}
	\label{fig:targets}
\end{figure}

So far, the printed boards are the most common targets and are widely supported by GCC tools.
They include the checkerboard, the AprilGrid \cite{mayeOnlineSelfcalibrationRobotic2016}, 
the circle grid,
the Charuco \cite{garrido-juradoAutomaticGenerationDetection2014} board, and the recent deltille board \cite{haDeltilleGridsGeometric2017}, etc., as shown in Fig.~\ref{fig:targets}.
Their properties are briefly described below.
There are also numerous customized calibration targets tailored for specific algorithms, e.g., 
the random pattern aggregated from noise at multiple scales in \cite{liMultiplecameraSystemCalibration2013},
 the pattern in \cite{schopsWhyHaving102020} with dense corners for generic models,
the Ecocheck board \cite{abelespeterBoofCV2011}, the PhotoModeler circle board \cite{photomodelertechnologiesPhotoModeler2022}.
A custom board can often be created by combining markers to disambiguate orientations, e.g., the AprilTag, 
and corners invariant to perspective and lens distortion, e.g., formed from repeating squares.
Lists of fiducial markers resilient to rotation can be found in \cite{sagitovARTagAprilTagCALTag2017, garrido-juradoAutomaticGenerationDetection2014}.

\subsubsection{Checkerboard}
The checkerboard is probably the most common calibration target.
It is also known as chessboard. We prefer the name checkerboard which is more general than chessboard.
Many checkerboard detection improvements have been proposed, such as \cite{rufliAutomaticDetectionCheckerboards2008, hengCamOdoCalAutomaticIntrinsic2013}.
The checkerboard requires that the corners inside the board are fully visible in an image so that their coordinates can be uniquely determined.
Though this weakness is reported to be remedied by a few recent methods \cite{geigerAutomaticCameraRange2012, fuersattelOCPADOccludedCheckerboard2016,  haDeltilleGridsGeometric2017, yanAutomaticCheckerboardDetection2018}, 
most current tools have not kept up.
To ensure that the pattern does not look the same after a 180$^\circ$ rotation, a checkerboard with odd rows and even columns or even rows and odd columns is usually used.

\subsubsection{Circle Grid}
A circle grid\cite{mengNewEasyCamera2003} usually consists of an array of circles, symmetrically or asymmetrically distributed (see Fig.~\ref{fig:targets}).
The circle centers are target points for calibration, and can be detected from images based on area, circularity, convexity, inertia \footnote{https://learnopencv.com/blob-detection-using-opencv-python-c/}, etc.
The circle grid has several downsides: first, all circles should be visible in each image; 
second, the detected circle centers suffer from the eccentricity error due to the perspective effect and lens distortion \cite{mallonWhichPatternBiasing2007}.
The eccentricity error is worth attention especially for lenses of large distortion.
Moreover, the symmetric circle grid also has the 180$^\circ$ ambiguity and thus asymmetric circle grid is generally preferred.

\subsubsection{Charuco}
The Charuco board\cite{anCharucoBoardbasedOmnidirectional2018} combines the checkerboard 
and the Aruco tags \cite{garrido-juradoAutomaticGenerationDetection2014} to deal with inaccurate corner positions and occlusions.
As shown in Fig.~\ref{fig:targets}(d), the white squares of checkerboards are occupied by uniquely identifiable Aruco tags.

\subsubsection{AprilGrid}
The Aprilgrid is an array of AprilTag markers \cite{olsonAprilTagRobustFlexible2011} connected by smaller black squares as shown in Fig.~\ref{fig:targets}(e), developed in the Kalibr package \cite{mayeOnlineSelfcalibrationRobotic2016}.
It is resilient to occlusion due to the AprilTag markers, and has accurate positions of corners which are surrounded by two black squares.

\subsubsection{Deltille Grid}
The Deltille grid is a pattern of adjacent regular triangles filled with alternating colors as shown in Fig.~\ref{fig:targets}(f).
It is the only other possible tiling with alternating colors besides the checkerboard tiling.
Its benefits compared to checkerboards are higher corner density and more accurate corner positions.
The wide use of Deltille grids is mainly hindered by the effort to adapt the interfaces of existing calibration tools.

\subsection{Calibration Algorithms}
This section gives a high-level overview of the calibration algorithms as implemented in GCC tools.
According to the used solver, GCC algorithms can be grouped into traditional geometric and learning-based ones.
Generally speaking, geometric approaches are explainable and accurate, whereas the learning-based approaches are intended to be more robust and flexible, e.g.,
\cite{jiCameraCalibrationGenetic2001, yaoResearchCameraCalibration2016, jeongSelfcalibratingNeuralRadiance2021}.

According to the type of calibration targets, GCC algorithms can be grouped into those based on artificial targets, those based on mapped natural scenes, and
self-calibration algorithms without targets.
Calibration with an artificial target is pretty standard and widely supported in GCC packages.
It is typically offline, and usually involves two phases, linear initialization and iterative nonlinear refinement.
Instances of linear initialization are DLT, \cite{ramalingamUnifyingModelCamera2017}, \cite{larssonRevisitingRadialDistortion2019}.
Iterative refinement is exemplified by \cite{tsaiVersatileCameraCalibration1987}, \cite{zhangFlexibleNewTechnique2000}, \cite{unalVariationalApproachProblems2007}, \cite{leeInflightCameraPlatform2010}, \cite{mayeOnlineSelfcalibrationRobotic2016}.
We refer to \cite{salviComparativeReviewCamera2002} for an overview of artificial-target-based methods.

Calibration with objects of known geometry includes
line-based undistortion methods \cite{brauer-burchardtNew2001, ahmedNonmetric2005, bukhariAutomatic2013, xueLearningCalibrateStraight2019}, 
infrastructure-based calibration methods \cite{hengCamOdoCalAutomaticIntrinsic2013, linInfrastructurebasedMulticameraCalibration2020}.
The goal of line-based methods is usually image rectification given a single image.
The infrastructure-based methods require an accurate 3D reconstruction of the site for calibration and rough values for intrinsic parameters and 
are suitable for camera systems with motion constraints.

Broadly speaking, camera self-calibration by using observations of opportunistic landmarks includes recursive refinement methods, methods that recover only camera intrinsic parameters, 
and methods that recover structure, motion and camera intrinsic parameters \cite{kenefick1972analytical}.
Methods in the first group recursively refine calibration parameters and have to start from coarse parameter values, e.g., \cite{keivanConstanttimeMonocularSelfcalibration2014, huaiObservabilityAnalysisKeyframebased2022}.
The second group includes methods like \cite{kenefick1972analytical, faugerasCameraSelfcalibrationTheory1992} and is reviewed in \cite{hemayedSurveyCameraSelfcalibration2003}.
Methods in the last group usually rely on bundle adjustment,
thus, they typically have the best accuracy among self-calibration methods and are commonly supported in SfM packages, e.g., colmap \cite{schonbergerStructurefrommotionRevisited2016}.

\begin{table*}[!htbp]
	\centering
	\caption{An (non-exhaustive) overview of geometric camera calibration (GCC) tools. Supported camera models are mostly explained in Section~\ref{subsec:cammodels}. rad: radial, tan: tangential, GUI: graphical user interface.}
	\label{tab:tools}
	\begin{tabular}{lllccccl}
		\hline
		\multicolumn{1}{c}{\textbf{\begin{tabular}[c]{@{}c@{}}GCC tools\end{tabular}}} &
		\multicolumn{1}{c}{\textbf{\begin{tabular}[c]{@{}c@{}}Supported\\ models\end{tabular}}} &
		\multicolumn{1}{c}{\textbf{\begin{tabular}[c]{@{}c@{}}Supported\\ targets\end{tabular}}} &
		\textbf{\begin{tabular}[c]{@{}c@{}}Outlier\\handling\end{tabular}} &
		\textbf{\begin{tabular}[c]{@{}c@{}}Param.\\ std. dev.\end{tabular}} &
		\textbf{Language} &
		\textbf{\begin{tabular}[c]{@{}c@{}}Open\\ source\end{tabular}} &
		\multicolumn{1}{c}{\textbf{Other features}} \\ \hline
		AprilCal \cite{olsonAprilTagRobustFlexible2011} &
		KB-8; pinhole rad tan &
		AprilTag grid &
		No &
		No &
		Java &
		Yes &
		GUI; interactive \\ \hline
		BabelCalib \cite{lochmanBabelCalibUniversalApproach2021} &
		\begin{tabular}[c]{@{}l@{}}division; division even; \\DS; EUCM; FOV; KB-8;\\ pinhole rad; UCM\end{tabular} &
		\begin{tabular}[c]{@{}l@{}}agnostic\end{tabular} &
		Huber &
		No &
		MATLAB &
		Yes & \begin{tabular}[c]{@{}l@{}}allow multiple\\planar targets\end{tabular}
		\\ \hline
		Basalt \cite{usenkoDoubleSphereCamera2018} &
		\begin{tabular}[c]{@{}l@{}}DS; EUCM; FOV;\\KB-8; UCM\end{tabular}
		&
		AprilGrid &
		Huber &
		No &
		C++ &
		Yes &
		\begin{tabular}[c]{@{}l@{}}efficient;\\GUI; modular\end{tabular}
		\\ \hline
		BoofCV \cite{abelespeterBoofCV2011} &
		\begin{tabular}[c]{@{}l@{}}KB-23; pinhole rad tan;\\Mei\end{tabular} &
		\begin{tabular}[c]{@{}l@{}}checkerboard\\circle grid\\Ecocheck\end{tabular} &
		No &
		No &
		Java &
		Yes &
		\begin{tabular}[c]{@{}l@{}}Android portable;\\ GUI\end{tabular} \\ \hline
		calib.io \cite{calib.ioCameraCalibrator2022} &
		\begin{tabular}[c]{@{}l@{}}B-spline model; division;\\DS; EUCM; FOV; KB-8;\\ pinhole rational rad tan\\thin prism\end{tabular} &
		\begin{tabular}[c]{@{}l@{}}Charuco;\\checkerboard\end{tabular} &
		Huber &
		Yes &
		? &
		No &
		\begin{tabular}[c]{@{}l@{}}correlation analysis;\\GUI; multi-camera; \\multiple targets\end{tabular} \\ \hline
		\begin{tabular}[c]{@{}l@{}}Calibu by ARPG\\ \cite{universityofcoloradoCalibu2022}\end{tabular} &
		\begin{tabular}[c]{@{}l@{}}FOV; KB-8;\\pinhole rational rad\end{tabular}
		 &
		circle grid &
		No &
		No &
		C++ &
		Yes &
		\\ \hline
		camcalib \cite{ivisogmbhCamcalib2022} &
		\begin{tabular}[c]{@{}l@{}}pinhole rad tan; \\ KB-8; DS\end{tabular} &
		\begin{tabular}[c]{@{}l@{}}AprilGrid;\\Charuco\end{tabular} &
		likely &
		No &
		? &
		No &
		\begin{tabular}[c]{@{}l@{}}GUI;\\multi-camera\end{tabular} \\ \hline
		Camodocal \cite{hengCamOdoCalAutomaticIntrinsic2013} &
		\begin{tabular}[c]{@{}l@{}}pinhole rad tan; Mei;\\KB-8\end{tabular} &
		checkerboard &
		Cauchy &
		No &
		C++ &
		Yes &
		\begin{tabular}[c]{@{}l@{}}Improved checkerboard \\ detector; modular;\\stereo camera\end{tabular} \\ \hline
		\begin{tabular}[c]{@{}l@{}}Colmap\\\cite{schonbergerStructurefrommotionRevisited2016}\end{tabular}
		&
		\begin{tabular}[c]{@{}l@{}}pinhole rad tan thin prism, \\KB-8; FOV\end{tabular} &
		N/A &
		\begin{tabular}[c]{@{}c@{}}soft L1, \\ Cauchy\end{tabular} &
		No &
		C++ &
		Yes &
		\begin{tabular}[c]{@{}l@{}}GUI; self-calibration;\\ well-documented\end{tabular} \\ \hline
		\begin{tabular}[c]{@{}l@{}}Generic camera\\calibration\cite{schopsWhyHaving102020} \end{tabular} &
		\begin{tabular}[c]{@{}l@{}}central generic; \\non-central generic;\\ pinhole rational rad tan\\thin prism;\end{tabular} &
		\begin{tabular}[c]{@{}l@{}}custom grid with\\an AprilTag\end{tabular} &
		Huber &
		No &
		C++ &
		Yes &
		\begin{tabular}[c]{@{}l@{}}accurate;\\board deformation\end{tabular}\\ \hline
		\begin{tabular}[c]{@{}l@{}}Learning CCS\\\cite{zhangLearningbased2022}\end{tabular} &
		pinhole rad &
		checkerboard &
		trim &
		No &
		Python &
		Yes &
		\\ \hline
		\begin{tabular}[c]{@{}l@{}}libomnical \cite{schonbeinCalibratingCenteringQuasicentral2014}\end{tabular} &
		\begin{tabular}[c]{@{}l@{}}a centered model;\\a geometric model;\\Mei; Scaramuzza\end{tabular} &
		\begin{tabular}[c]{@{}l@{}}checkerboard\end{tabular} &
		No &
		No &
		MATLAB &
		Yes &
		\begin{tabular}[c]{@{}l@{}}\end{tabular} \\ \hline
		\begin{tabular}[c]{@{}l@{}}Kalibr \cite{mayeOnlineSelfcalibrationRobotic2016} /\\ TartanCalib \cite{duisterhofTartanCalibIterativeWideangle2023}\end{tabular} &
		\begin{tabular}[c]{@{}l@{}}DS; EUCM; FOV;\\pinhole equidistant;\\ pinhole rad tan; Mei\end{tabular} &
		\begin{tabular}[c]{@{}l@{}}AprilGrid;\\checkerboard\end{tabular} &
		trim &
		Yes &
		C++ &
		Yes &
		\begin{tabular}[c]{@{}l@{}}multi-camera;\\board deformation\end{tabular} \\ \hline
		\begin{tabular}[c]{@{}l@{}}MATLAB camera\\calibrator \cite{scaramuzzaToolboxEasilyCalibrating2006}\end{tabular} &
		\begin{tabular}[c]{@{}l@{}}
		pinhole rad tan\\Scaramuzza\end{tabular} &
		\begin{tabular}[c]{@{}l@{}}AprilTag grid;\\checkerboard;\\circle grid\end{tabular} &
		likely &
		Yes &
		MATLAB &
		Yes &
		\begin{tabular}[c]{@{}l@{}}GUI; modular;\\stereo camera;\\
			well-documented\end{tabular} \\ \hline
\begin{tabular}[c]{@{}l@{}}MC-Calib \cite{rameauMCCalibGenericRobust2022}\end{tabular} &
		KB-8; pinhole rad tan &
		\begin{tabular}[c]{@{}l@{}}Charuco\end{tabular} &
		Huber &
		No &
		C++ &
		Yes &
		\begin{tabular}[c]{@{}l@{}}board deformation;\\multi-camera\end{tabular} \\ \hline
		\begin{tabular}[c]{@{}l@{}}Metashape calibrator\\ \cite{agisoftllcAgisoftMetashape2022}\end{tabular} &
		\begin{tabular}[c]{@{}l@{}}a custom fisheye model;\\pinhole rad tan\end{tabular} &
		checkerboard &
		? &
		Yes &
		\begin{tabular}[c]{@{}c@{}}C++ /\\ Python\end{tabular} &
		No & \begin{tabular}[c]{@{}l@{}}correlation analysis;\\GUI\end{tabular} \\ \hline
		Metashape \cite{agisoftllcAgisoftMetashape2022} &
		\begin{tabular}[c]{@{}l@{}}a custom fisheye model;\\pinhole rad tan\end{tabular}&
		N/A &
		Yes &
		No &
		C++ &
		No &
		GUI; self-calibration \\ \hline
		mrcal \cite{mrcal} &
		\begin{tabular}[c]{@{}l@{}}pinhole rational rad tan\\thin prism;\\splined stereographic \end{tabular} &
		checkerboard &
		trim &
		No &
		\begin{tabular}[c]{@{}c@{}}C++ /\\ Pyrhotn\end{tabular} &
		Yes &
		\begin{tabular}[c]{@{}l@{}}board deformation;\\ uncertainty analysis\end{tabular} \\ \hline
		\begin{tabular}[c]{@{}l@{}}MRPT camera calib \\ \cite{universityofmalagaMRPTCameraCalib2022} / \\ OpenCV\end{tabular} &
		pinhole rad tan &
		checkerboard &
		No &
		No &
		C++ &
		Yes &
		\begin{tabular}[c]{@{}l@{}}GUI; multi-\\checkerboard detector\end{tabular} \\
		\hline
		\begin{tabular}[c]{@{}l@{}}Omni calibrator\\ \cite{meiSingleViewPoint2007} \end{tabular} &
		\begin{tabular}[c]{@{}l@{}}Mei\end{tabular} &
		\begin{tabular}[c]{@{}l@{}}checkerboard\end{tabular} &
		\begin{tabular}[c]{@{}l@{}}No\end{tabular} &
		Yes &
		MATLAB &
		Yes &
		\begin{tabular}[c]{@{}l@{}}GUI; manual\\checkerboard detection\end{tabular} \\ \hline
		\begin{tabular}[c]{@{}l@{}}ROS camera calibrator\\ \cite{rosROSCameraCalibration2022} /\\ OpenCV\end{tabular} &
		\begin{tabular}[c]{@{}l@{}}KB-8; pinhole rational rad \\tan thin prism; Mei\end{tabular} &
		\begin{tabular}[c]{@{}l@{}}Charuco\\checkerboard\\circle grid\end{tabular} &
		\begin{tabular}[c]{@{}l@{}}No, but\\trim (Mei)\end{tabular} &
		No &
		Python &
		Yes &
		\begin{tabular}[c]{@{}l@{}}efficient; GUI;\\modular; stereo camera;\\well-documented\end{tabular} \\ \hline
		\begin{tabular}[c]{@{}l@{}}PhotoModeler calibrator\\ \cite{photomodelertechnologiesPhotoModeler2022}\end{tabular}   &
		backward pinhole rad tan &
		\begin{tabular}[c]{@{}l@{}}customized\\circle grid\end{tabular} &
		likely &
		Yes &
		? &
		No &
		\begin{tabular}[c]{@{}l@{}}correlation analysis;\\GUI\end{tabular} \\ \hline
		Pix4DMapper \cite{pix4ds.a.PIX4Dmapper2022}&
		\begin{tabular}[c]{@{}l@{}}pinhole rad tan;\\ adapted Scaramuzza\end{tabular} &
		N/A &
		Yes &
		No &
		C++ &
		No &
		GUI; self-calibration \\ \hline
		SCNeRF \cite{jeongSelfcalibratingNeuralRadiance2021} & KB-6 & N/A & trim & No & Python & Yes & self-calibration 
		\\ \hline
		vidar \cite{fangSelfsupervisedCameraSelfcalibration2022} &
		DS; EUCM; UCM &
		N/A &
		N/A &
		No &
		Python &
		Yes &
        self-calibration\\ \hline
	\end{tabular}%
\end{table*}

\section{GCC Tools}
\label{sec:tools}
This section reviews tools developed for wide-angle camera calibration.
These tools mainly implement algorithms using artificial targets or target-free bundle adjustment.
Several learning-based GCC tools are also cited as examples from this active research field.
Since our focus is on intrinsic calibration, tools solely for extrinsic calibration are left out, e.g.,
\cite{linInfrastructurebasedMulticameraCalibration2020, Liu2016, oliveiraGeneralApproachExtrinsic2019}. 
An extensive list of GCC tools to our knowledge is given in Table~\ref{tab:tools}.
For brevity, the table only list a few photogrammetric software tools which unanimously allow self-calibration.
This table can serve as a reference in choosing a proper GCC tool and hopefully can help prevent duplication of development effort.

We assess a GCC tool based on characteristics which are grouped into accessibility and quality evaluation.
For accessibility, these characteristics include supported camera models and targets, stereo / multiple camera support, the user interface, source availability, and the coding language.
Usually, a graphical user interface (GUI) is more accessible than a command line interface to an average user.
When a tool is open-source or modular, it is easy to extend it to other camera models and calibration targets.
The coding language usually implies the execution efficiency and the community support.

From quality evaluation, we look at the outlier strategy and the availability of covariance output.
The outlier strategy dictates how to handle outliers in detected corners which may deviate from their true positions by a few pixels.
For quality check, all calibration tools output some metric based on reprojection errors, such as the mean reprojection error and the root mean square (RMS) reprojection error.
However, these metrics are highly dependent on the used image corners, and thus are inadequate to compare results from different methods \cite{duisterhofTartanCalibIterativeWideangle2023}.
The covariance output is an quality indicator besides these metrics, and directly links to the correlation analysis \cite{luhmannCloseRangePhotogrammetry2006}.
Next, we describe several popular calibration tools in terms of these characteristics.

\subsection{BabelCalib}
The monocular camera calibrator, BabelCalib, employs a back-projection model as a proxy for a variety of radial-symmetric forward camera models, including the pinhole radial distortion model \eqref{eq:pinhole-radtan}, DS \eqref{eq:ds}, EUCM \eqref{eq:eucm}, FOV \eqref{eq:fov}, KB-8 \eqref{eq:kb8}, and UCM \eqref{eq:ucm}.
In practice, the back-projection model, a two-parameter division model of even degrees \eqref{eq:division}, can be obtained by linear solvers, 
and then the desired camera models can be regressed from the division model.
BabelCalib is agnostic to the calibration targets, 
supports calibration with multiple targets, 
and handles outliers with the Huber loss.

\subsection{Basalt}
The Basalt package \cite{usenkoDoubleSphereCamera2018} can carry out monocular camera calibration, supporting camera models including DS, EUCM, FOV, KB-8, and UCM.
Its default calibration target is the AprilGrid.
A Levenberg-Marquardt algorithm is implemented in Basalt for robust calibration with the Huber loss.
With neat use of C++ templates, it is a lean and fast tool.

\subsection{calio.io}
The commercial calibration tool by calio.io comes with an intuitive GUI, supports a variety of camera models, including the pinhole rational radial tangential model with the thin prism effect \eqref{eq:pinhole-radtan-prism},
the division model \eqref{eq:division}, DS, KB-8, FOV, EUCM, and a B-spline camera model, and supports many calibration targets including the checkerboard and the Charuco board.
Moreover, it allows calibrating multiple cameras with multiple targets,
and optimizing the target points to deal with board deformation, and deals with outliers with the Huber loss.

\subsection{Camodocal}
The Camodocal package supports monocular and stereo GCC with models including the pinhole radial tangential model \eqref{eq:pinhole-radtan}, KB-8, and Mei \eqref{eq:mei}.
By default, it supports the checkerboard, but it is relatively easy to extend to other targets.
It uses the Cauchy loss to deal with outliers.

\subsection{Kalibr}
Kalibr is a popular GCC tool that can select informative images for calibration \cite{mayeOnlineSelfcalibrationRobotic2016}.
It supports projection models including pinhole projection \eqref{eq:pinhole}, UCM, EUCM, and DS, and distortion models including radial tangential distortion, equidistant distortion, and FOV.
As mentioned for \eqref{eq:kb8}, the KB-8 model in Kalibr discards points of non-positive depth $Z_c$.
The supported targets include checkerboards and AprilGrids.
Outliers are handled by removing corners of reprojection errors exceeding a certain threshold.
This tool has been extended to deal with the rolling shutter effect \cite{huaiContinuoustimeSpatiotemporalCalibration2022}, 
and to better detect corners in images of high distortion lenses \cite{duisterhofTartanCalibIterativeWideangle2023}.

\subsection{MATLAB Camera Calibrator}
The MATLAB camera calibrator \cite{mathworksinc.MATLABComputerVision2021} supports both monocular and stereo camera calibration with both the pinhole radial tangential model \eqref{eq:pinhole-radtan} and the Scaramuzza model \eqref{eq:scaramuzza}.
It can be seen as a superset of \cite{bouguetCameraCalibrationToolbox2001} and \cite{scaramuzzaToolboxEasilyCalibrating2006}.
The supported targets by default are checkerboards, circle grids, and AprilTag grids.
With its modular design, it is easy to use other calibration targets, e.g., the AprilGrid.
The MATLAB calibrator has an easy-to-follow GUI and many visualization functions.

\subsection{ROS Camera Calibrator}
The OpenCV library provides functions for calibrating monocular and stereo cameras with 
the pinhole rational radial tangential model with the thin prism effect \eqref{eq:pinhole-radtan-prism}, the KB-8 model for fisheye cameras, and the Mei model for omnidirectional cameras.
The omnidirectional module in OpenCV also supports a multi-camera setup and can be seen as a reimplementation of the MATLAB tool in \cite{liMultiplecameraSystemCalibration2013}.
The current KB-8's realization in OpenCV does not support points of non-positive depth.
The calibration functions in OpenCV do not have outlier handling schemes, but its omnidirectional module removes images of large total reprojection errors in calibration.

Several programs have been developed on top of OpenCV, such as the ROS camera calibrator \cite{rosROSCameraCalibration2022} and 
the MRPT camera calibrator \cite{universityofmalagaMRPTCameraCalib2022}.
The ROS camera calibrator is a thin wrap of OpenCV calibration functions, can run in both interactive and batch mode, and supports checkerboards, circle grids, and Charuco boards.
Besides wrapping the OpenCV functions, the MRPT camera calibrator extends the checkerboard detection to support multiple checkerboards.

\subsection{Self-Calibration Tools with SfM}
Self-calibration is usually based on a SfM pipeline which is realized in commercial software or open source programs.
For space, we limit the discussion to several representatives of the two groups.
Professional photogrammetric packages usually support self-calibration, for instance, the Metashape by Agisoft \cite{agisoftllcAgisoftMetashape2022},
the calibrator in PhotoModeler \cite{photomodelertechnologiesPhotoModeler2022}, and the Pix4D mapper \cite{pix4ds.a.PIX4Dmapper2022}.
The Metashape realizes both checkerboard-based calibration and self-calibration using natural landmarks within its SfM pipeline.
Both methods support the pinhole radial tangential model and a customized fisheye model that is made of
the equidistant projection and the radial tangential distortion.
The calibration tool in PhotoModeler adopts the inverse pinhole radial tangential model \eqref{eq:pinhole-radtan-inv}, and 
supports target-based calibration with either multiple boards each of five RAD (Ringed Automatically Detected) tags or 
a single board of a circle grid with four non-ringed coded tags.
When the scene to be reconstructed is much larger than the printed targets, a self-calibration of the camera in the field may be conducted with PhotoModeler.
The Pix4D mapper can also estimate the camera intrinsic parameters with a collection of images of natural scenes.
It supports the pinhole radial tangential model and an adapted Scaramuzza model.

The open-source SfM packages also widely support camera self-calibration, such as the popular colmap, 
and the recent Self-Calibration package based on the Neural Radiance Field, SCNeRF \cite{jeongSelfcalibratingNeuralRadiance2021}.
Based on geometric bundle adjustment, colmap \cite{schonbergerStructurefrommotionRevisited2016} supports camera models including the pinhole radial tangential model with the thin prism distortion, KB-8, and FOV.
The learning-based SCNeRF considers both geometric and photometric consistency in constructing the implicit scene geometry and estimating the camera parameters.

\section{Evaluation of Target-Based GCC Tools}
\label{sec:results}
This section evaluates six popular target-based GCC tools on simulated and real wide-angle camera data acquired by cameras of varying AOVs, 
to show their extensibility and repeatability.

\subsection{Data Acquisition}
The real data were captured by an UI-3251LE-M-GL camera of a 1/1.8'' sensor 
from the IDS Imaging, fitted with six fixed focus lenses listed in Table~\ref{tab:lenses}, leading to varying camera DAOVs from 90$^\circ$ to 194$^\circ$.
Notably, in focal length, the 90$^\circ$ lens resembles lenses on smartphones whose actual focal lengths are about 4 mm.
Also, empirically, the calibrated focal lengths are close to the physical focal lengths (from datasheets) in Table~\ref{tab:lenses} in pixels.
The camera can capture grayscale images at 25 frames/second and resolution 1600$\times$1200 in global shutter mode.
Prior to data capture, the exposure time was set to 5 ms to reduce motion blur.
For each lens, the camera was gently moved in front of an AprilGrid, passing through a variety of poses.
We chose the AprilGrid since it is accurate \cite{dossantoscesarEvaluationArtificialFiducial2015}, widely used, and resilient to occlusions, 
among the reviewed calibration targets.
Three sequences each of a minute were recorded for each lens.
From each sequence, three subsequences each of 200 frames were uniformly drawn without replacement.
This resulted in $54 = 6\times3\times3$ calibration sequences for six lenses.

\begin{table}[!htp]
	\centering
	\caption{Lenses used on the industrial camera UI3251LE-M-GL which has a CMOS sensor of 1600$\times$1200 4.5 $\mu$m square pixels. The camera's DAOV is computed using the captured images.}
	\label{tab:lenses}
	\begin{tabular}{lccccc}
		\hline
		\multicolumn{1}{c}{\multirow{2}{*}{Lens}} & \multirow{2}{*}{\begin{tabular}[c]{@{}c@{}}Max. \\ sensor\end{tabular}} & \multirow{2}{*}{f-num} & \multirow{2}{*}{\begin{tabular}[c]{@{}c@{}}DAOV\\ ($^\circ$)\end{tabular}} & \multicolumn{2}{c}{Focal length} \\ \cline{5-6} 
		\multicolumn{1}{c}{}                      &                                                                                &                         &                                                                            & \multicolumn{1}{c}{(mm)}  & (px) \\ \hline
		FocVis S04525                               & 1/1.8''                                                                        & 2.5                   & 90                                                                       & \multicolumn{1}{c}{4.5}   & 1000 \\ \hline
		Matrix Vision E1M3518                       & 1/2.5''                                                                        & 1.8                   & 94                                                                       & 
		\multicolumn{1}{c}{3.5}   & 778  \\ \hline
		Lensagon BM4218                             & 1/3''                                                                          & 1.8                   & 103                                                                       & \multicolumn{1}{c}{4.2}   & 933  \\ \hline
		Lensagon BM4018S118                         & 1/1.8''                                                                        & 1.8                   & 127                                                                        & \multicolumn{1}{c}{4.0}   & 889  \\ \hline
		Lensagon BT2120                             & 1/3''                                                                          & 2.0                   & 164                                                                        & \multicolumn{1}{c}{2.1}   & 467  \\ \hline
		ZLKC MTV185IR12MP                           & 1/1.8''                                                                        & 2.0                   & 194                                                                        & \multicolumn{1}{c}{1.85}  & 411  \\ \hline
	\end{tabular}
\end{table}

\begin{figure}[!tbp]
	\centering
	\includegraphics[width=0.48\columnwidth]{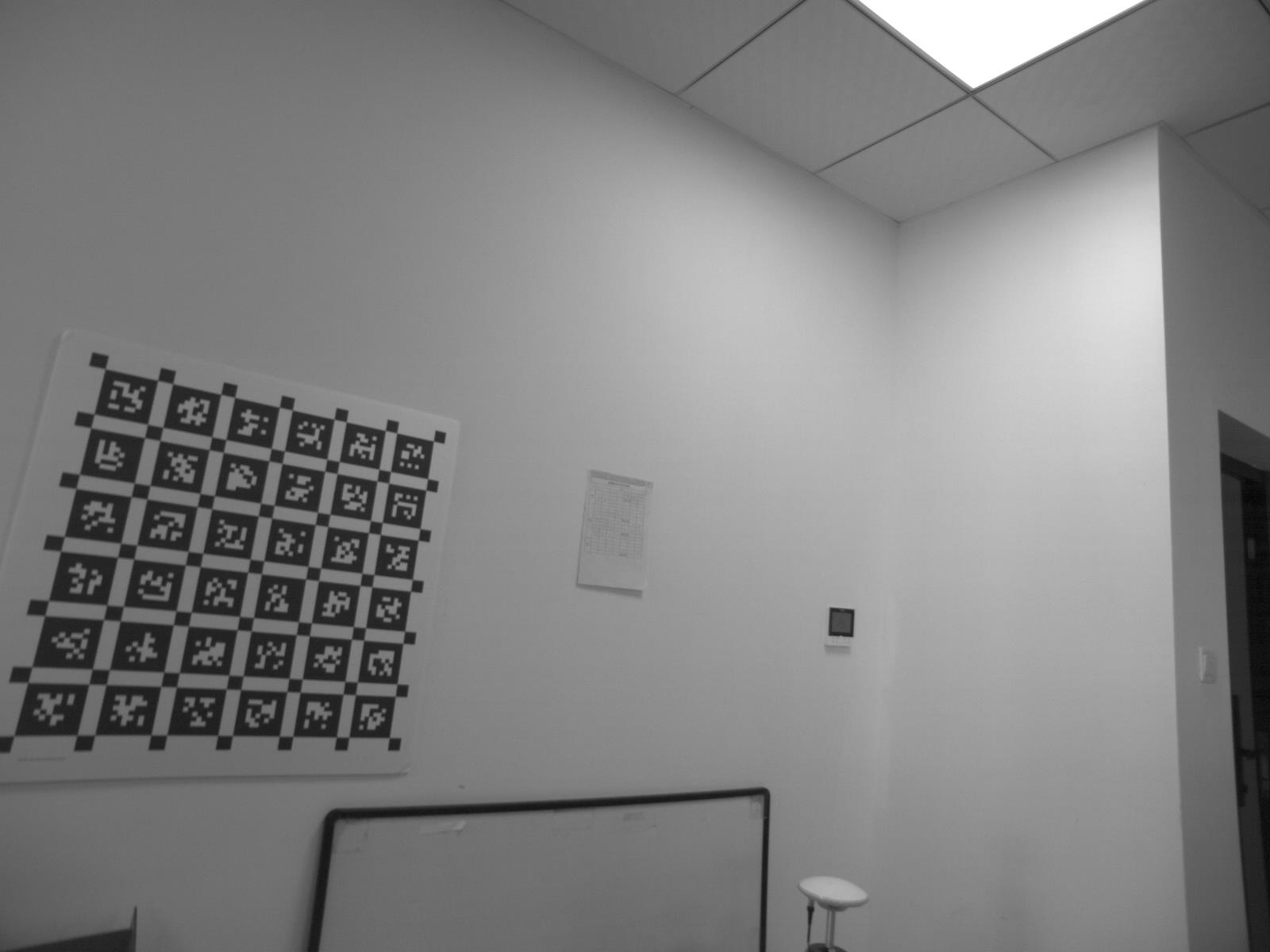}
	\includegraphics[width=0.48\columnwidth]{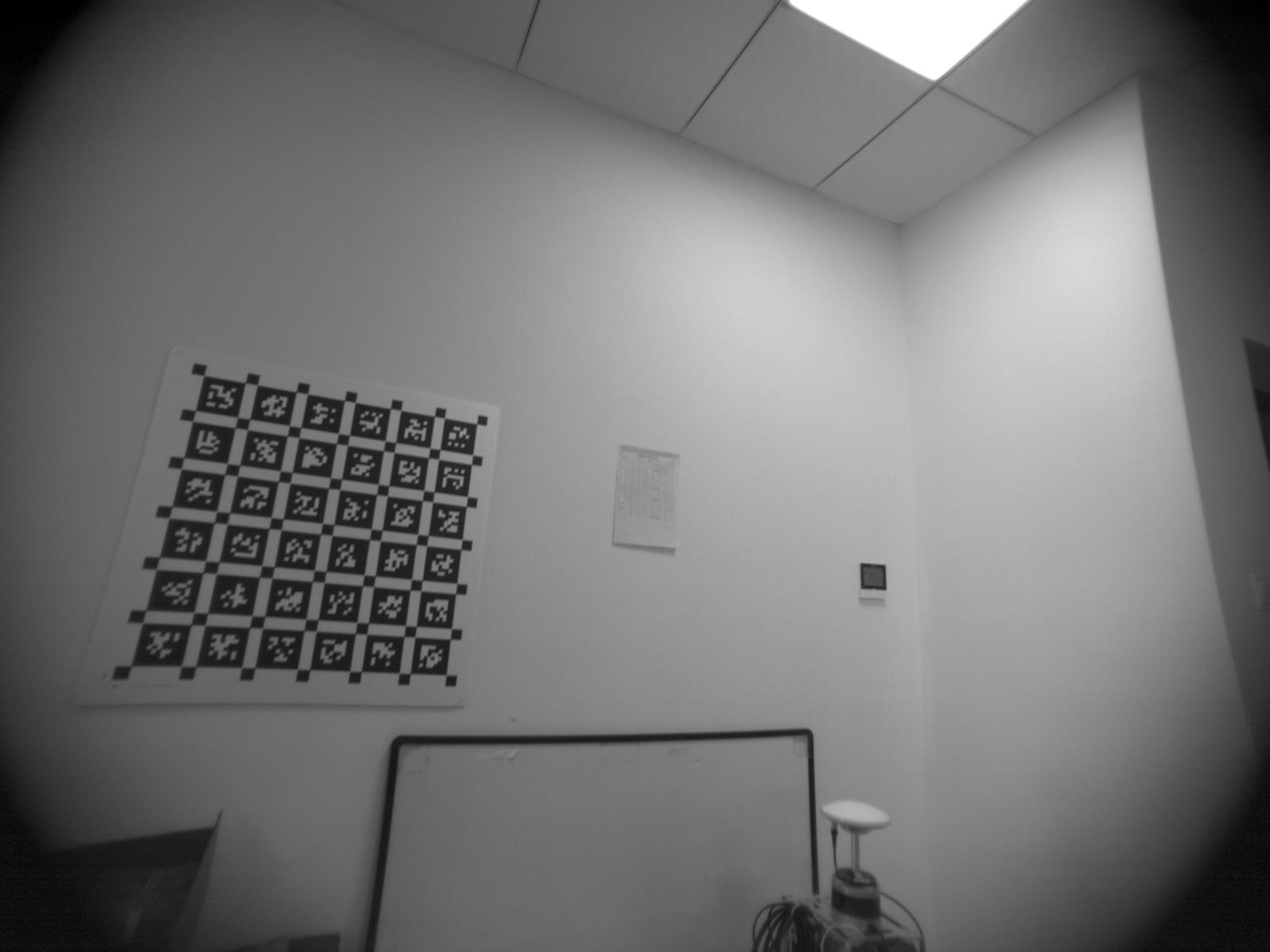} \\ \includegraphics[width=0.48\columnwidth]{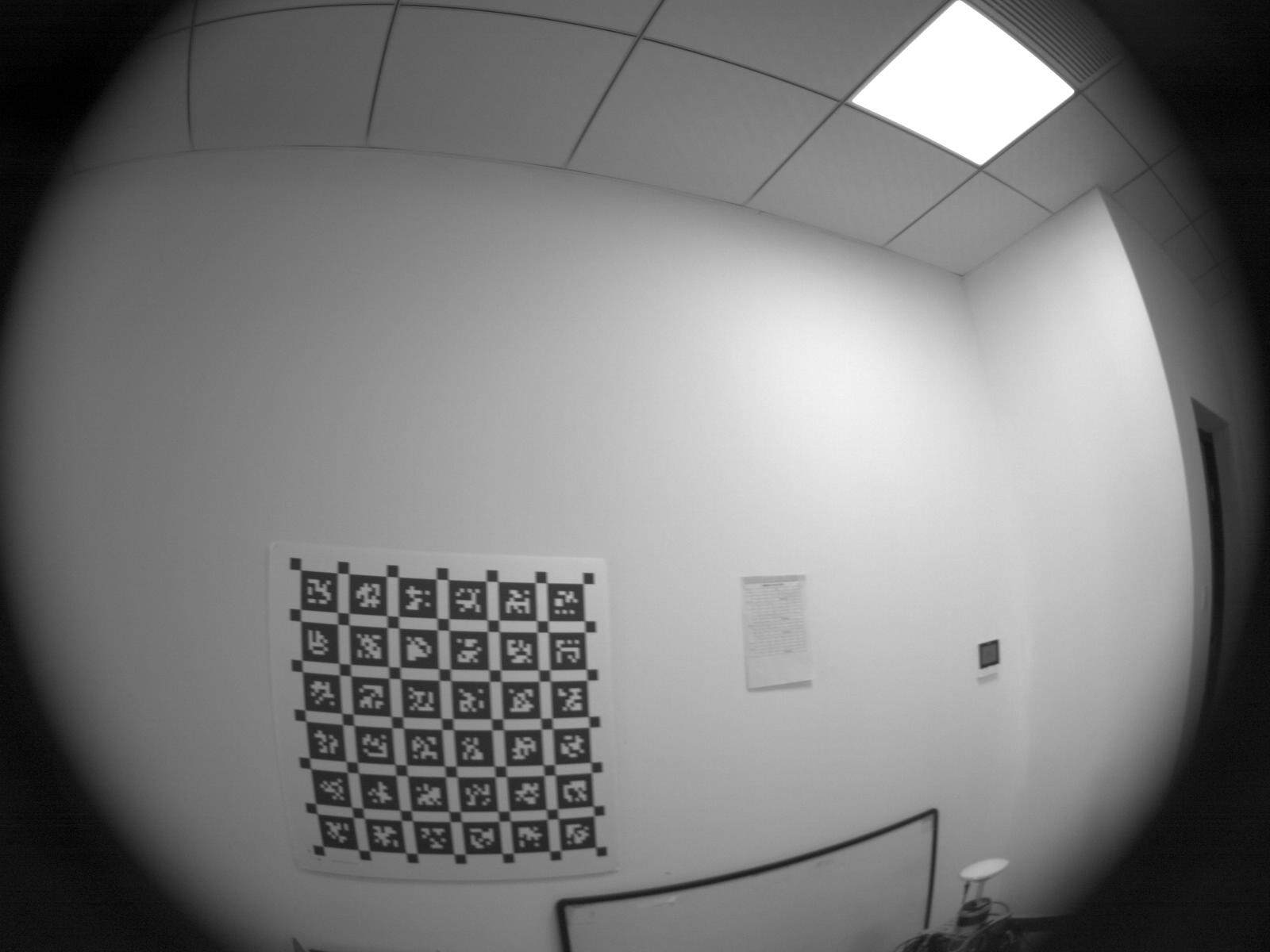}
	\includegraphics[width=0.48\columnwidth]{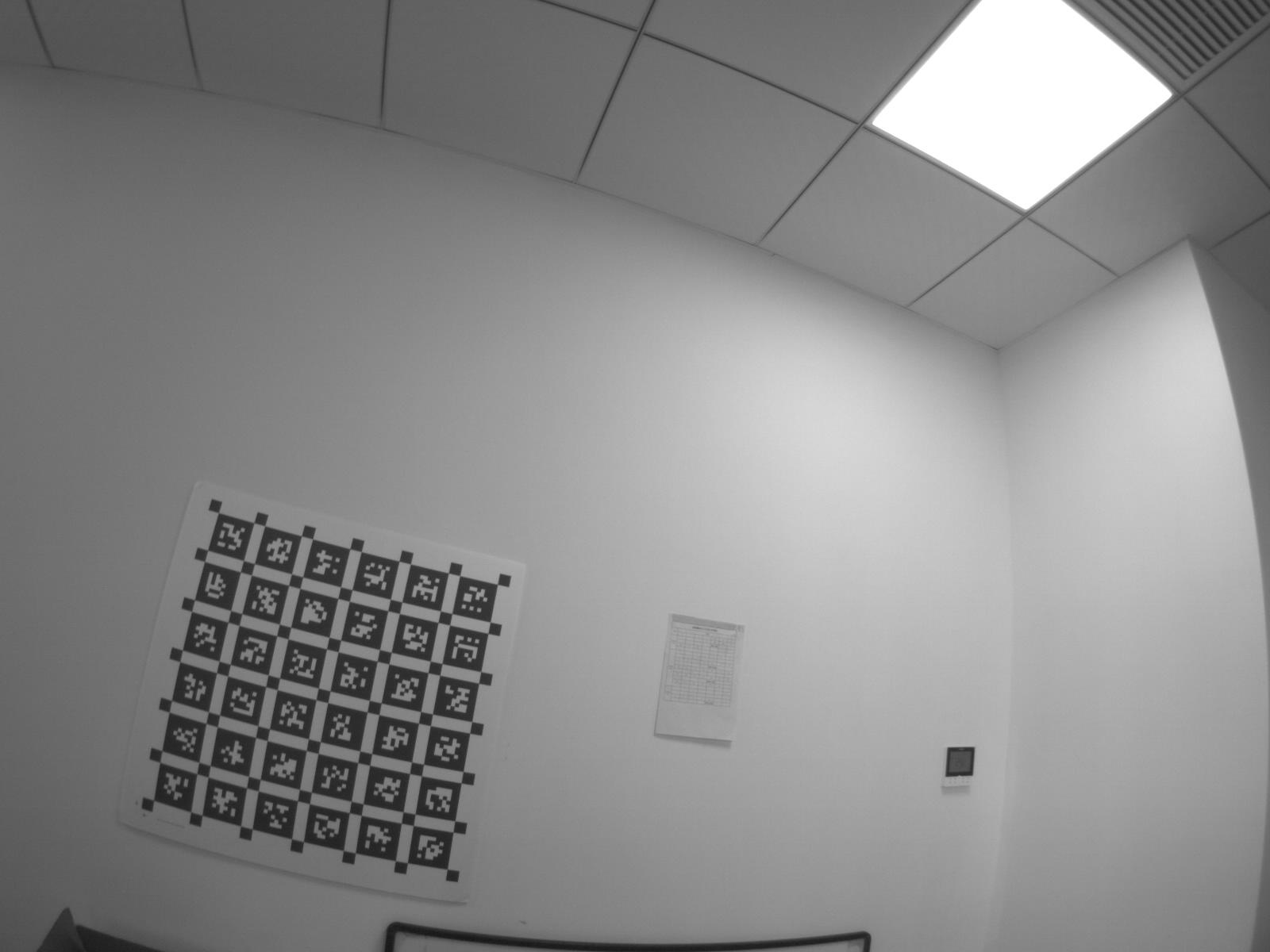} \\
	\includegraphics[width=0.48\columnwidth]{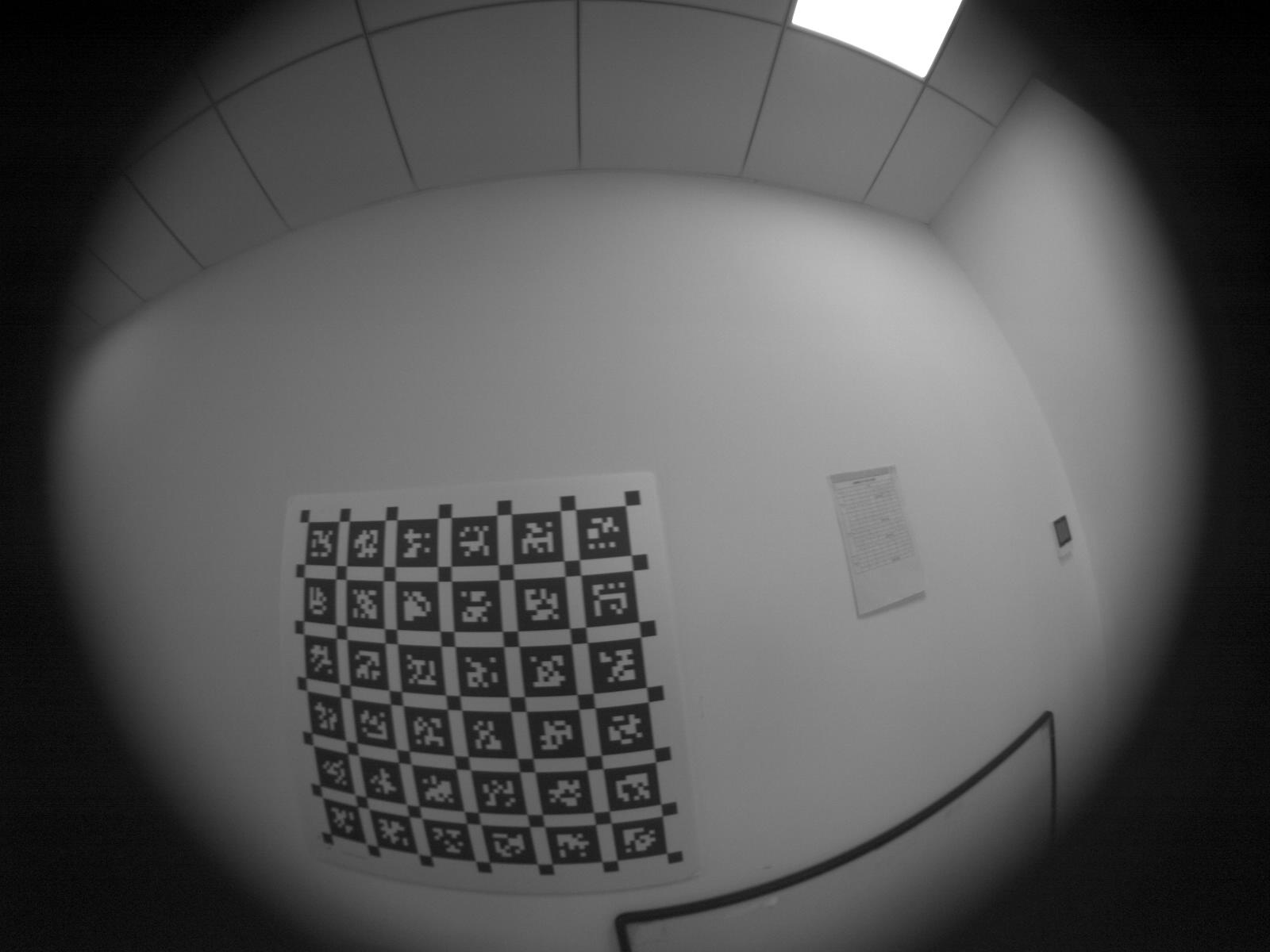}
	\includegraphics[width=0.48\columnwidth]{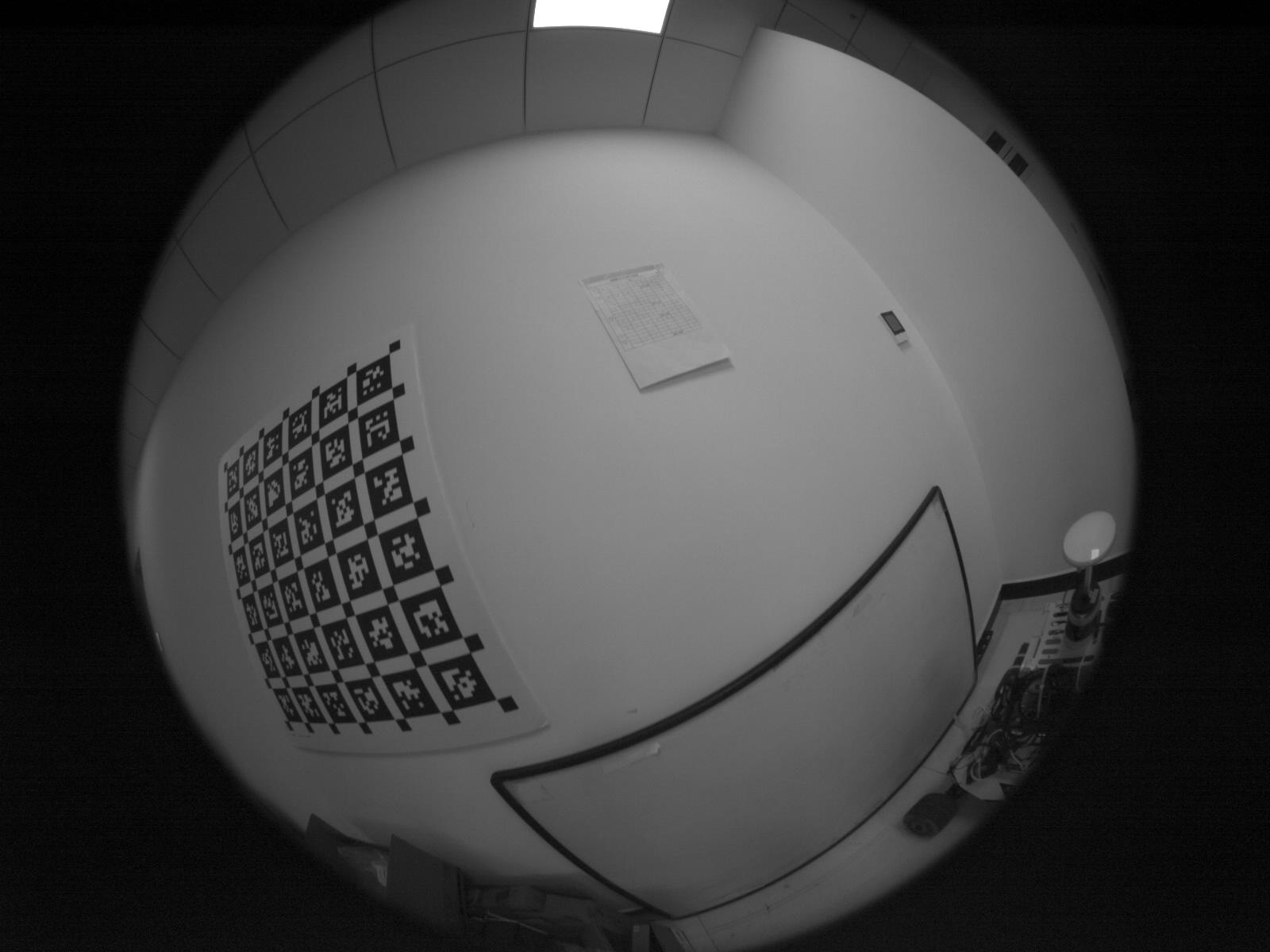} \\
	\caption{Sample images for S04525, E1M3518, BM4218, BM4018, BT2120, MTV185, in row-major order.}
	\label{fig:sample-images}
\end{figure}

We evaluated six GCC tools on Ubuntu 20.04, including BabelCalib \cite{lochmanBabelCalibUniversalApproach2021}, Basalt \cite{usenkoDoubleSphereCamera2018}, Camodocal \cite{hengCamOdoCalAutomaticIntrinsic2013}, 
TartanCalib \cite{duisterhofTartanCalibIterativeWideangle2023} (Kalibr with enhanced corner detection),
the Matlab calibrator \cite{mathworksinc.MATLABComputerVision2021}, and the ROS calibrator \cite{rosROSCameraCalibration2022} based on OpenCV.
which were chosen for their wide use and easy extension to an alternative type of target and data input.
Within these tools, we evaluated several camera models, the pinhole model with the radial tangential distortion for wide-angle cameras, KB-8 for fisheye cameras, and Mei / EUCM for omnidirectional cameras, 
which were chosen mainly for their wide support by GCC tools and downstream applications.
The test plan is shown in Table~\ref{tab:experiments} which lists GCC tools and camera models for processing particular data.
In general, the pinhole model with distortion was used for cameras with a DAOV \textless 120$^\circ$, 
KB-8 for cameras with a DAOV $\ge$ 100$^\circ$, and Mei / EUCM for cameras with a DAOV $\ge$ 120$^\circ$.

The simulation data were generated from the real data with the workflow shown in Fig.~\ref{fig:workflow} (bottom).
We first processed the real data by the TartanCalib with proper models according to the test plan.
Thus, we obtained the frames of detected corners and their poses,
and the estimated calibration parameters, from TartanCalib.
As an exception, for simulating observations of the KB-8 model on MTV185 sequences, 
we first processed them by TartanCalib with the Mei model to get the frame poses, and then estimated the KB-8 parameters by Camodocal on the used corners by TartanCalib.
In any case, these frame poses and camera parameters were then used to simulate the corners in images 
by projecting the target landmarks and adding a Gaussian noise of 0.7 px at both $x$ and $y$-axis.
These camera parameters served as the reference in evaluation.

\subsection{Data Processing}
\label{subsec:dataproc}
\begin{figure}[!tbp]
	\centering
	\includegraphics[width=0.90\columnwidth]{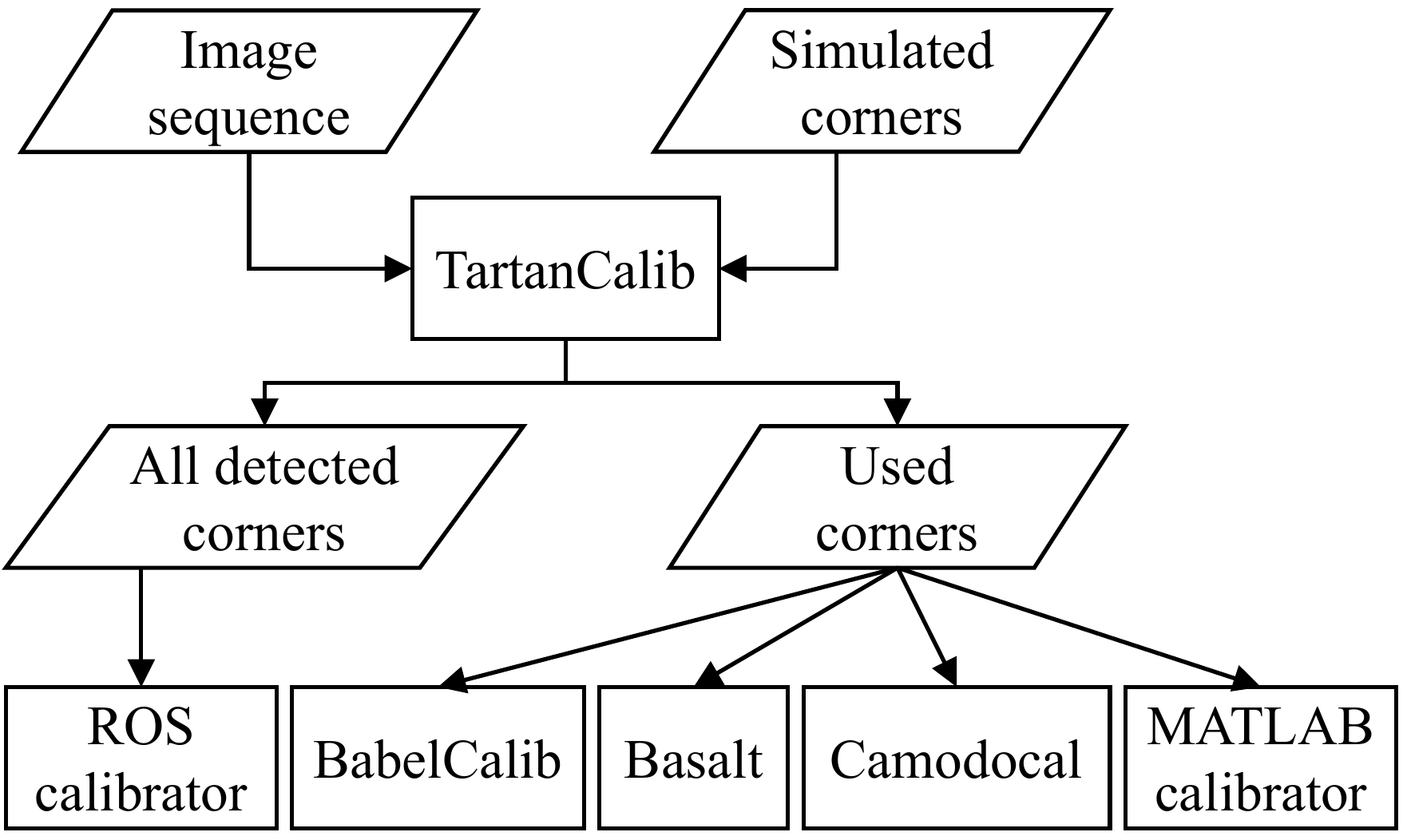}\\
	\includegraphics[width=0.85\columnwidth]{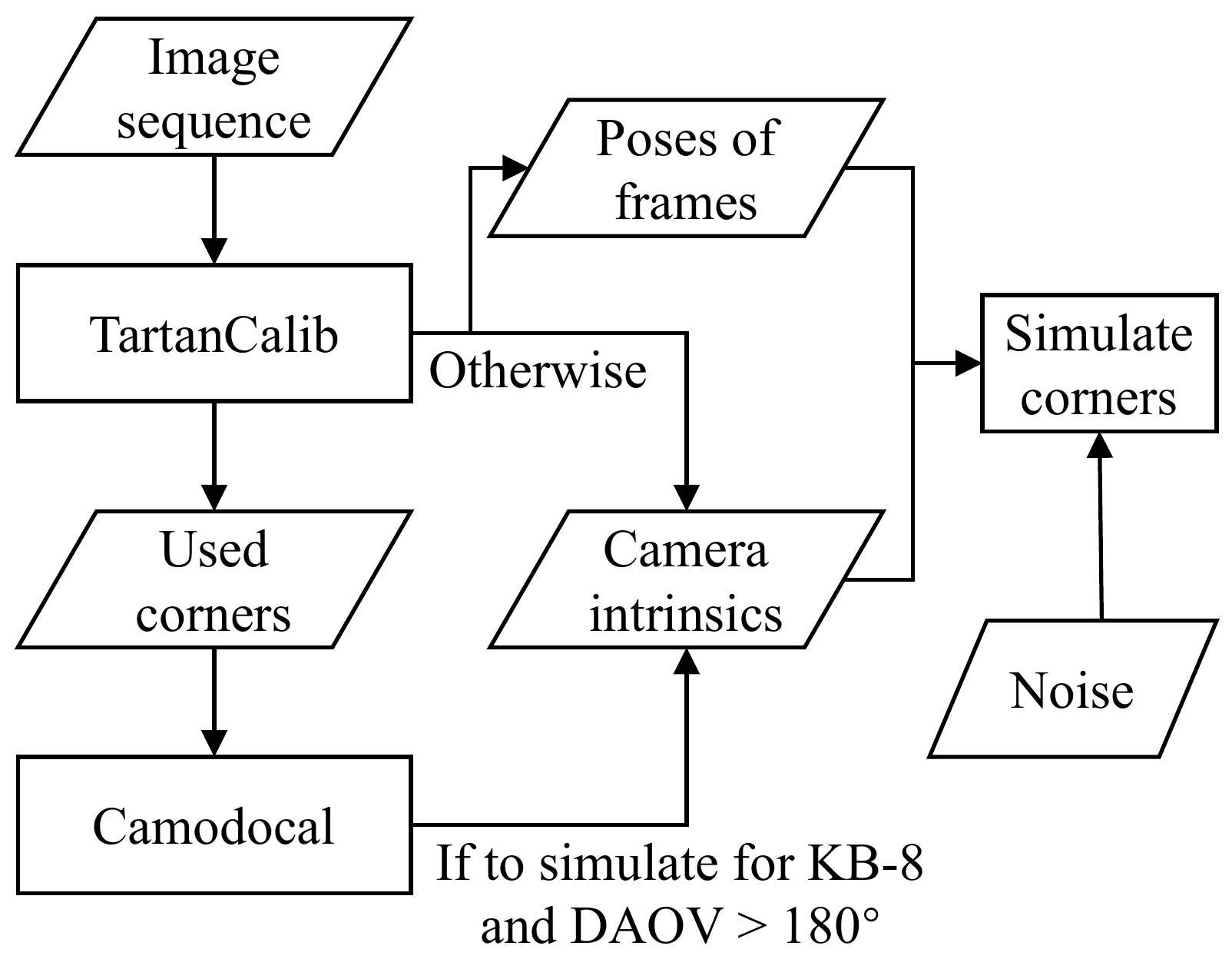}
	\caption{(Top) The calibration workflow for either real data or simulated data.
		(Bottom) The workflow to simulate image points from real data.}
	\label{fig:workflow}
\end{figure}

For either real or simulated data, the evaluation pipeline is shown in Fig.~\ref{fig:workflow} (top).
For better comparison, all tools except for the ROS calibrator used the same corners.
Specifically, we first ran TartanCalib on a (real or simulated) sequence,
and save the frames with detected corners and mark the frames used by TartanCalib.
The TartanCalib was chosen to extract corners from AprilGrid images 
since it could identify sufficient corners under large distortion \cite{duisterhofTartanCalibIterativeWideangle2023}.
All frames of corners were provided to the ROS calibrator.
But only frames of corners used by TartanCalib were given to the four methods, BabelCalib, Basalt, Camodocal, and the MATLAB calibrator.
For these tools, we wrote necessary data loading functions and adapted the calibration initialization with the AprilGrid if needed.
Note that TartanCalib / Kalibr always failed for the MTV185 sequences, we gave these four tools the corners of TartanCalib with the Mei model for these sequences.

Feeding the four tools by TartanCalib had several other reasons.
First, empirically, Kalibr usually chose $\le$ 40 informative frames for calibration. 
This coincided the assertion that global camera models were usually well constrained with 40 frames in \cite{bergamascoCanFullyUnconstrained2013}.
Second, BabelCalib often failed to find a solution with too many frames (e.g., $\ge$ 100), especially for the pinhole model with radial distortion.
Third, the MATLAB calibrator took up to an hour to solve for the Scaramuzza model with 100 frames.

For the ROS calibrator, we ran it five times, each with a sample of 40 randomly chosen frames without replacement,
and kept the run of the minimum RMS reprojection error as the final result.
The exclusive treatment of the ROS calibrator was because
the OpenCV calibration functions hardly dealt with outliers and often gave poor results on corners used by TartanCalib.

Apart from the above, we ran these six calibration tools with their default parameter settings.

A test run was considered failed if no solution was found or 
the recovered focal lengths deviated from the nominal values (for real data) or the reference (in simulation) by $\ge$ 100 px.
Failures in real and simulated tests are marked in Table~\ref{tab:experiments}, for BabelCalib, Basalt, Kalibr, and the ROS calibrator.
BabelCalib failed once when it converged to a wrong focal length.
Basalt failed for either converging to a wrong focal length or 
unable to converge in 100 iterations.
Kalibr's failures were due to the poorly implemented equidistant model for large FOV cameras.
The ROS calibrator was affected by outliers and largely unsuccessful on MTV185 sequences for the poorly implemented pinhole equidistant model.
Oddly, it always aborted with ill-conditioned matrices on sequences of 103$^\circ$ and 127$^\circ$ DAOV cameras, perhaps unable to initialize in such cases.

\begin{table*}[!htbp]
	\centering
	\caption{Tests carried out on real and simulated data. `r' and `s' values denote numbers of failures out of 9 sequences in real data tests and simulation, respectively. rad: radial, tan: tangential, equi: equidistant. }
	\label{tab:experiments}
	\begin{tabular}{ccccccc}
		\hline
		Lens &
		BabelCalib &
		Basalt &
		Camodocal &
		\begin{tabular}[c]{@{}c@{}}Kalibr /\\ TartanCalib\end{tabular} &
		MATLAB &
		\begin{tabular}[c]{@{}c@{}}ROS\\ calibrator\end{tabular} \\ \hline
		S04525  & pinhole rad, s: 1 & N/A              & pinhole rad tan & pinhole rad tan          & pinhole rad tan & pinhole rad tan            \\ \hline
		E1M3518 & pinhole rad       & N/A              & pinhole rad tan & pinhole rad tan          & pinhole rad tan & pinhole rad tan            \\ \hline
		BM4218  & pinhole rad       & N/A              & pinhole rad tan & pinhole rad tan          & pinhole rad tan & pinhole rad tan, r: 4, s: 3 \\ \hline
		BM4218  & KB-8              & KB-8, r: 7, s: 4 & KB-8            & pinhole equi             & Scaramuzza      & pinhole equi, r: 9, s: 9   \\ \hline
		BM4018  & KB-8              & KB-8, s: 7       & KB-8            & pinhole equi             & Scaramuzza      & pinhole equi, r: 9, s: 9   \\ \hline
		BT2120  & KB-8              & KB-8, s: 4       & KB-8            & pinhole equi             & Scaramuzza      & pinhole equi               \\ \hline
		MTV185  & KB-8              & KB-8, r: 7, s: 7 & KB-8            & pinhole equi, r: 9, s: 9 & Scaramuzza      & pinhole equi, r: 8, s: 9   \\ \hline
		BM4018  & EUCM               & EUCM, s: 2      & Mei             & Mei                      & N/A             & Mei                        \\ \hline
		BT2120  & EUCM               & EUCM, r: 2, s: 2   & Mei             & Mei                      & N/A             & Mei                        \\ \hline
		MTV185  & EUCM               & EUCM, r: 8, s: 9   & Mei             & Mei                      & N/A             & Mei                        \\ \hline
	\end{tabular}%
\end{table*}

Next, we evaluated the GCC tools by looking at the consistency of estimated camera parameters and the RMS reprojection errors for both simulated and real data.
The RMS reprojection errors are computed by these tools on all inlier observations.
The RMS values should be viewed lightly when comparing across tools since the inlier sets may vary slightly even for the same data.

\subsection{Simulation Results}
\begin{figure}[!htbp]
	\centering
	\includegraphics[width=0.49\columnwidth]{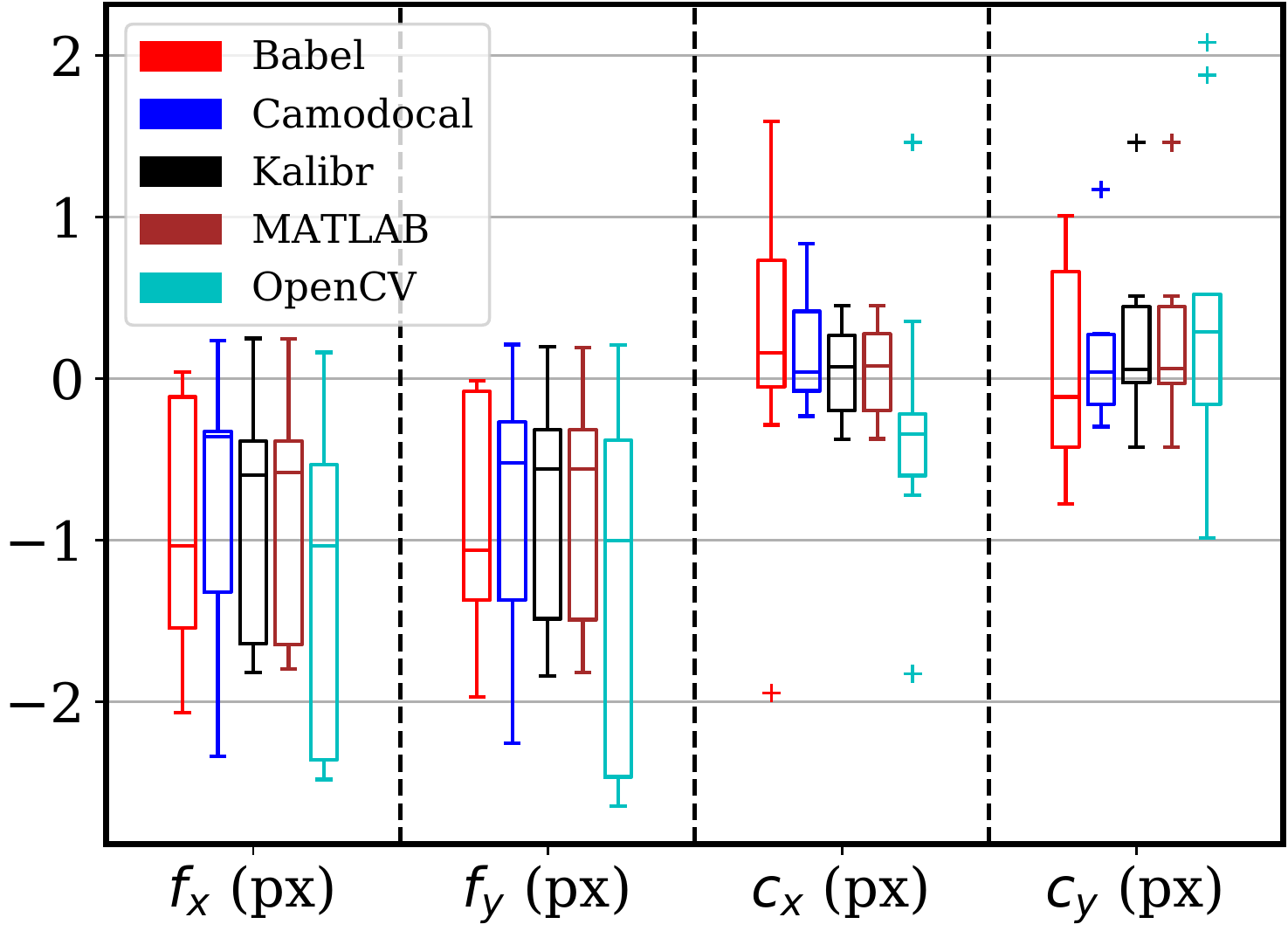}
	\includegraphics[width=0.48\columnwidth]{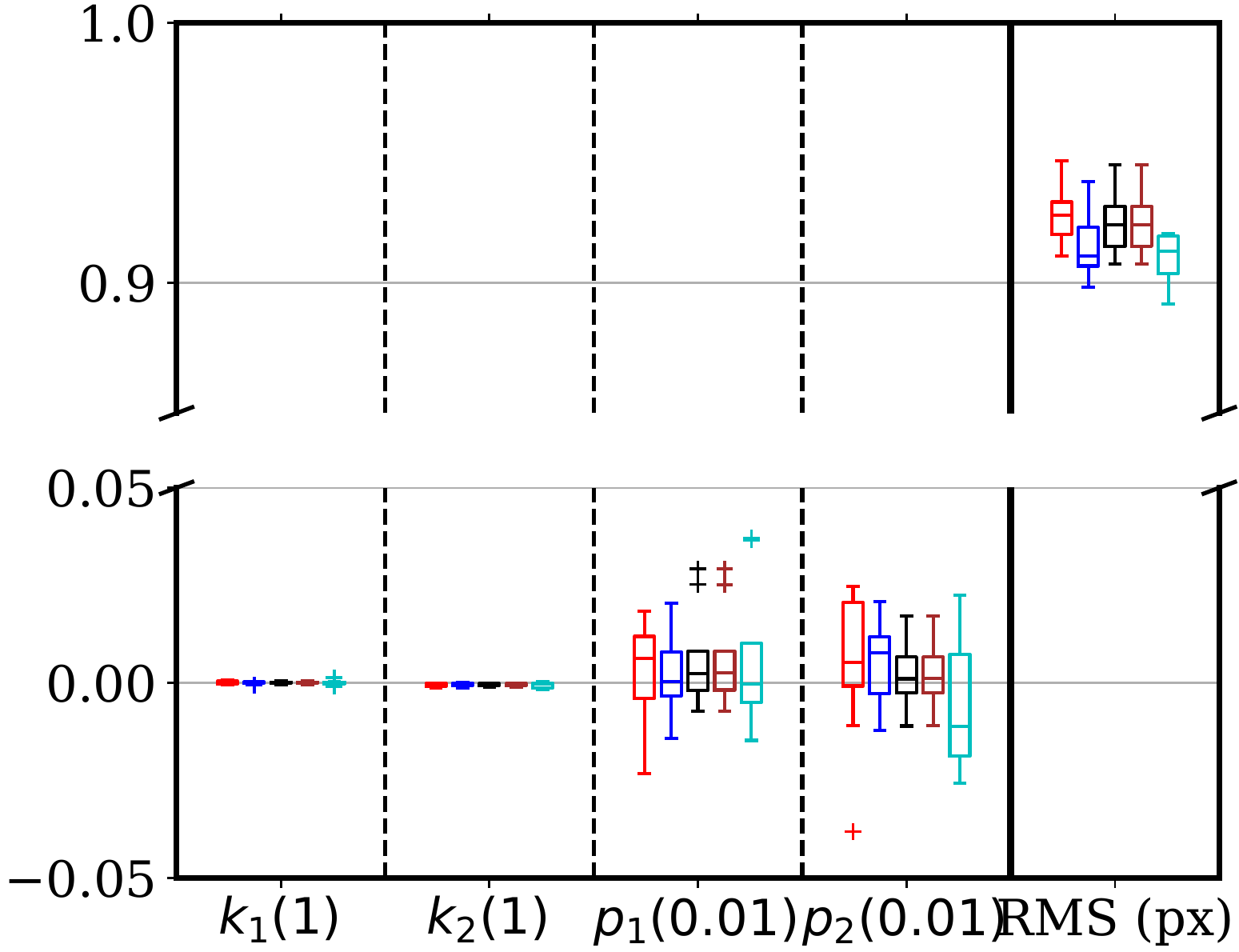} \\
	(S04525) \\
	\includegraphics[width=0.49\columnwidth]{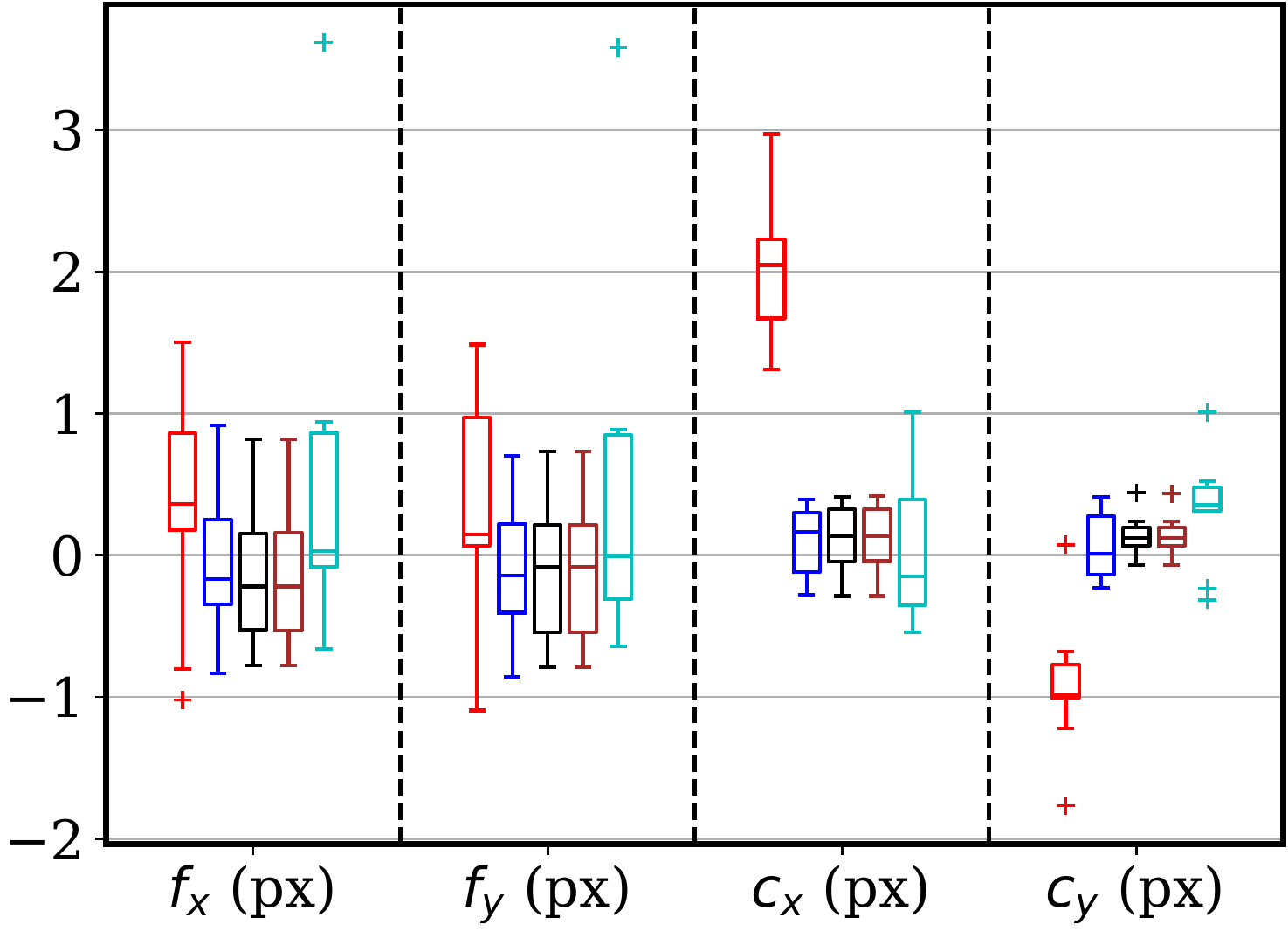}
	\includegraphics[width=0.48\columnwidth]{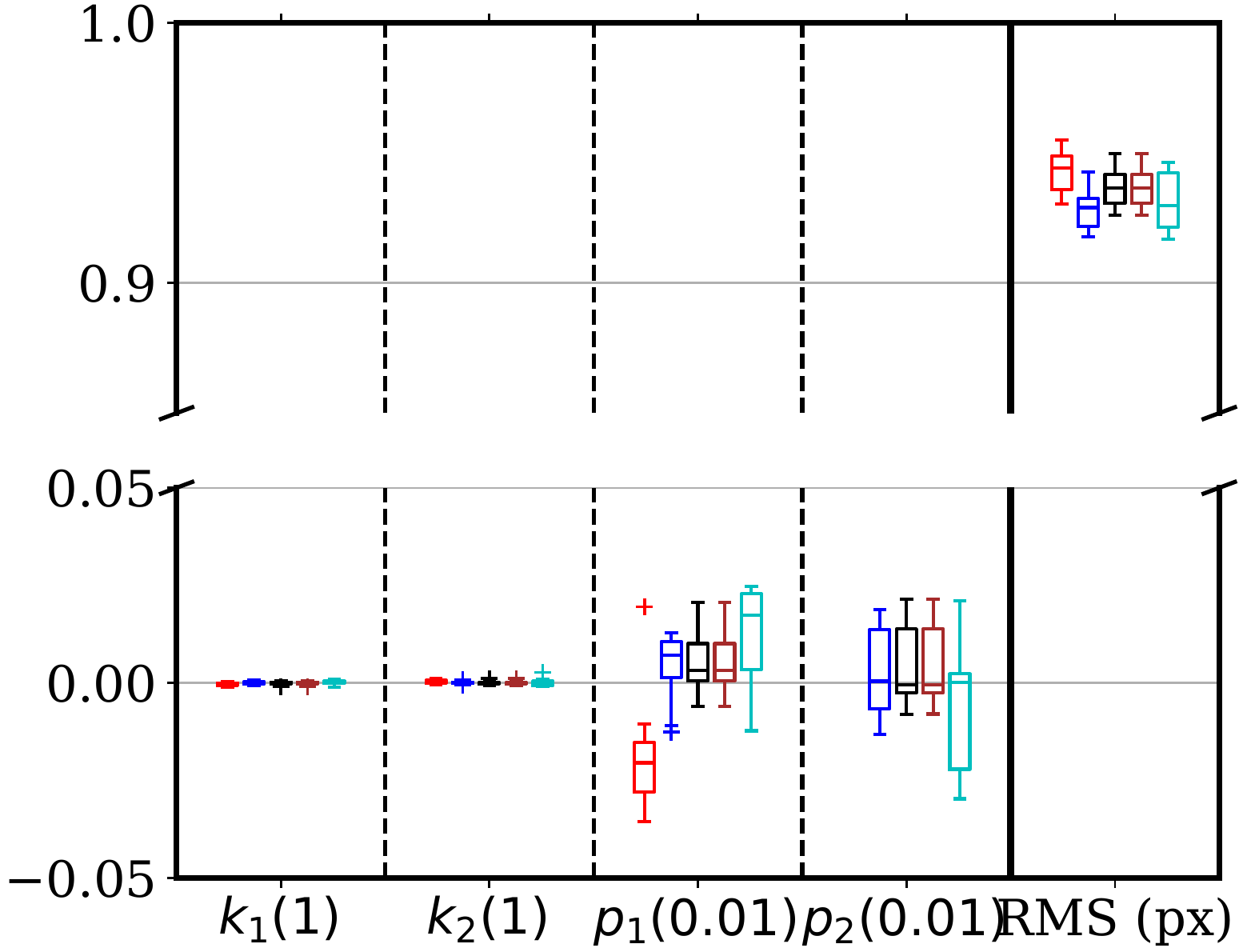} \\
	(E1M3518)\\
	\includegraphics[width=0.49\columnwidth]{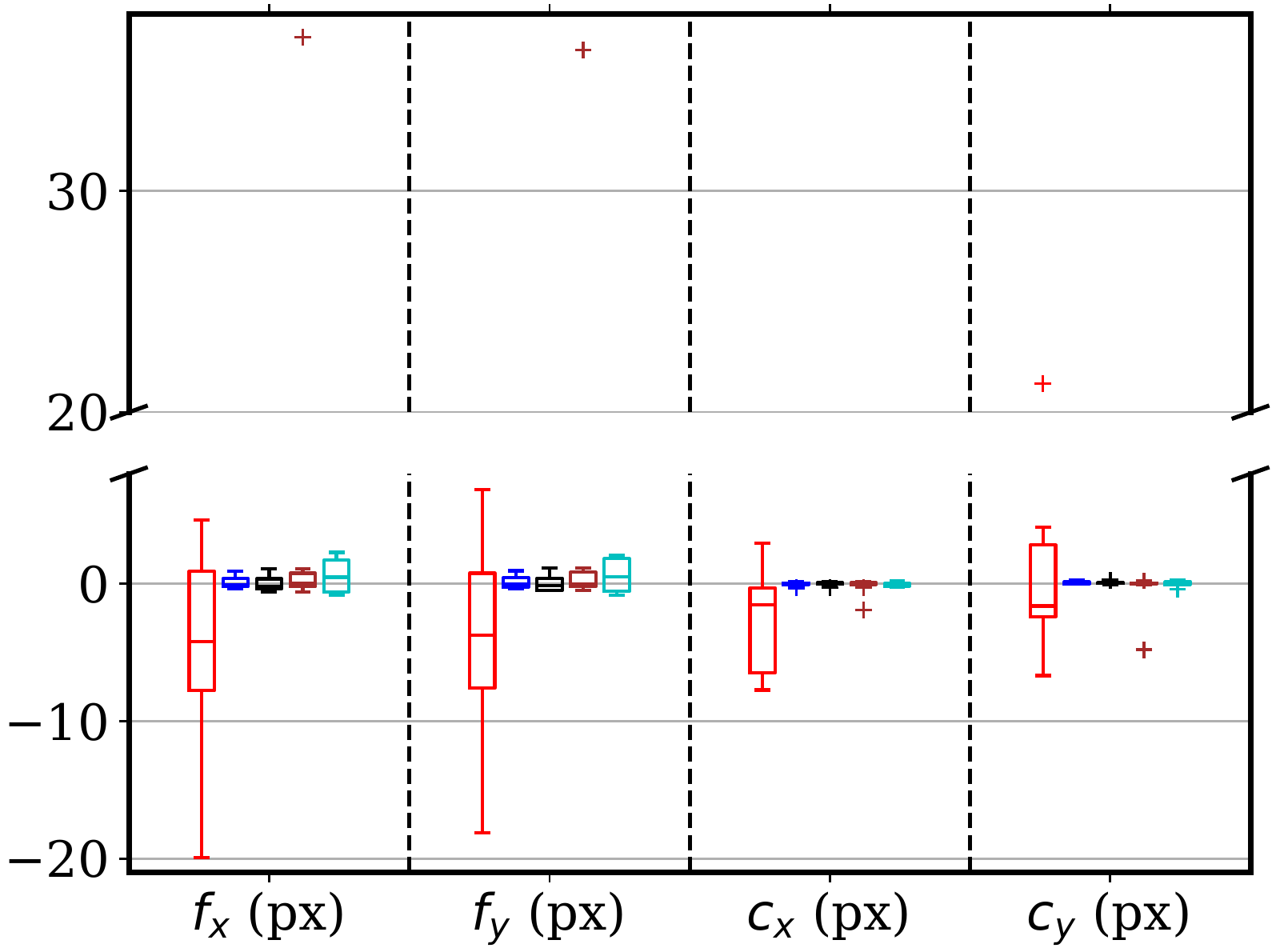}
	\includegraphics[width=0.48\columnwidth]{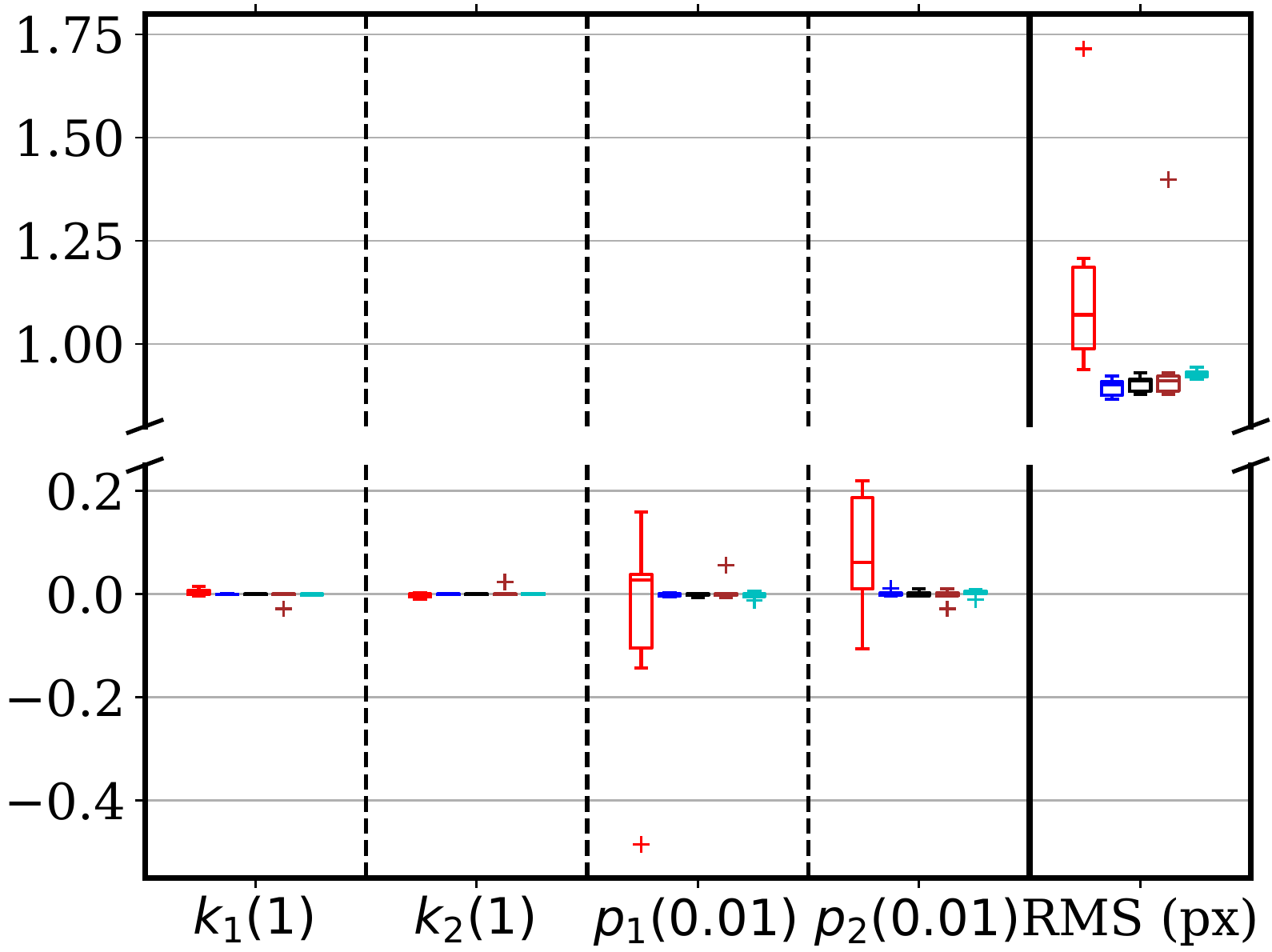} \\
	(BM4218)
	\caption{Error distributions of pinhole radial tangential model parameters and the root mean square (RMS) reprojection errors, by five geometric camera calibration tools, 
		BabelCalib, Camodocal, Kalibr / TartanCalib, the MATLAB calibrator, and the ROS / OpenCV calibrator,
		on the simulated data for S04525, E1M3518, and BM4218 lenses, each of 9 sequences.
	Note that BabelCalib used a pinhole radial model, resulting in zero values in $p_1$ and $p_2$, which are shown relative to the true values.
	}
	\label{fig:sim-pinhole}
\end{figure}

The simulated data were processed as described above.
The data from cameras with S04525, E1M3518, and BM4218 lenses, were processed by five tools with the pinhole radial tangential model
except Basalt which did not support the model.
The camera parameter errors and the RMS reprojection errors are shown in Fig.~\ref{fig:sim-pinhole}, where failed tests were excluded in drawing the box plots.
The units are specified in parentheses for all box plot figures.
These tools generally gave very similar results close to the reference.
The focal lengths and principal points were usually within (-2, 2) px of the true values.
Since BabelCalib did not consider the tangential distortion, its estimates had larger errors than other methods, especially for BM4218 sequences of 103$^\circ$ DAOV.
The RMS reprojection errors slightly above 0.9 were resulted from the Gaussian noise of $\sigma = \sqrt{2}\times 0.7 = 0.99$.
The ROS calibrator based on OpenCV had slightly larger error dispersions, likely due to corners of large reprojection residuals.
For a BM4218 sequence, the MATLAB calibrator converged to a focal length off by 37 px for no apparent reason.

\begin{figure}[!htbp]
	\centering
	\includegraphics[width=0.48\columnwidth]{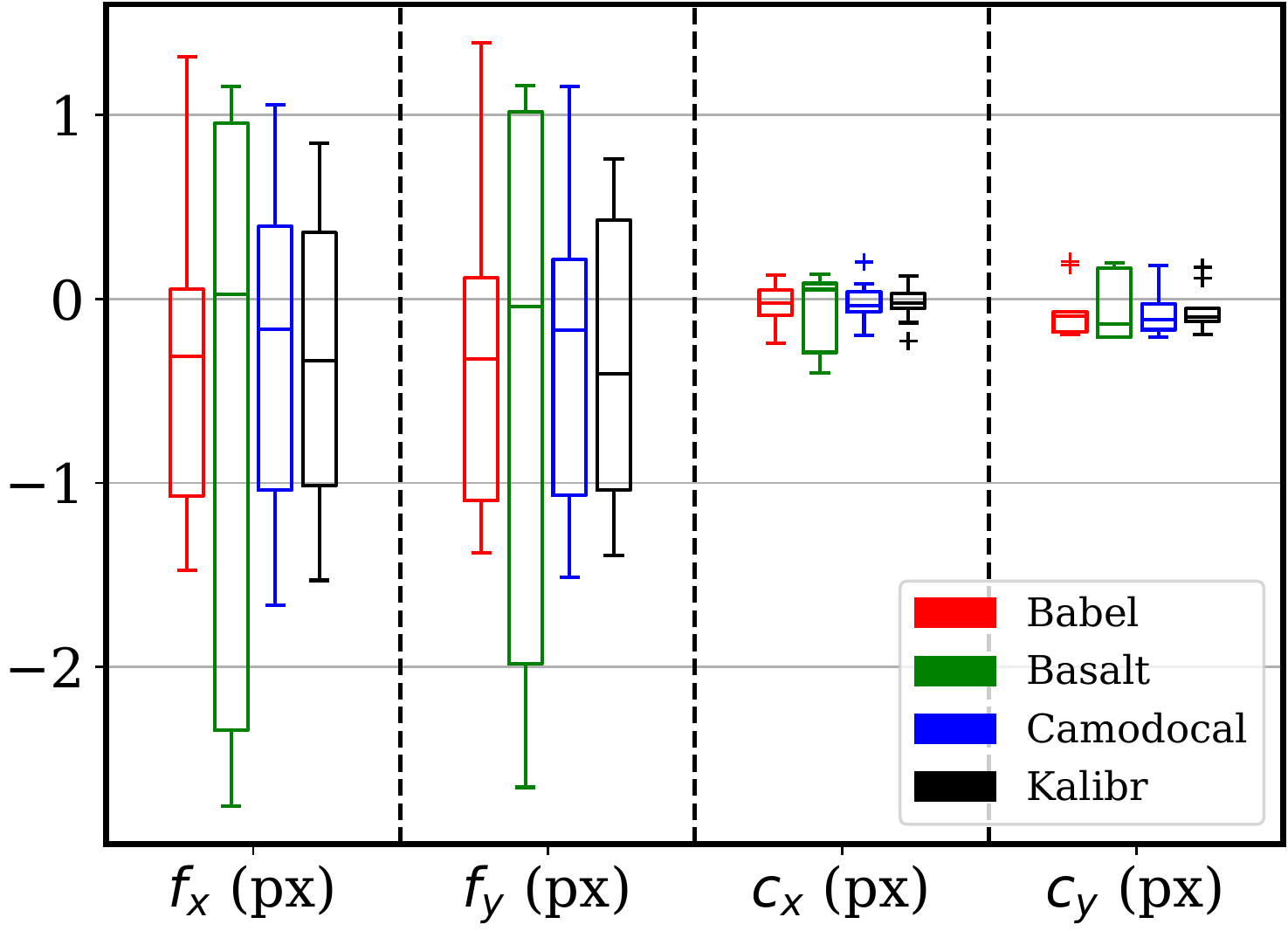}
	\includegraphics[width=0.48\columnwidth]{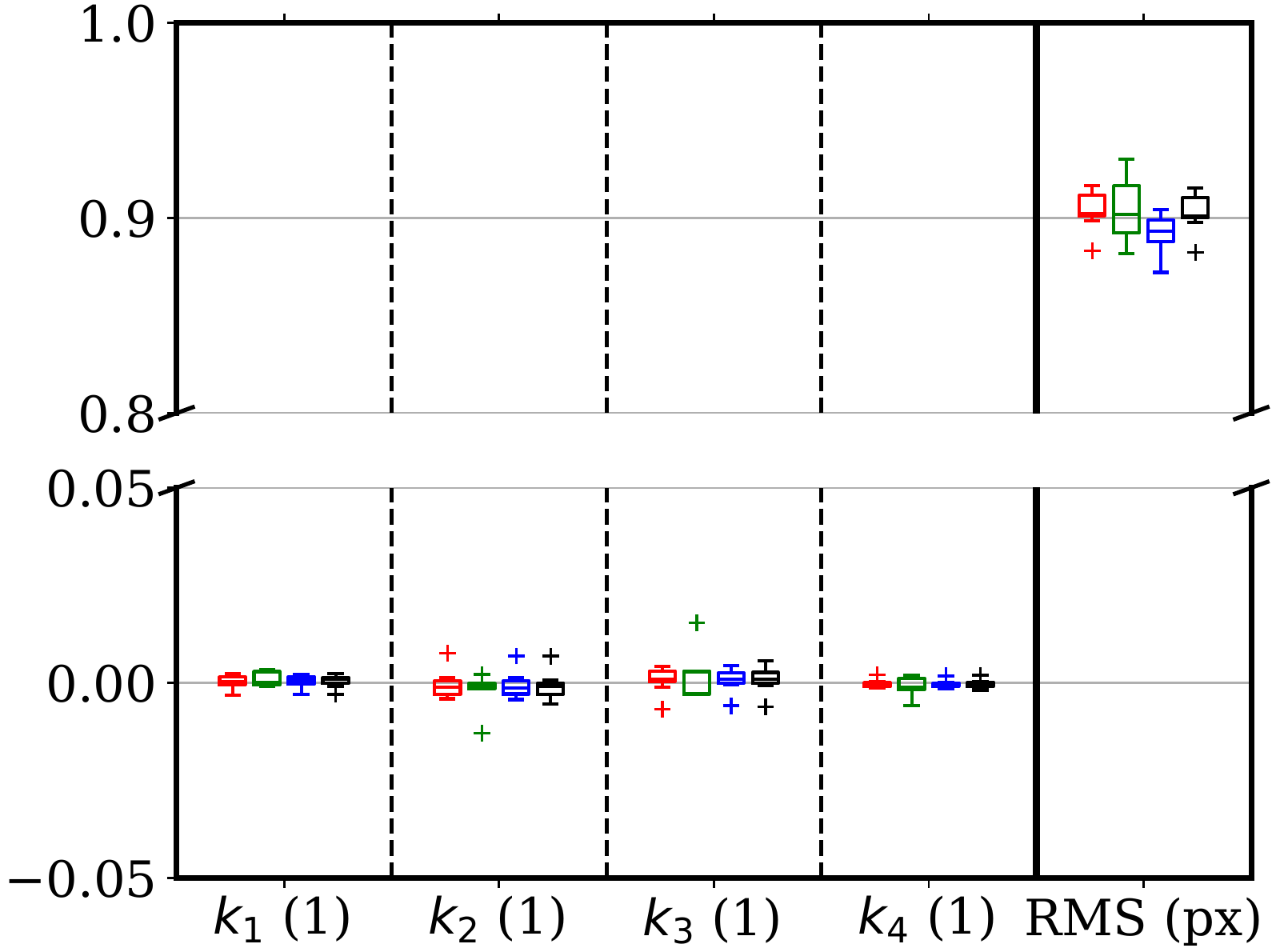} \\
	(BM4218) \\
	\includegraphics[width=0.48\columnwidth]{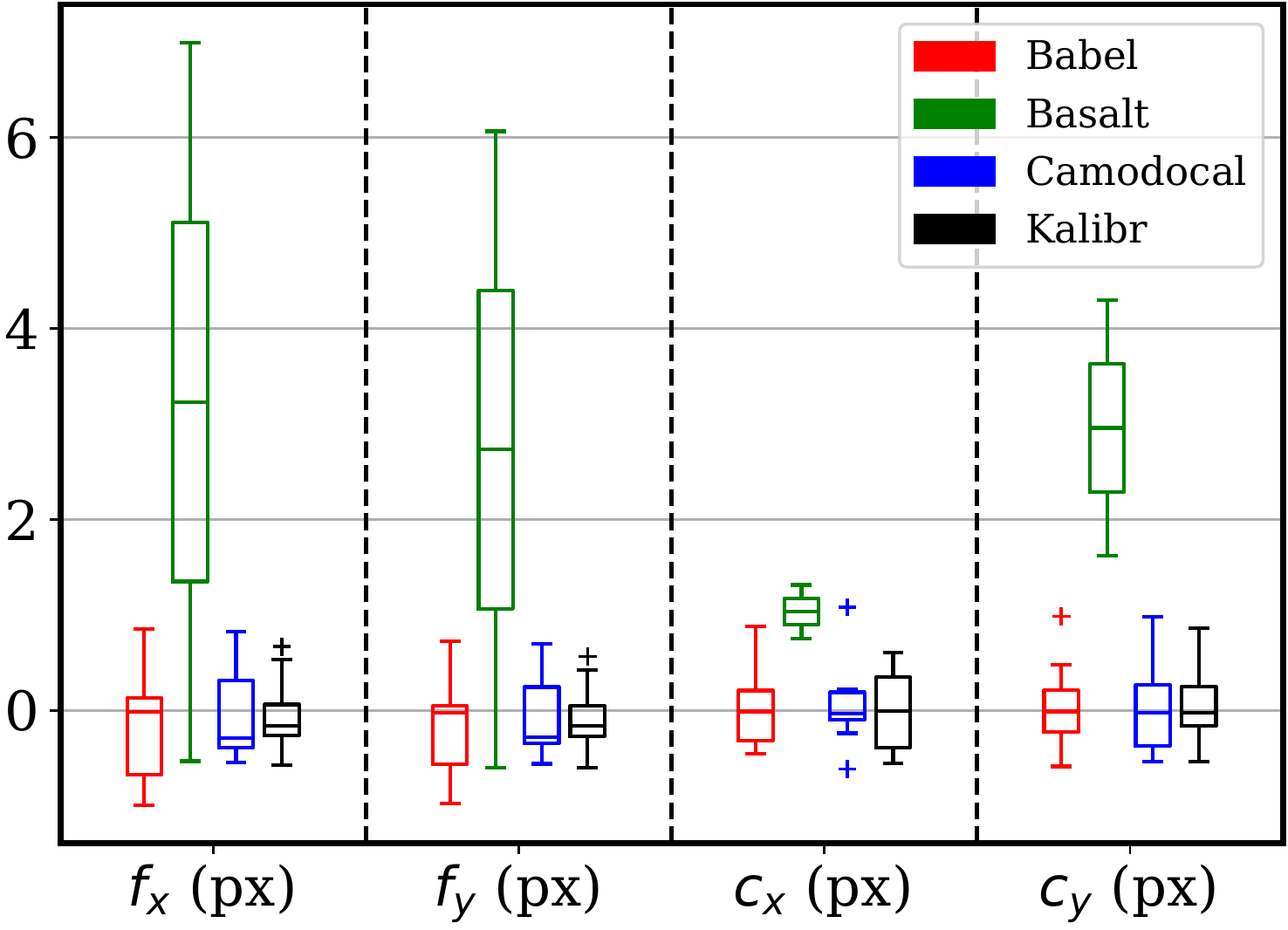}
	\includegraphics[width=0.48\columnwidth]{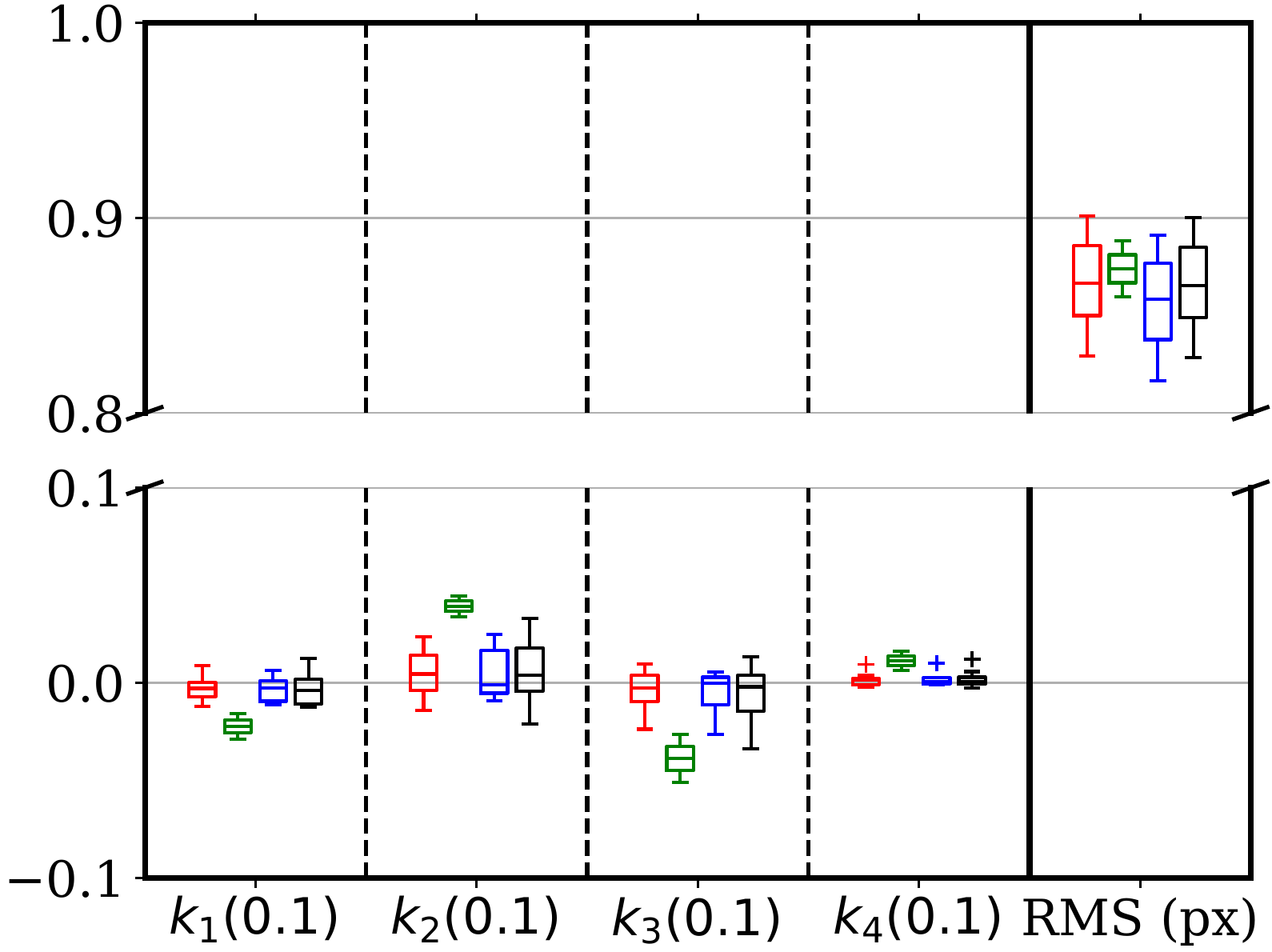} \\
	(BM4018)\\
	\includegraphics[width=0.5\columnwidth]{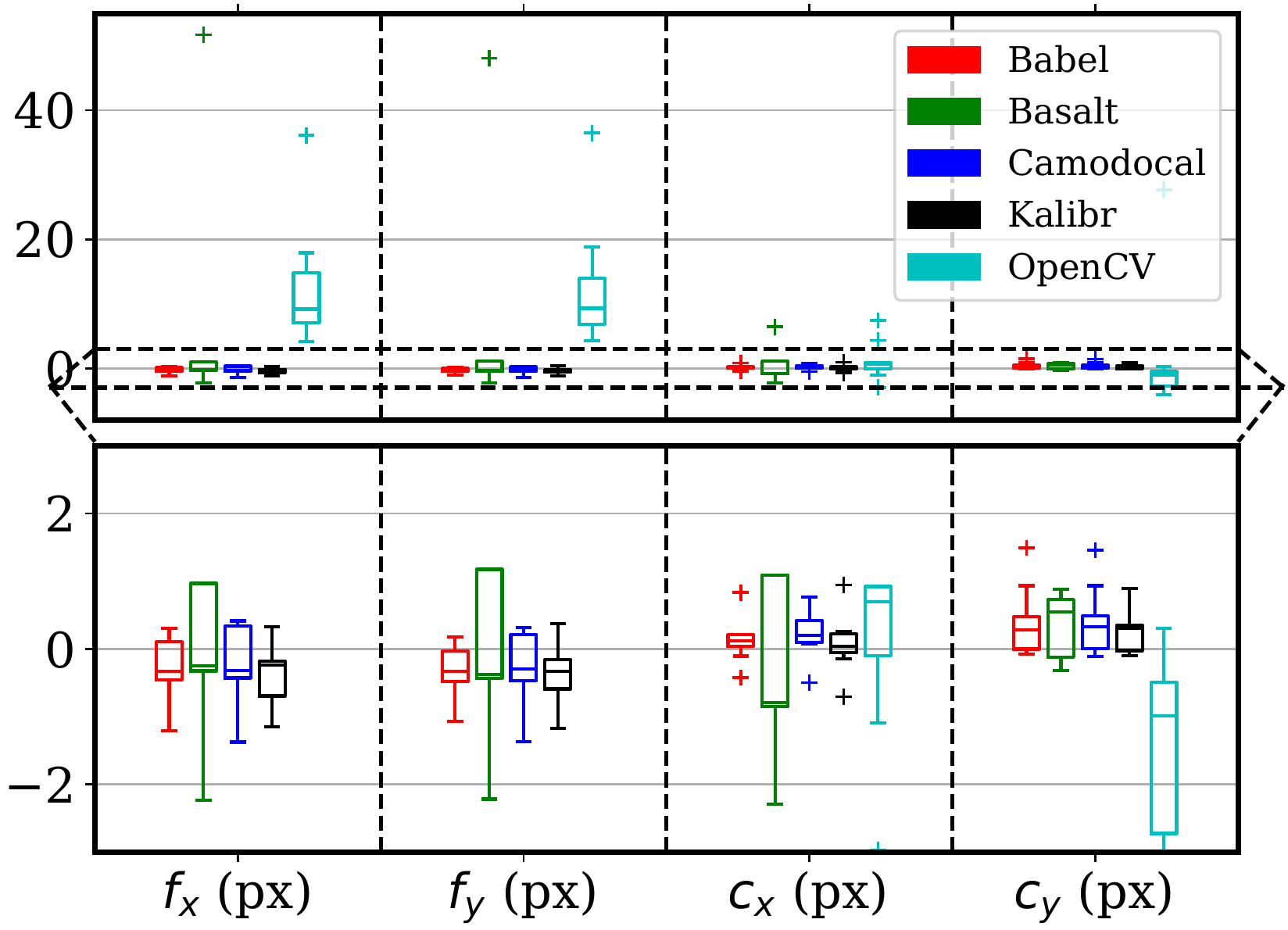}
	\includegraphics[width=0.48\columnwidth]{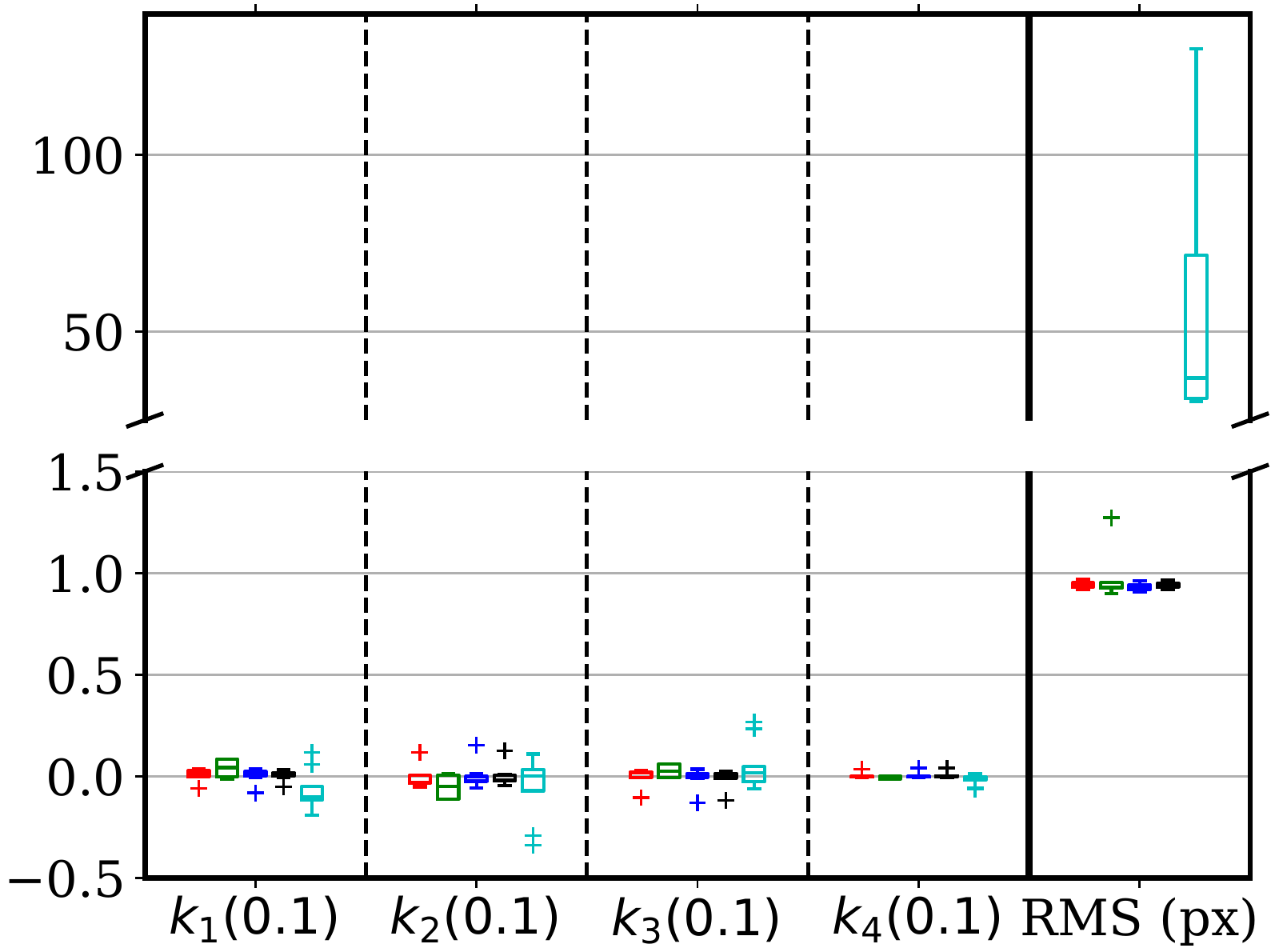} \\
	(BT2120)\\
	\includegraphics[width=0.5\columnwidth]{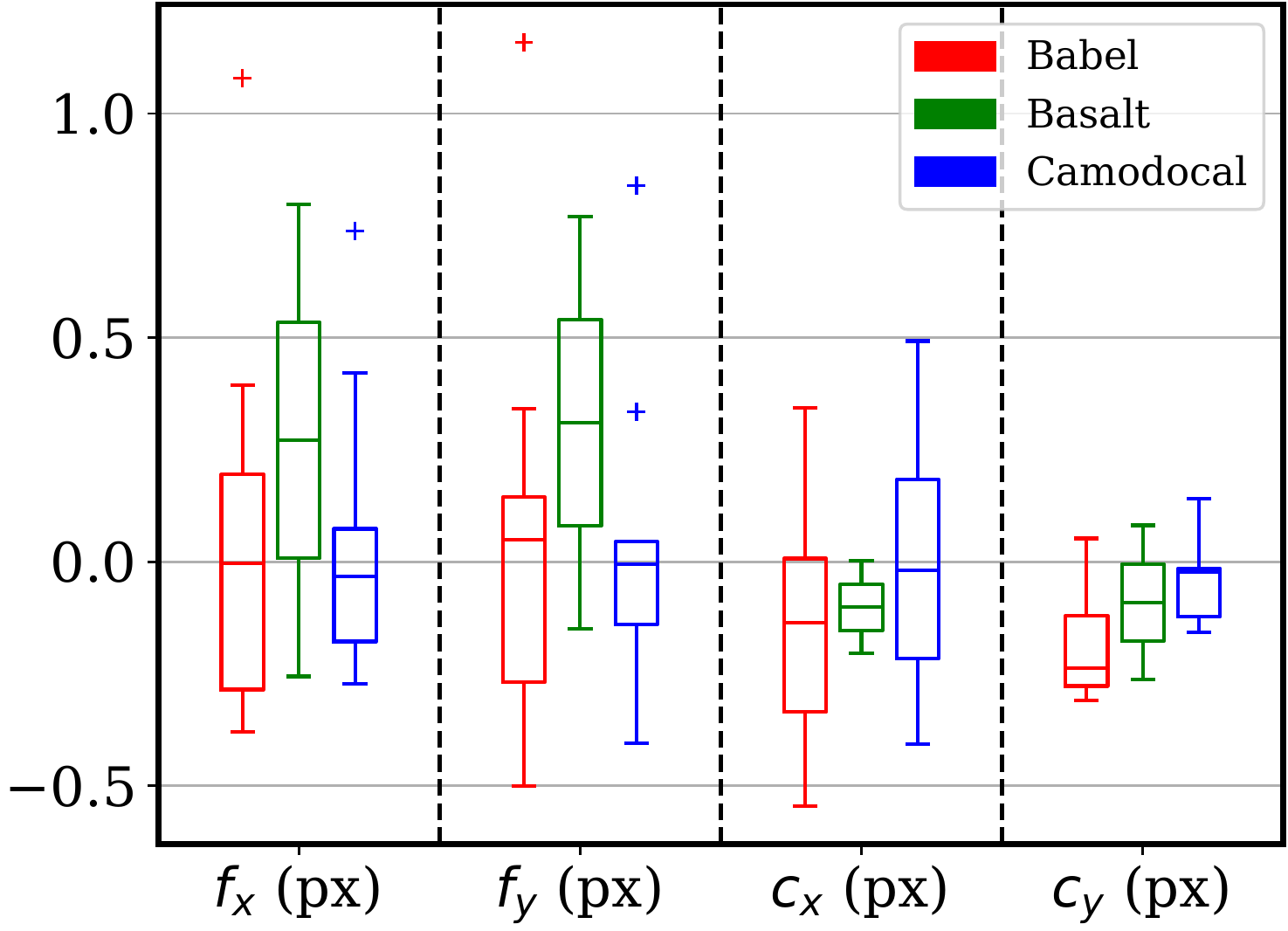}
	\includegraphics[width=0.48\columnwidth]{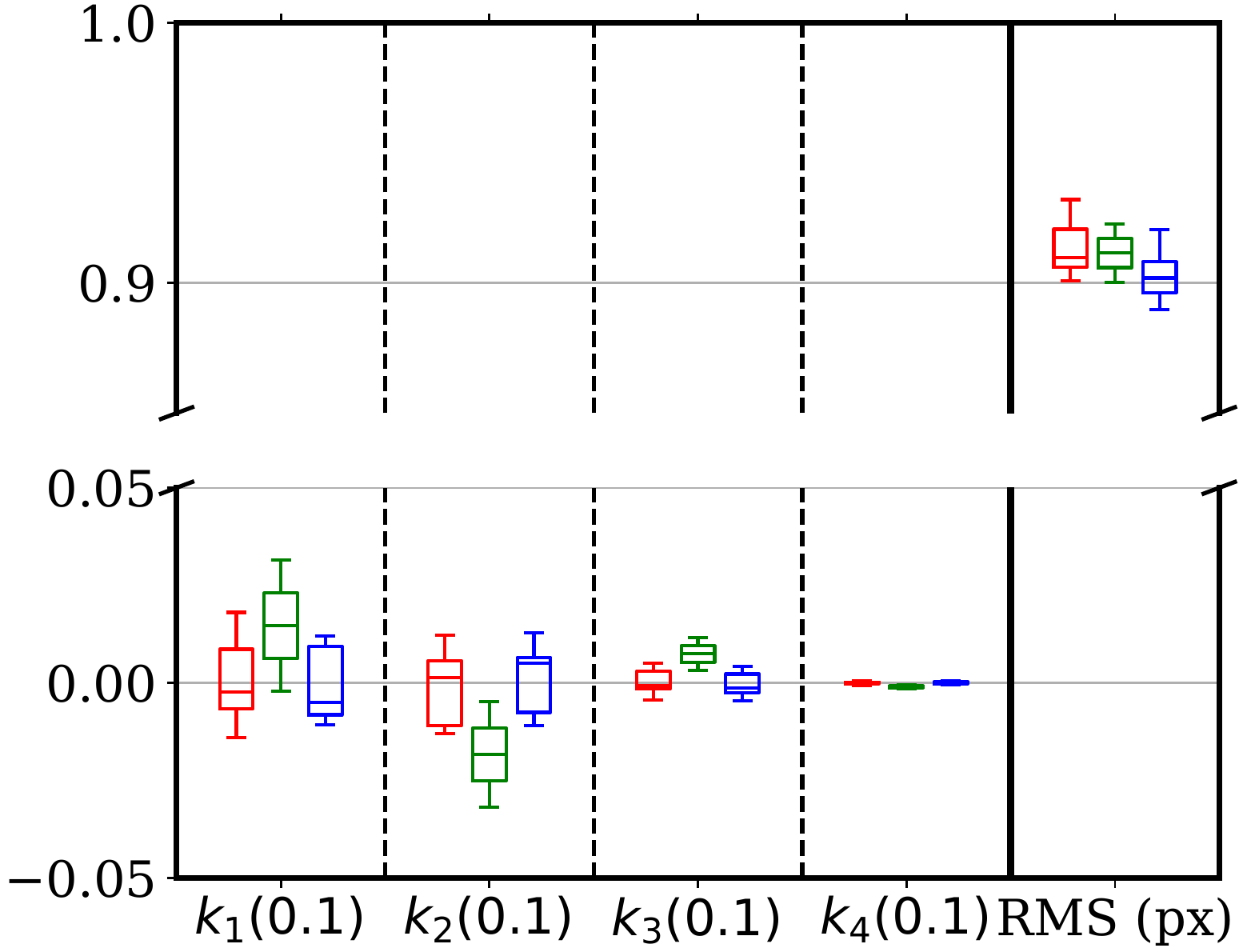}\\
	(MTV185)
\caption{Error distributions of KB-8 parameters and the root mean square (RMS) reprojection errors,
by five geometric camera calibration tools, BabelCalib, Basalt, Camodocal, Kalibr, and the OpenCV-based ROS calibrator,
on the simulated data for BM4218, BM4018, BT2120, and MTV185 lenses, each of 9 sequences.
Results of the ROS calibrator for BM4218, BT4018, and MTV185 lenses 
and the MATLAB calibrator for the MTV185 lens, were excluded for persistent failures.}
\label{fig:sim-kb8}
\end{figure}

For cameras with a DAOV $\ge100^\circ$, the KB-8 model was solved for by using five tools except for the MATLAB calibrator which does not support KB-8.
The parameter errors and RMS errors are shown in Fig.~\ref{fig:sim-kb8}.
The ROS calibrator results for the BM4218, BM4018, and MTV185 lenses, and the MATLAB results for the MTV185 lens,
were excluded for consistent failures explained in Section~\ref{subsec:dataproc}.
Among these tools, we see that the Basalt and the OpenCV-based ROS calibrator sometimes converged to focal lengths of large errors $>$5 px.
Other tools consistently estimated the focal lengths and principal points within (-2, 2) px as well as the distortion parameters.

\begin{figure}[!tbp]
	\centering
	\includegraphics[width=0.48\columnwidth]{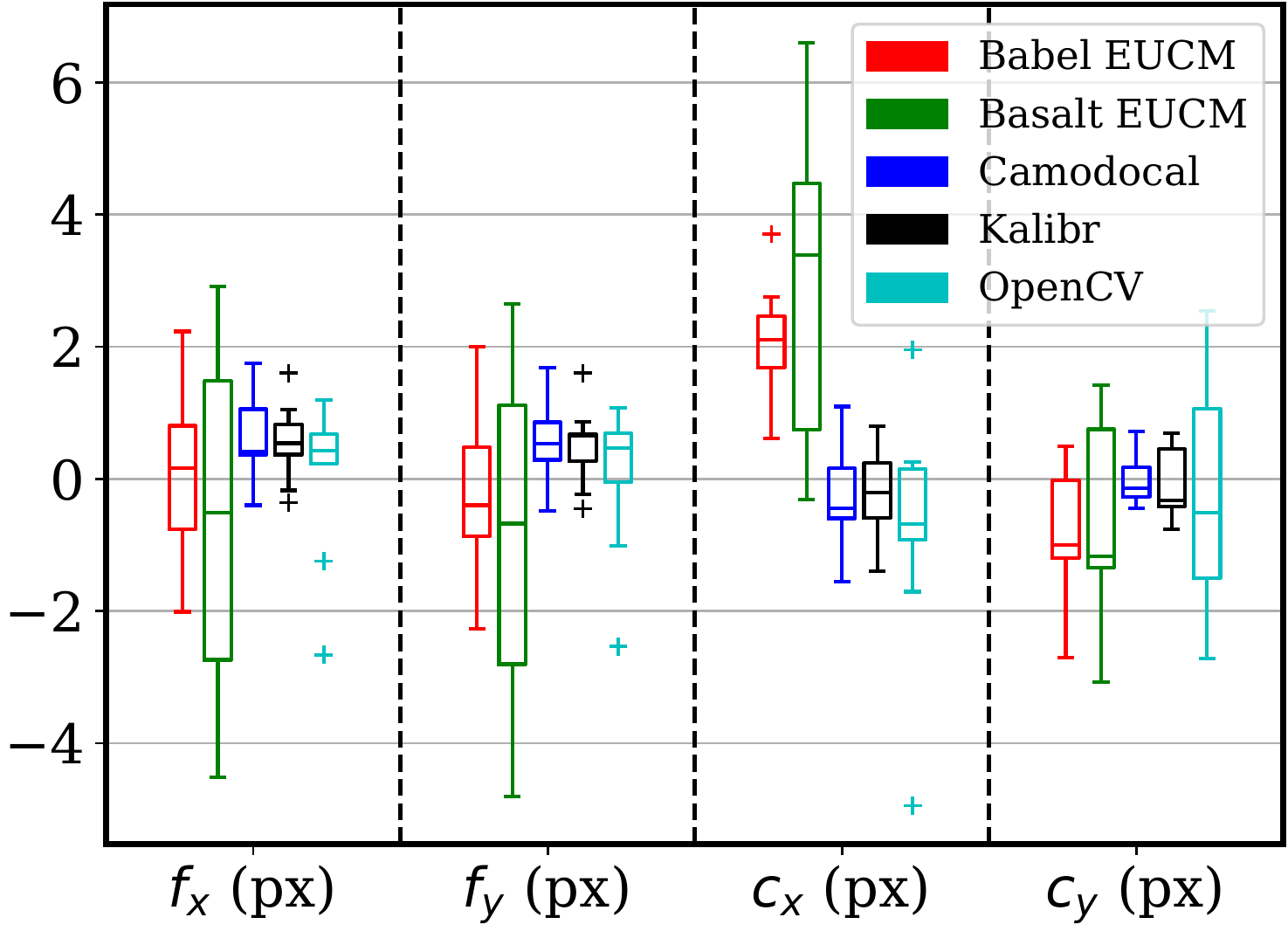}
	\includegraphics[width=0.48\columnwidth]{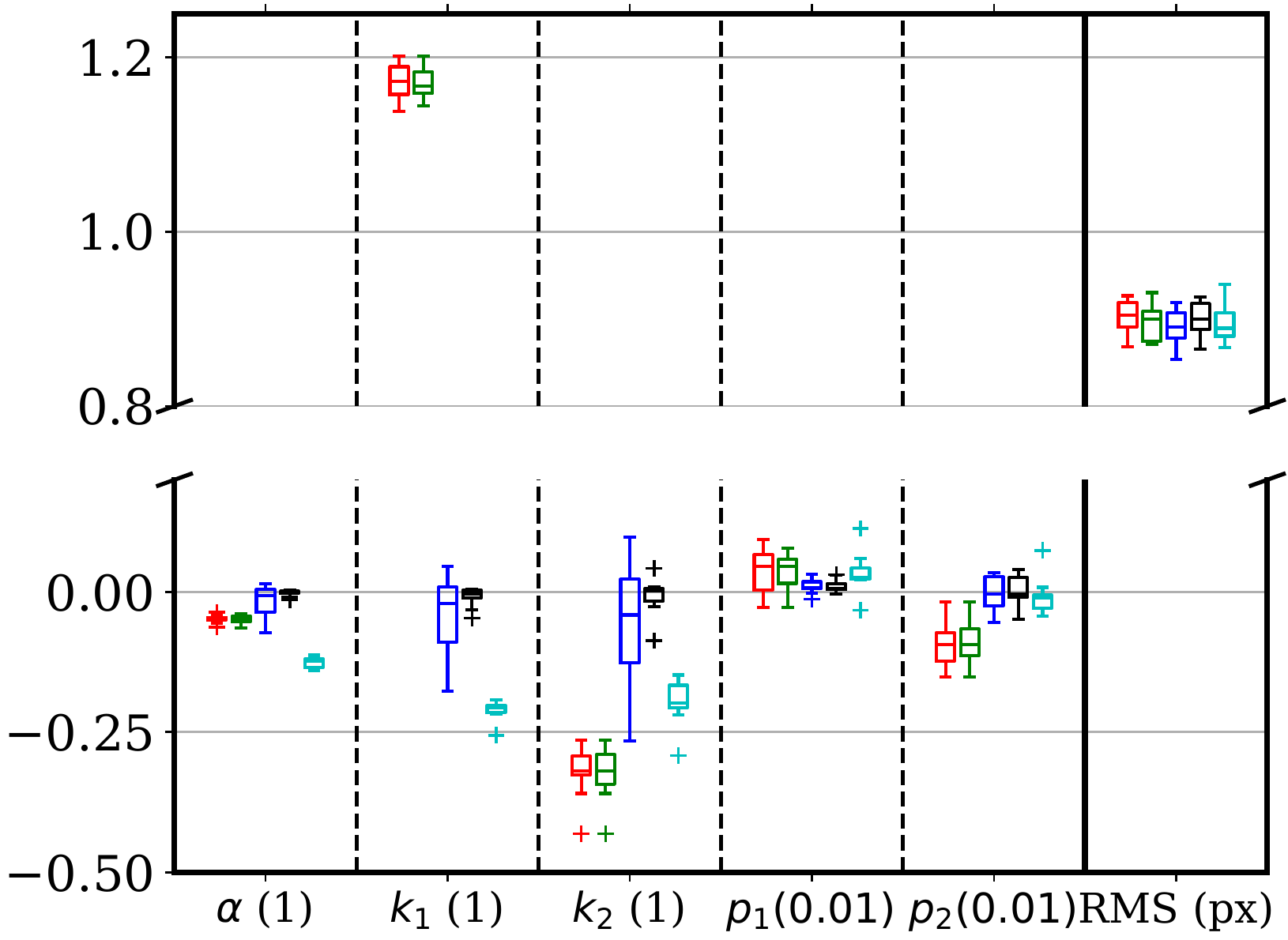} \\
	(BM4018) \\
	\includegraphics[width=0.48\columnwidth]{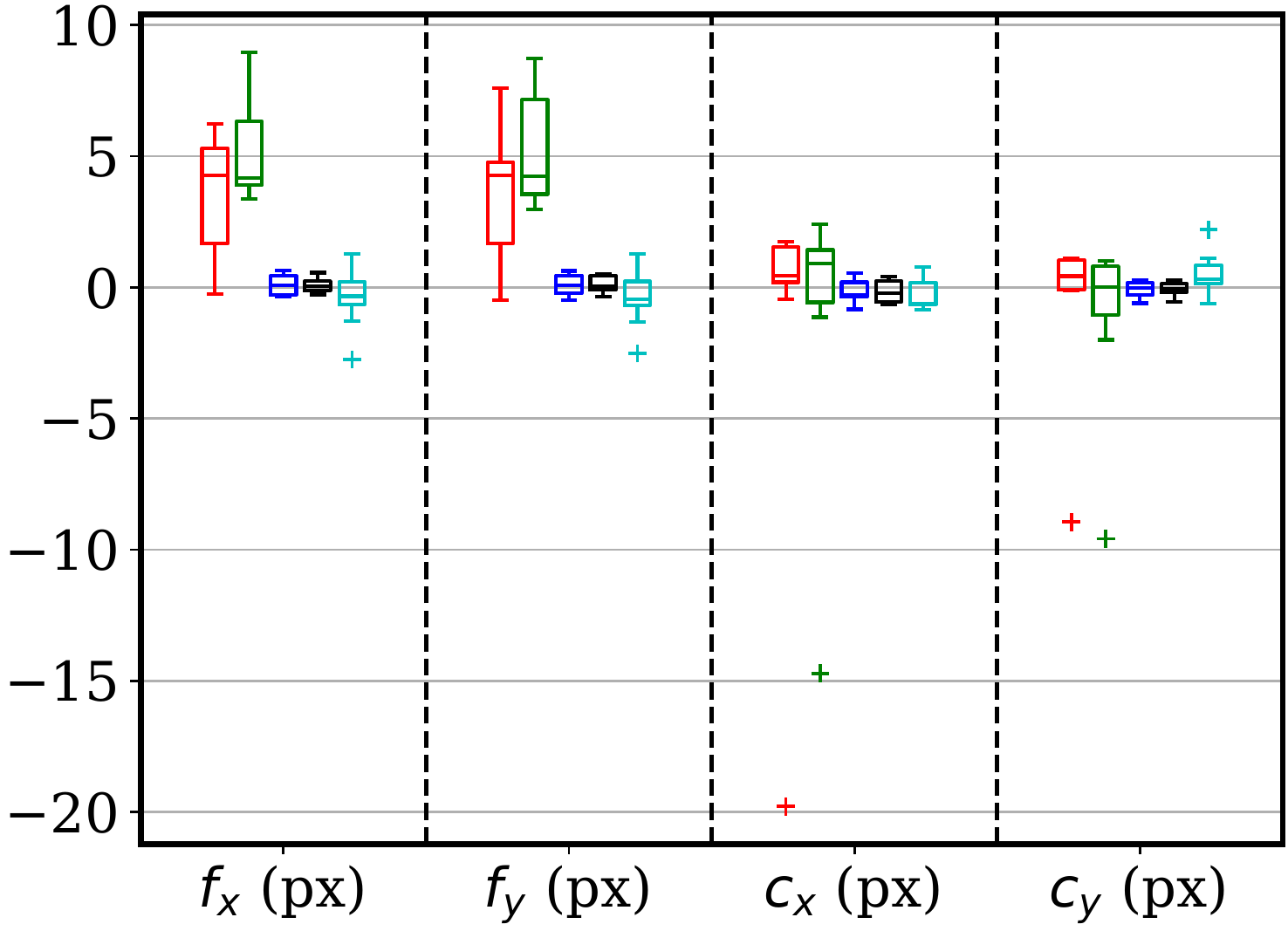}
	\includegraphics[width=0.48\columnwidth]{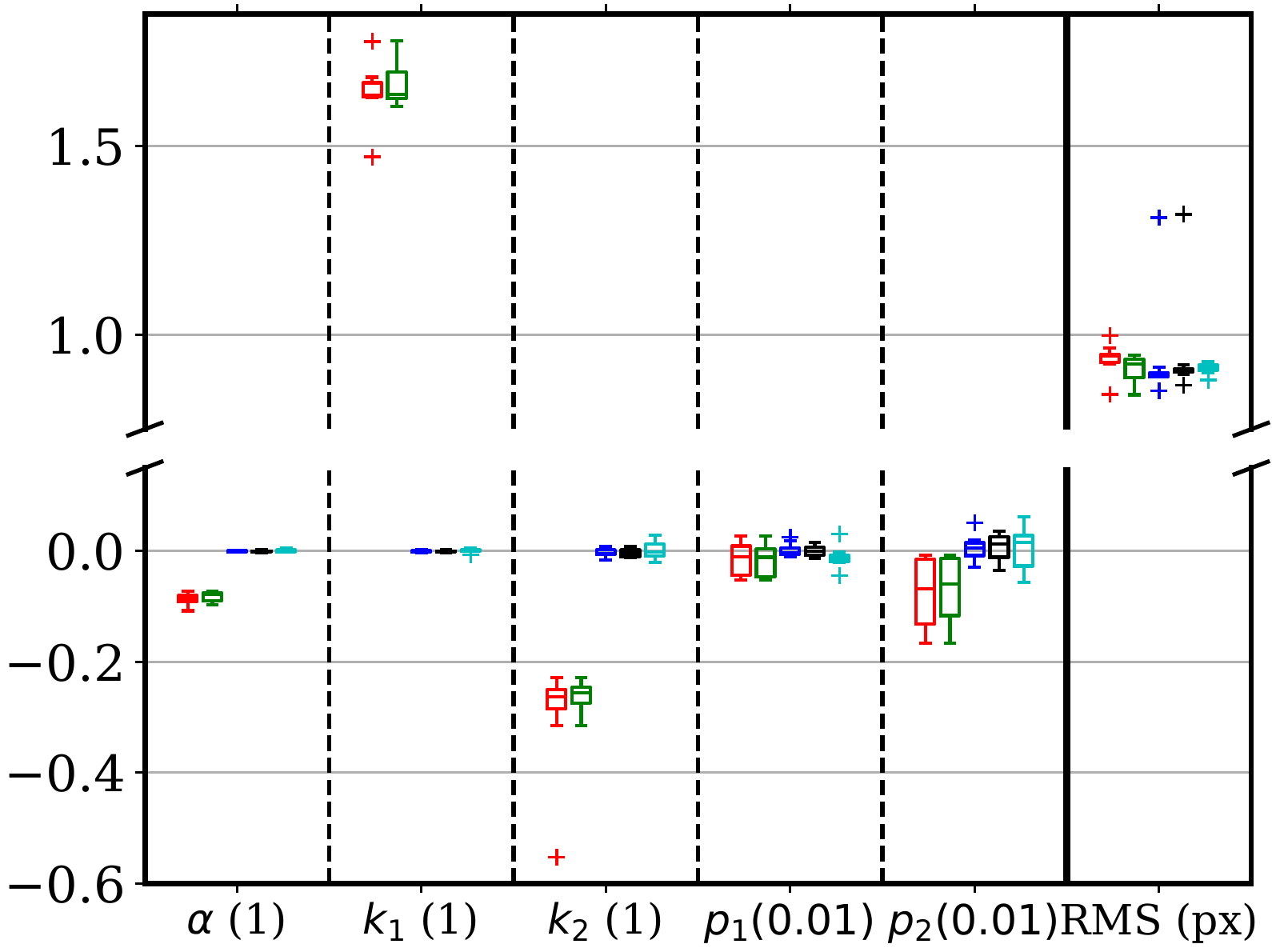} \\
	(BT2120) \\
	\includegraphics[width=0.48\columnwidth]{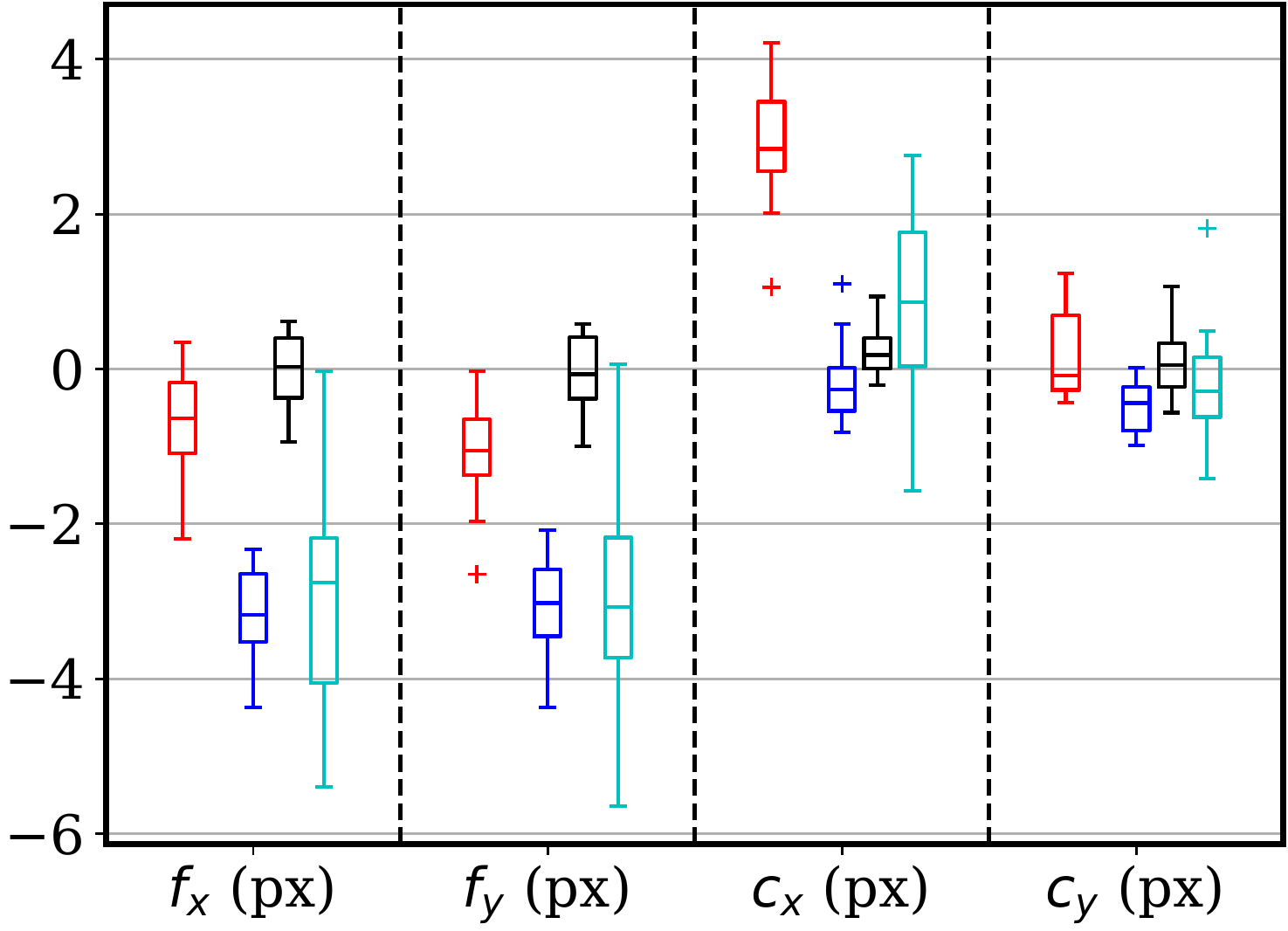}
	\includegraphics[width=0.48\columnwidth]{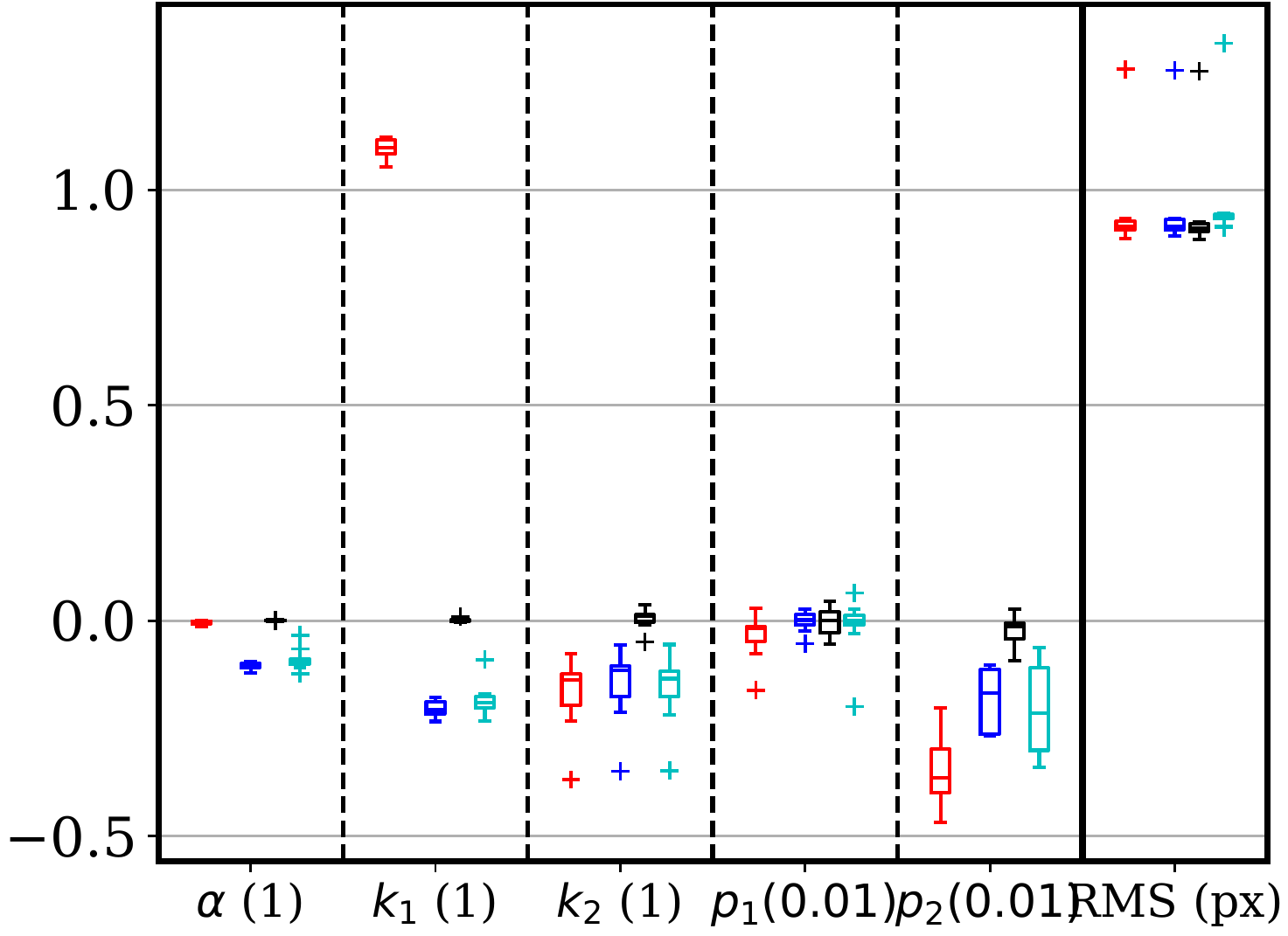} \\
	(MTV185)
	\caption{Error distributions of Mei parameters and the root mean square (RMS) reprojection errors, by geometric camera calibration tools, BabelCalib, Basalt, Camodocal, Kalibr, and the OpenCV-based ROS camera calibrator,
	on the simulated data for BM4018, BT2120, and MTV185 lenses, each of 9 sequences.
	BabalCalib and Basalt used the extended unified camera model (EUCM), while others used the Mei model.
	Note that $\beta$ of the EUCM is shown together with $k_1$ of the Mei model relative to $k_1$'s true value.
	A parameter unavailable to the EUCM, e.g., $k_2$, is zero by default, and shown relative to the parameter's true value.
	Basalt failed all MTV185 sequences.
	}
	\label{fig:sim-mei}
\end{figure}

For sequences with lenses, BM4018, BT2120, and MTV185, 
three tools including Kalibr, Camodocal, and the ROS / OpenCV calibrator were used to solve for the Mei parameters.
For comparison, we also solved for the EUCM model by BabelCalib and Basalt.
The parameter errors and reprojection errors are shown in Fig.~\ref{fig:sim-mei}.
where we used $(f_x, f_y)$ instead of $(\gamma_x, \gamma_y)$ as the latter has large variance caused by that of $\xi$.
For both the BM4018 and BT2120 sequences, the three methods with the Mei model gave similar results.
Overall, Kalibr gave the best estimates, notably on the MTV185 sequences.
The ROS calibrator tended to have larger variances in focal lengths and principal points but their errors were within (-2, 2) px.
For the MTV185 sequences, the Camodocal and OpenCV results showed about 3 px errors in focal lengths and about 0.7 errors in $\xi$, but with reasonable RMS errors.
We attribute this to two reasons. First, the UCM model is numerical unstable. Second, $k_1$ in the Mei model is redundant.
As for the EUCM models, the BabelCalib achieved smaller dispersions in $f_x$ and $f_y$ than Basalt, although the data were simulated with the Mei model.

\begin{figure}[!htbp]
	\centering
	\includegraphics[width=0.48\columnwidth]{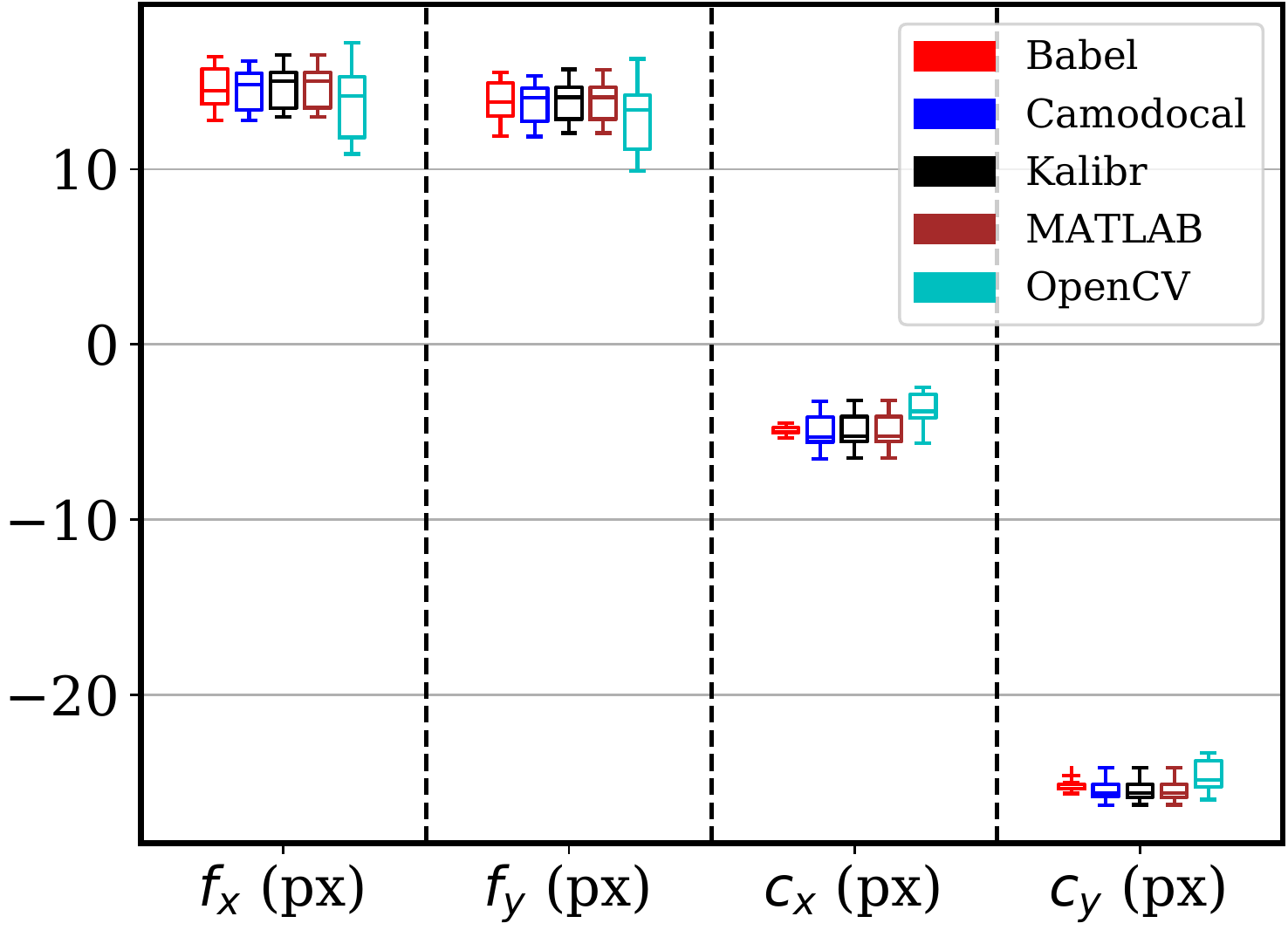}
	\includegraphics[width=0.48\columnwidth]{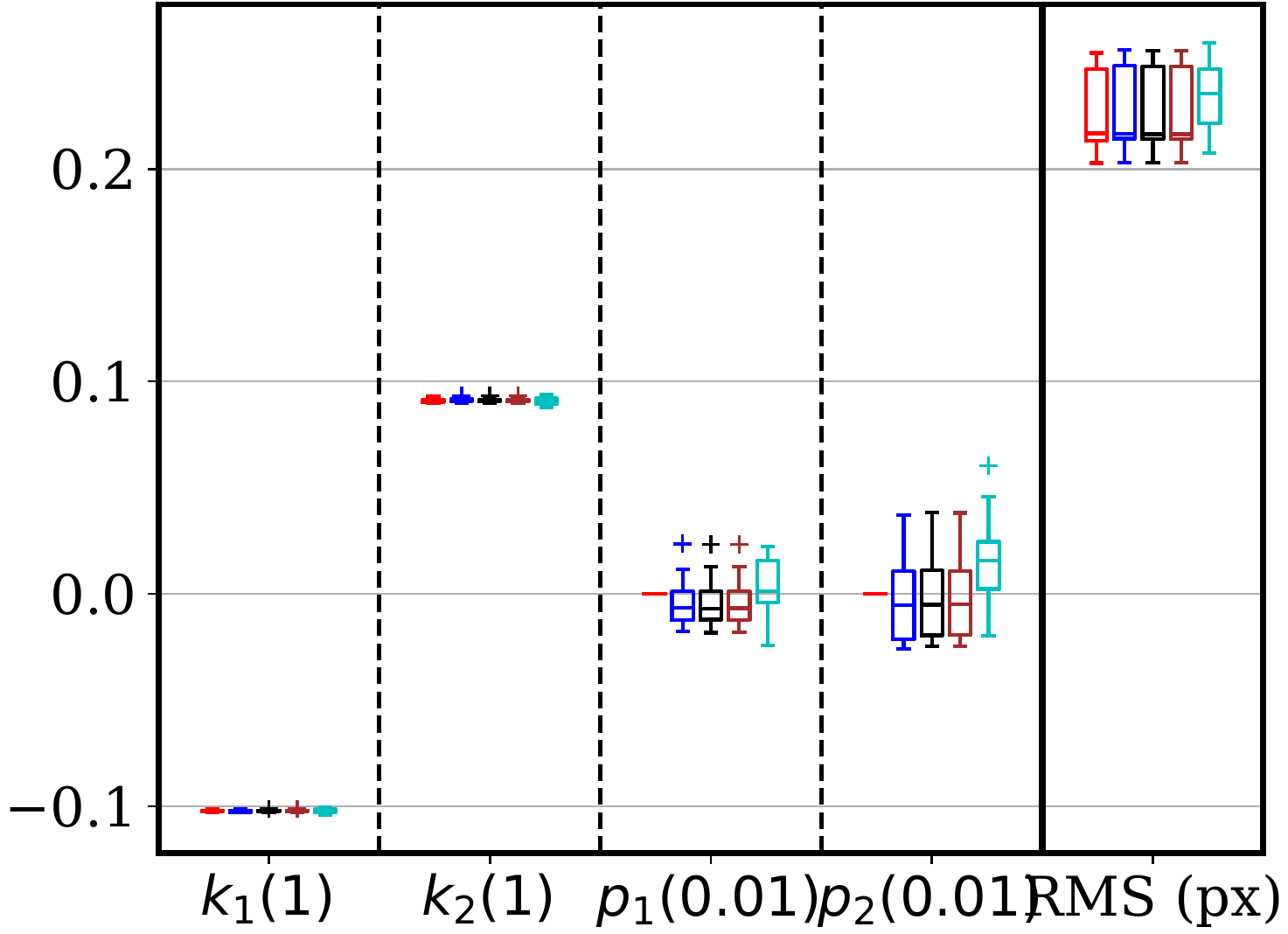} \\
	(S04525)\\
	\includegraphics[width=0.48\columnwidth]{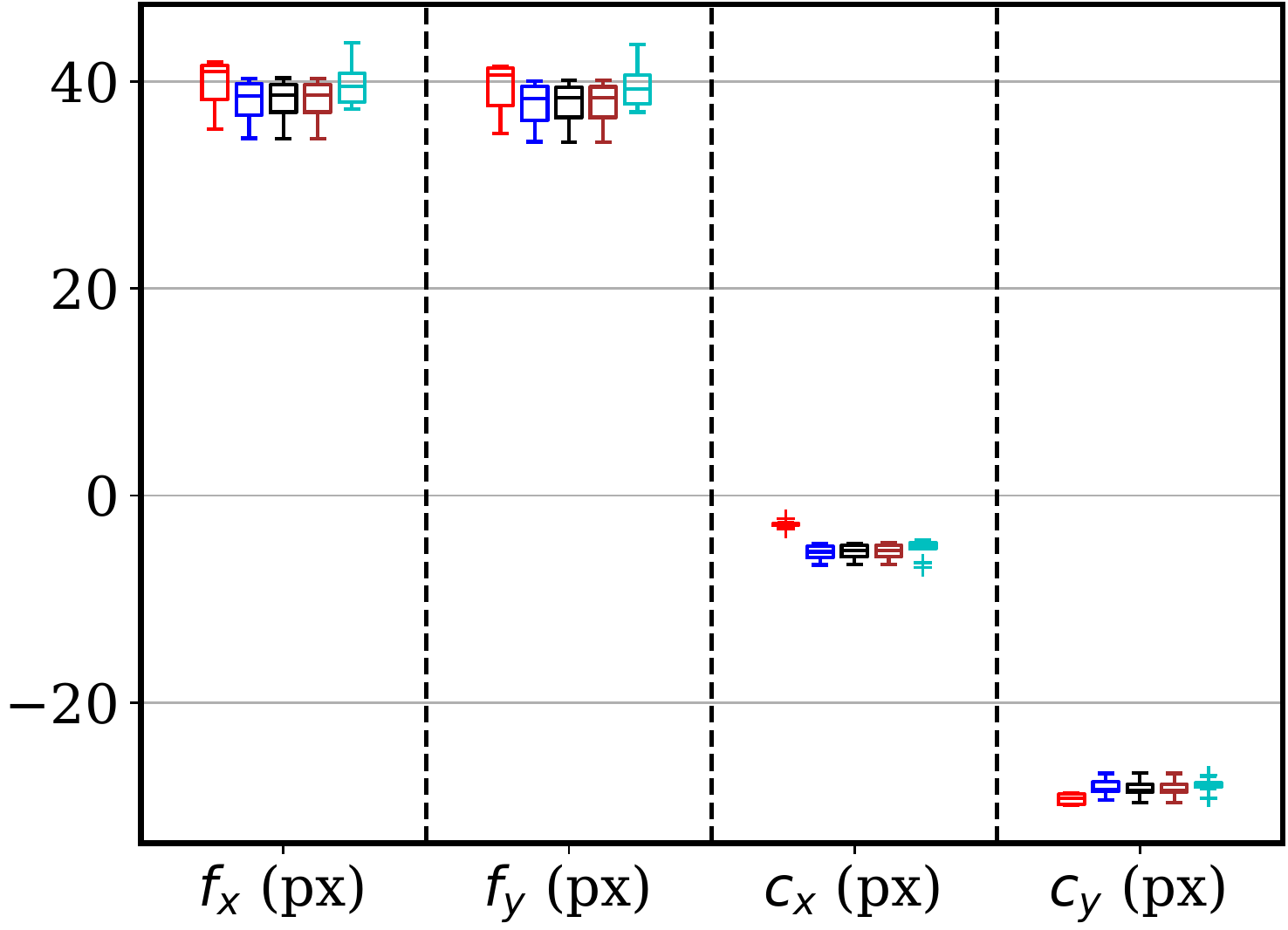}
	\includegraphics[width=0.48\columnwidth]{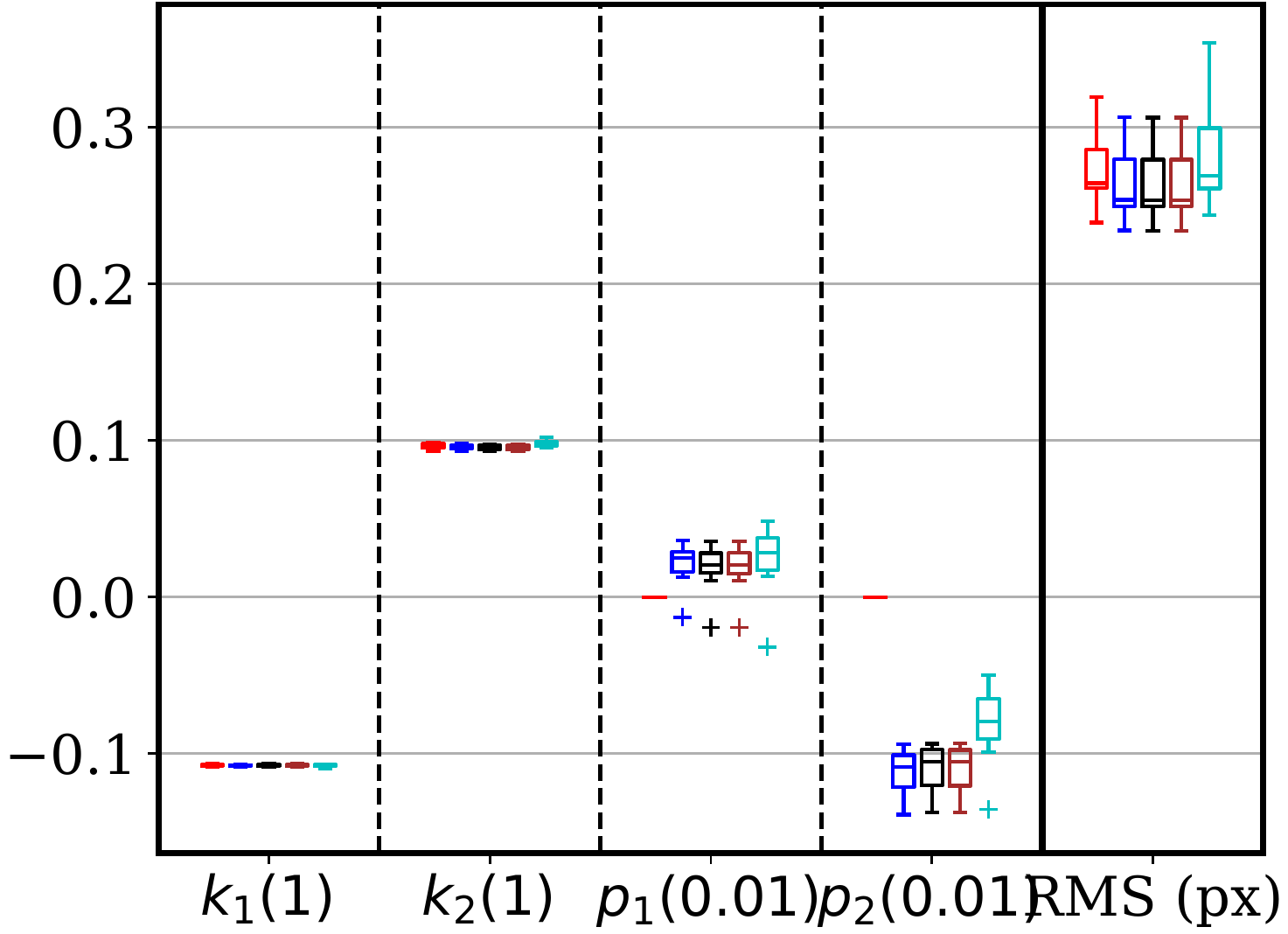}\\
	(E1M3518)\\
	\includegraphics[width=0.48\columnwidth]{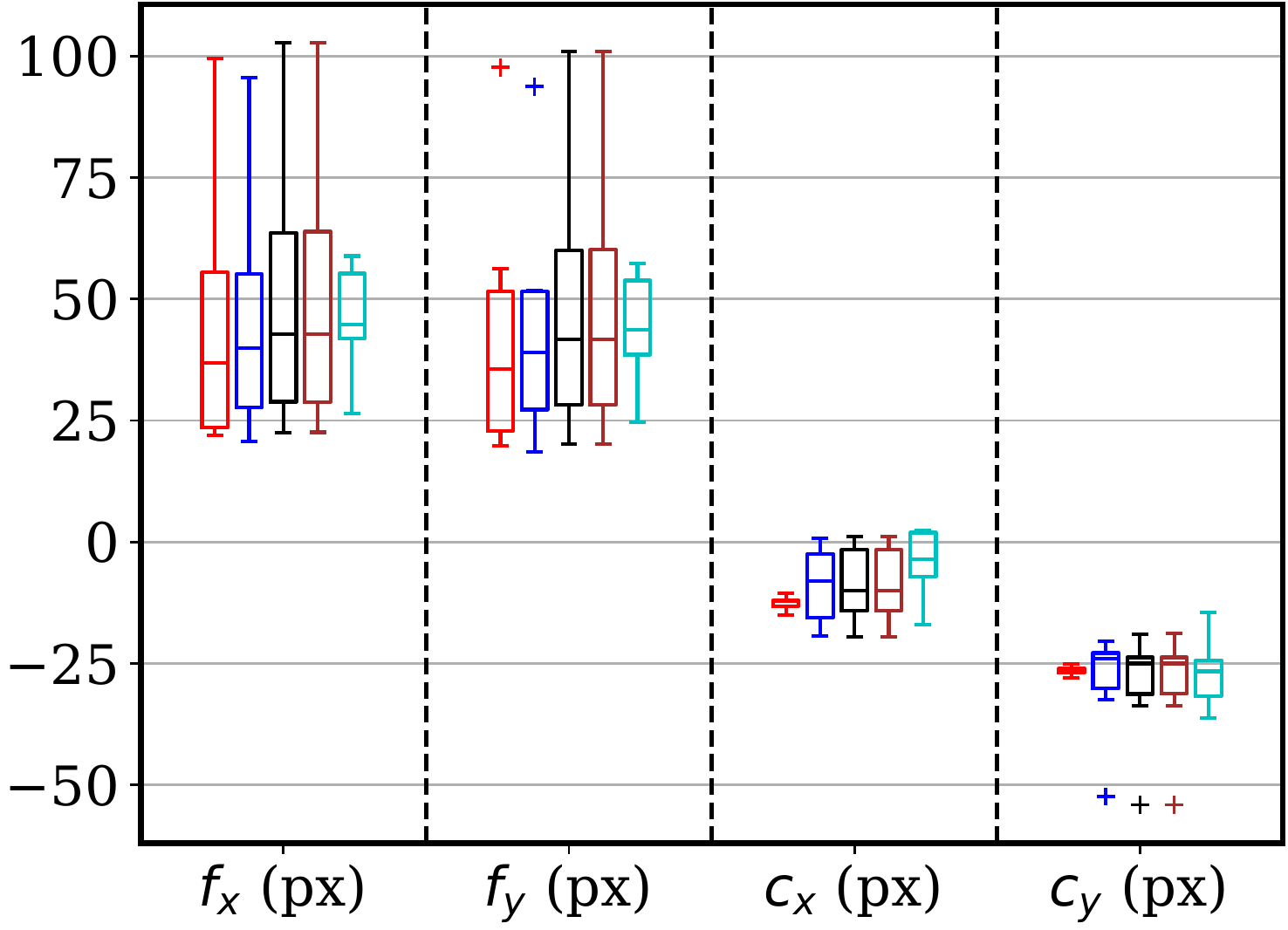}
	\includegraphics[width=0.48\columnwidth]{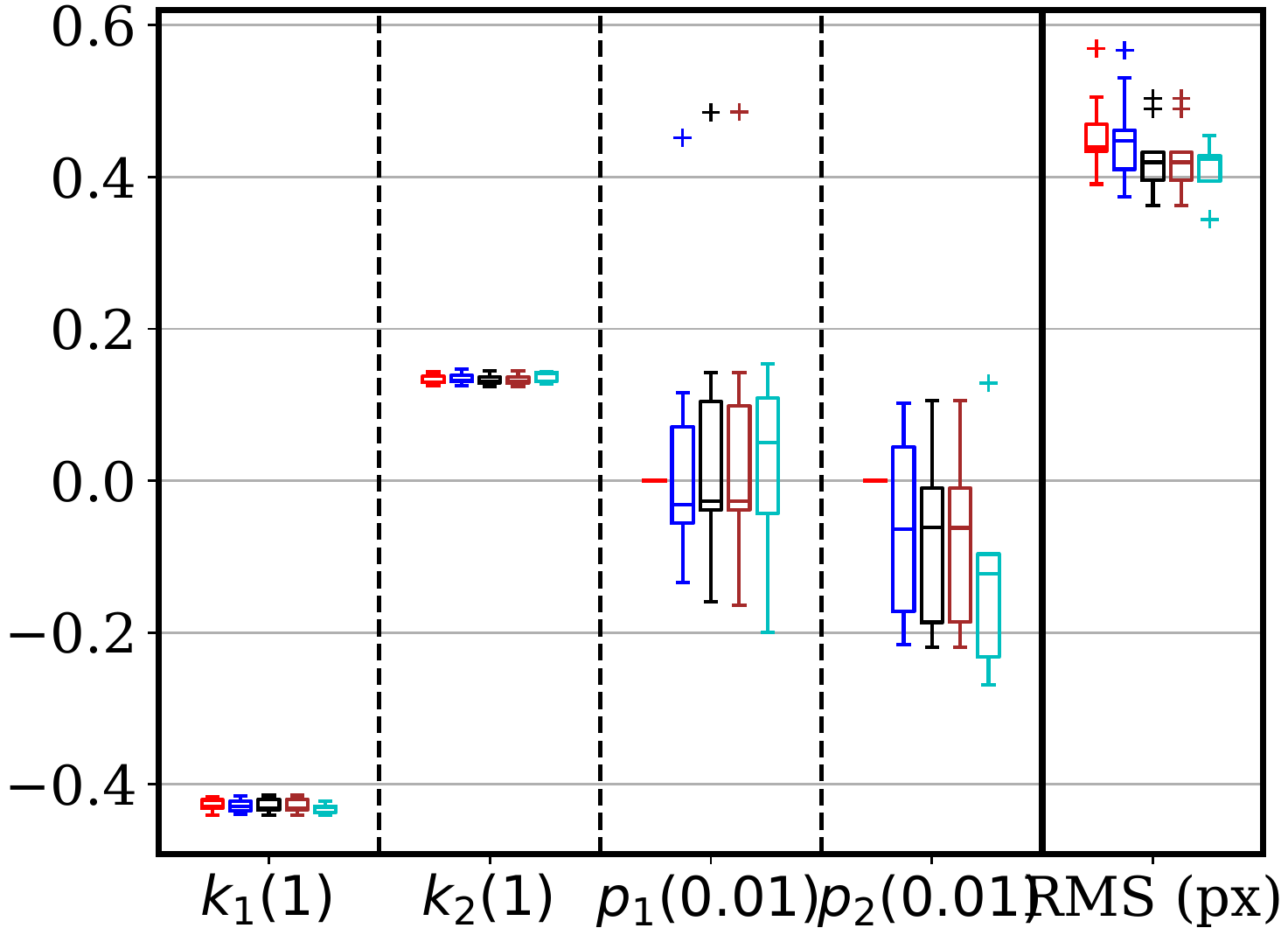} \\
	(BM4218)\\
\caption{Pinhole radial tangential model parameters and the root mean square (RMS) reprojection errors,
by geometric camera calibration tools, BabelCalib, Camodocal, Kalibr, the MATLAB calibrator, and the ROS / OpenCV camera calibrator,
on the real datasets captured with S04525, E1M3518, and BM4218 lenses, each of 9 sequences.
Note that BabelCalib adopted a pinhole radial model.
The ROS calibrator failed four out of 9 times for the BM4218 dataset.
}
\label{fig:boxplot-pinhole}
\end{figure}

\subsection{Real Data Results}
We processed the real data according to Table~\ref{tab:experiments} and analyzed the estimated parameters and RMS reprojection errors.
For clarity, the nominal focal lengths from Table~\ref{tab:lenses} and the coordinates (800, 600) are subtracted from their estimates in plots.

\begin{figure}[!htbp]
	\centering
	\includegraphics[width=0.48\columnwidth]{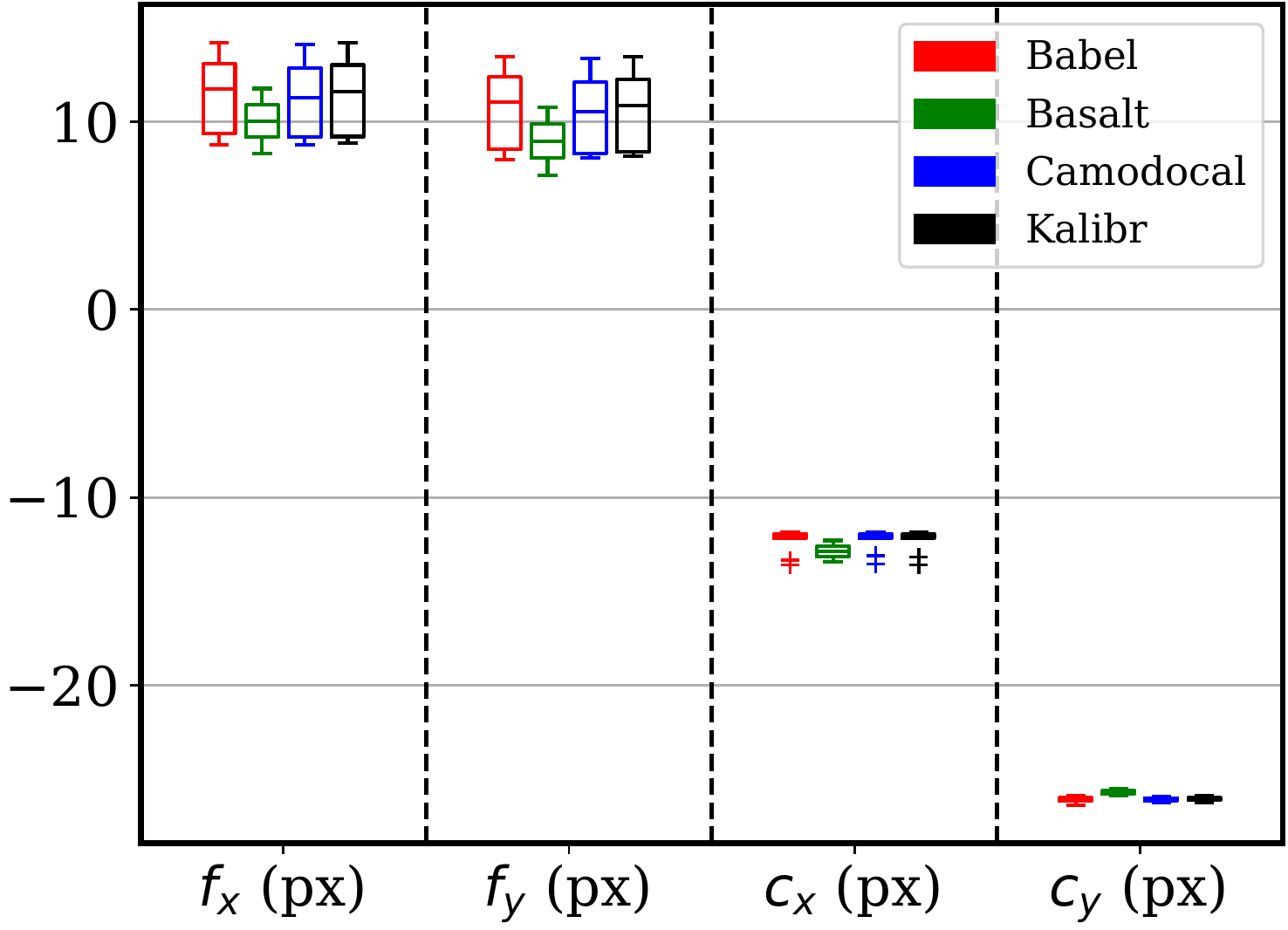}
	\includegraphics[width=0.48\columnwidth]{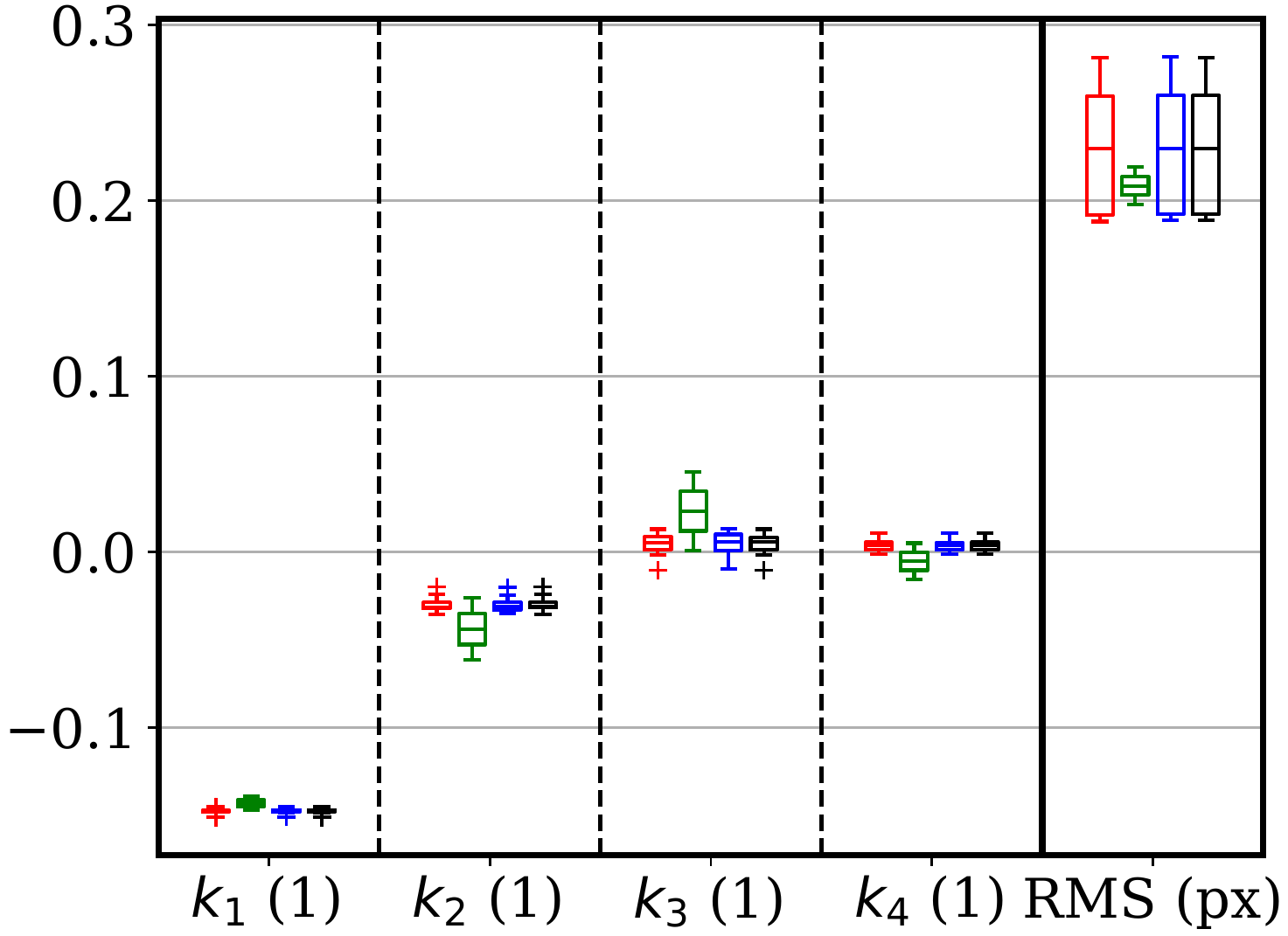} \\
	(BM4218) \\
	\includegraphics[width=0.48\columnwidth]{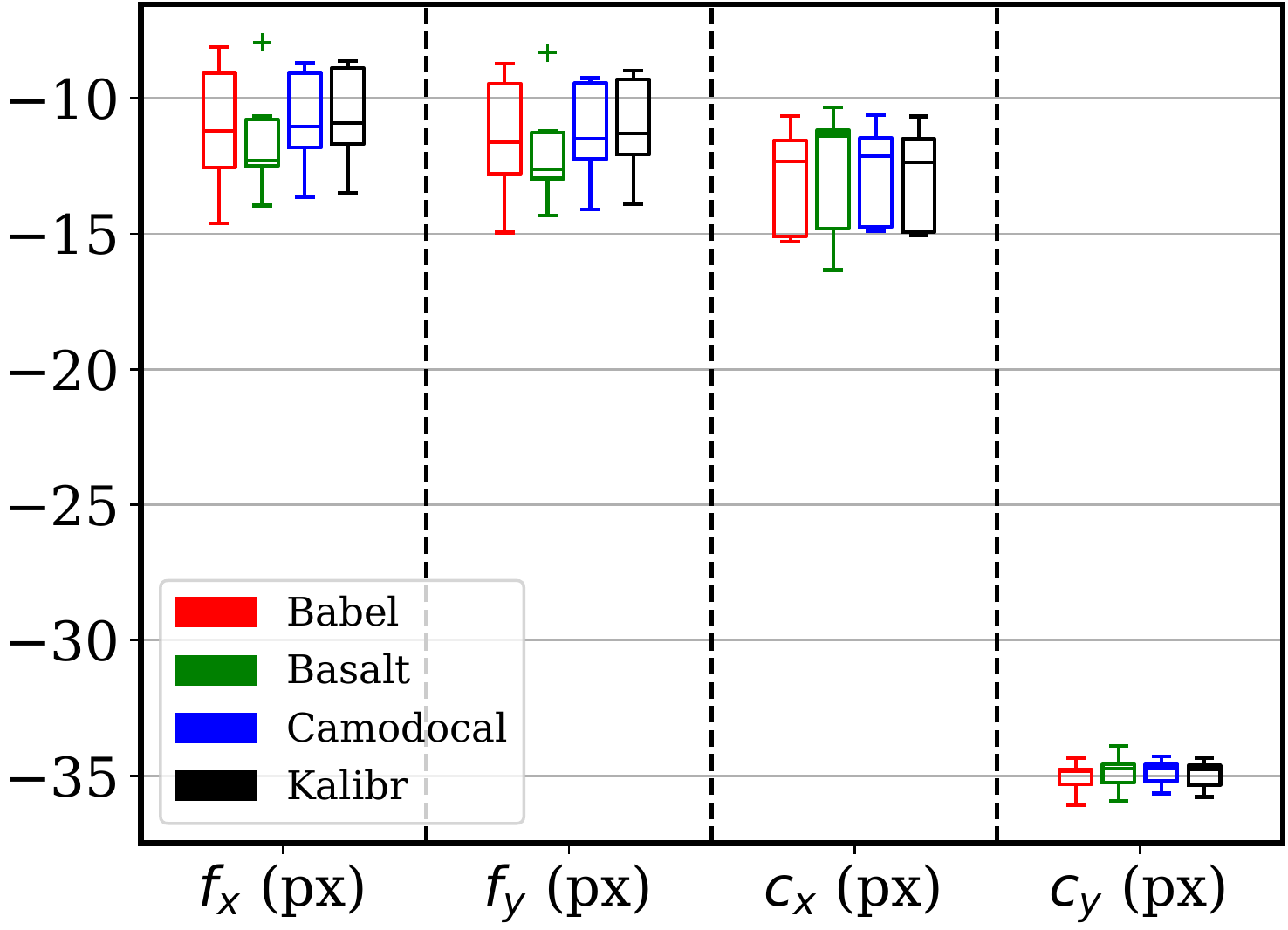}
	\includegraphics[width=0.48\columnwidth]{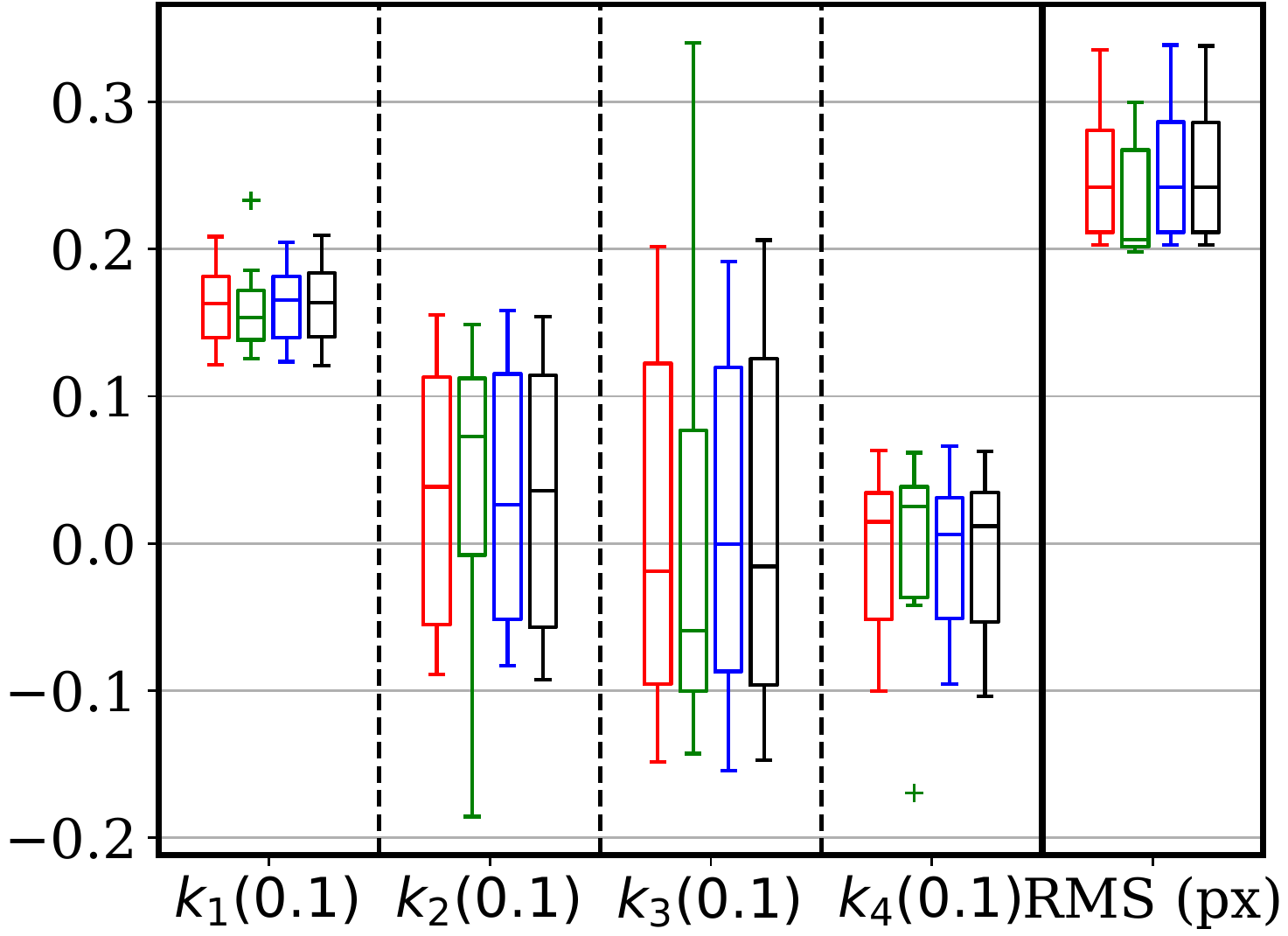} \\
	(BM4018)\\
	\includegraphics[width=0.48\columnwidth]{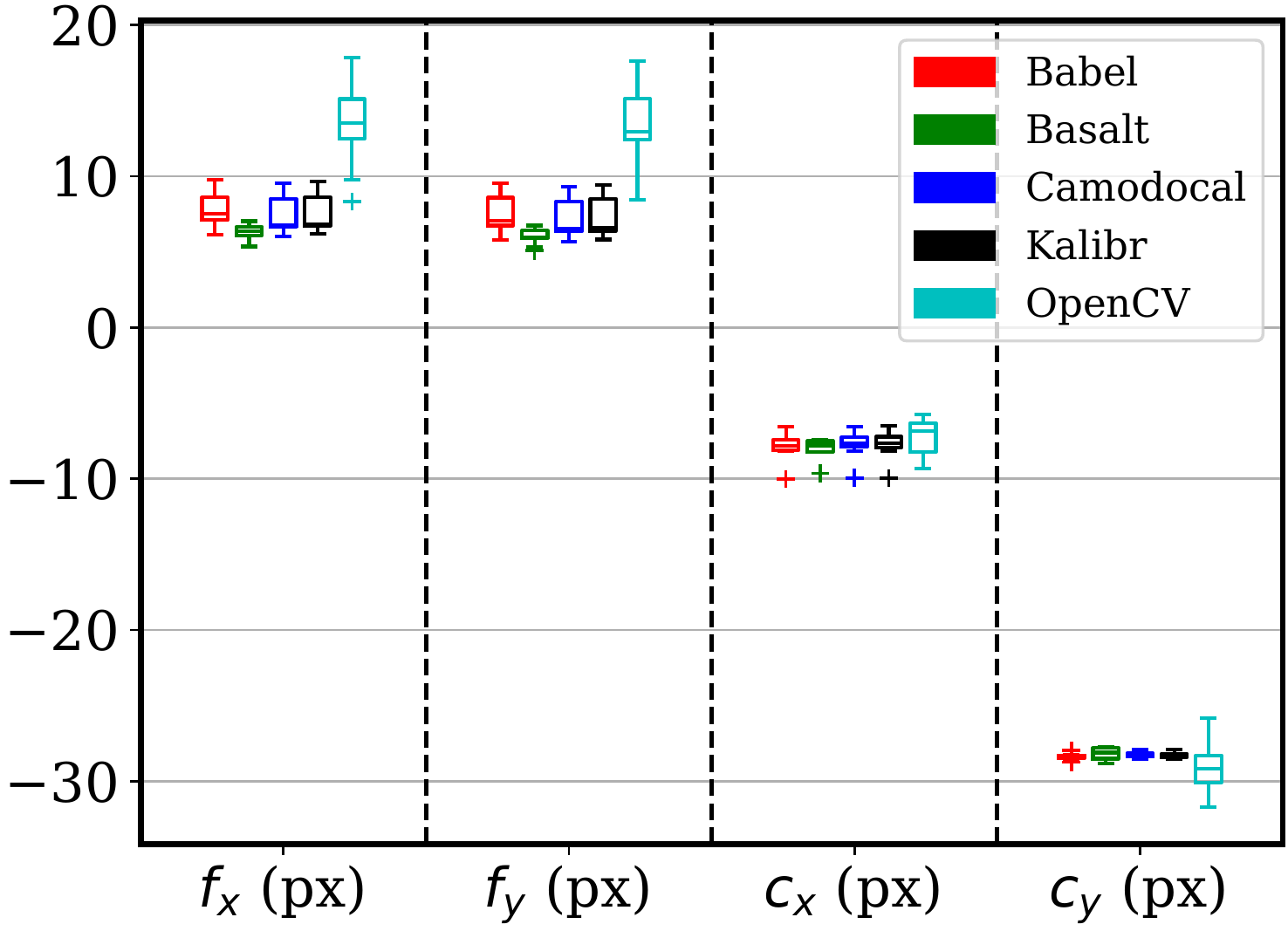}
	\includegraphics[width=0.48\columnwidth]{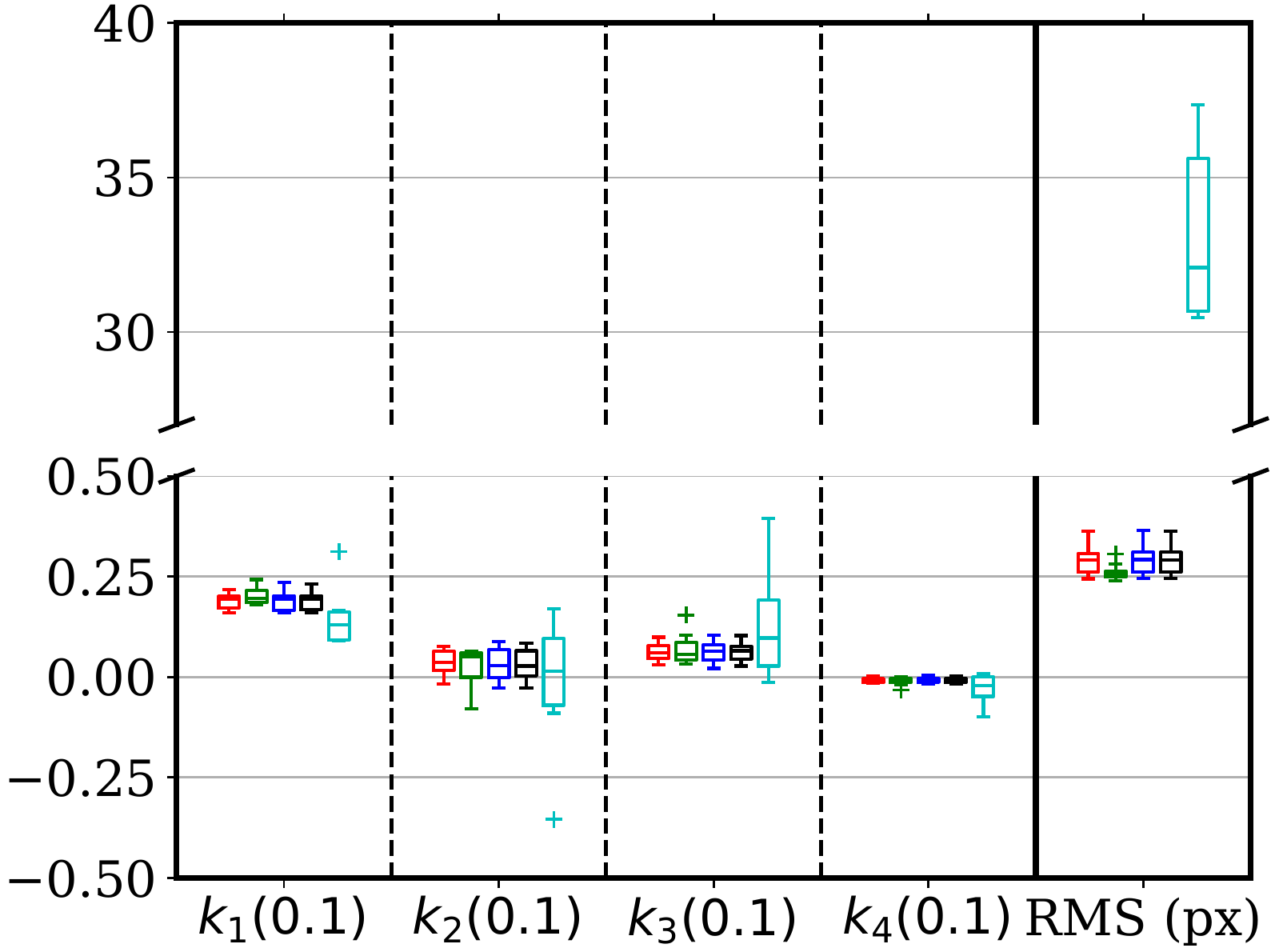} \\
	(BT2120)\\
	\includegraphics[width=0.48\columnwidth]{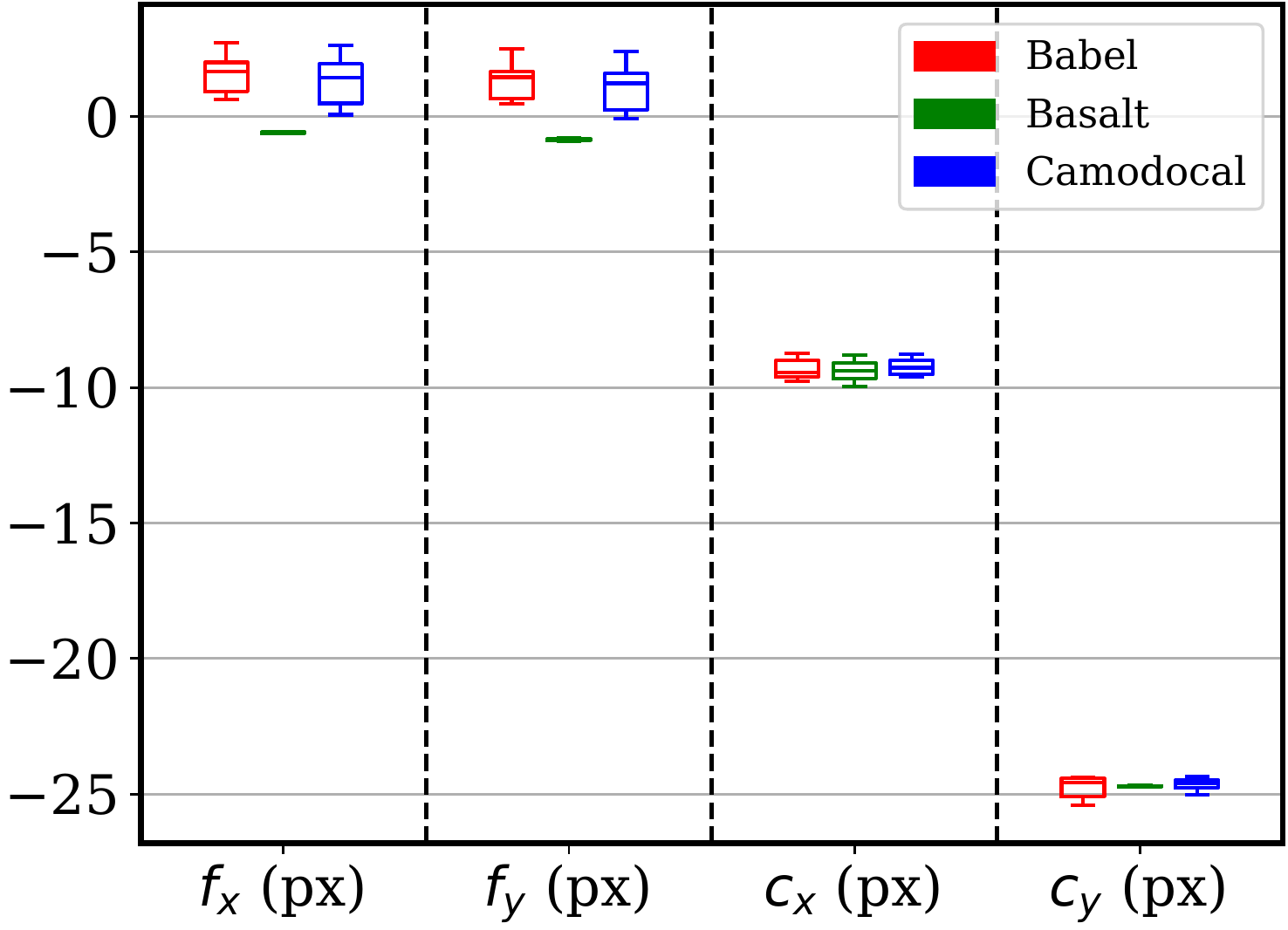}
	\includegraphics[width=0.48\columnwidth]{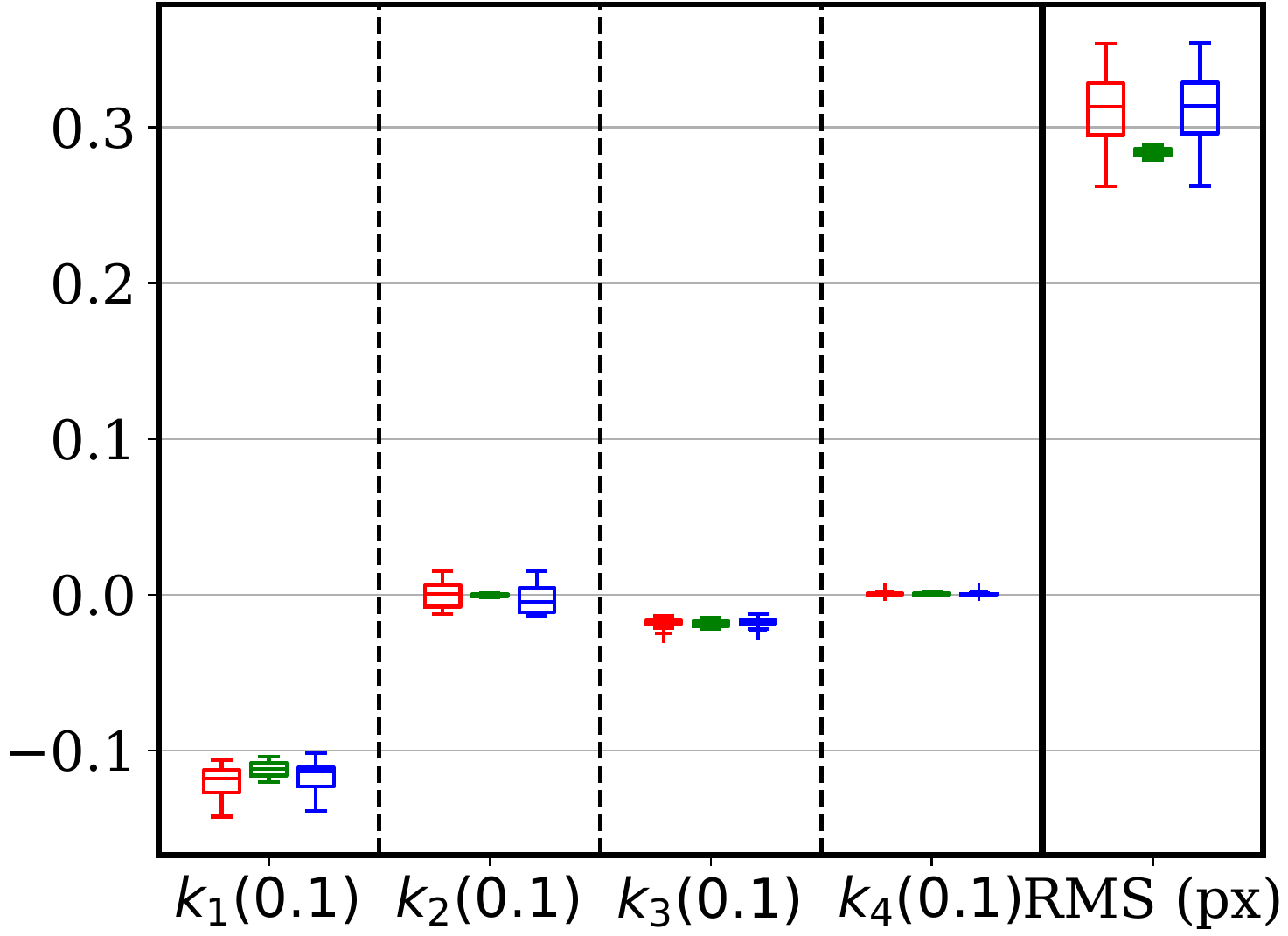}\\
	(MTV185)
	\caption{KB-8 parameters and the root mean square (RMS) reprojection errors 
	by geometric camera calibration tools, BabelCalib, Basalt, Camodocal, Kalibr, and the ROS / OpenCV calibrator,
	on the real datasets captured with BM4218, BM4018, BT2120, and MTV185 lenses, each of 9 sequences.
	Basalt failed 7 out of 9 times for both BM4218 and MTV185 datasets, leading to its small variances.
}
	\label{fig:boxplot-kb8}
\end{figure}

The sequences of S04525, E1M3518, and BM4218 lenses were processed by five tools except for Basalt.
The calibration parameters and the RMS reprojection errors are shown in Fig.~\ref{fig:boxplot-pinhole}, where the failed cases are not included in the box plots.
These five tools had fairly similar results.
The difference in principal points for BabelCalib was caused by its model that ignored tangential distortion.
The large dispersion in projection parameters of BM4218 sequences was likely because 
the pinhole radial tangential model was somewhat improper for the camera with this lens as implied in Fig.~\ref{fig:boxplot-kb8}.

For the KB-8 model, the sequences of BM4218, BM4018, BT2120, and MTV185 lenses were processed by five tools except the MATLAB calibrator.
As shown in Fig.~\ref{fig:boxplot-kb8}, these tools achieved very similar calibration results in general.
Basalt failed frequently for the BM4218 and BT2120 sequences, leading to apparent small parameter dispersions.
Comparing the dispersion of focal lengths for BM4218 in Fig.~\ref{fig:boxplot-pinhole} and \ref{fig:boxplot-kb8}, we think that the KB-8 model is more suitable than the pinhole radial tangential model for the BM4218 sequences.
The OpenCV-based ROS calibrator aborted on BM4218 and BM4018 sequences perhaps for failed initialization.
Both Kalibr and the ROS calibrator did not handle MTV185 sequences of a $194^\circ$ DAOV.
For the BT2120 data, we see that the results by OpenCV were affected by outliers leading to large RMS reprojection errors and parameters slightly deviated from other methods.

\begin{figure}[!tbp]
	\centering
	\includegraphics[width=0.48\columnwidth]{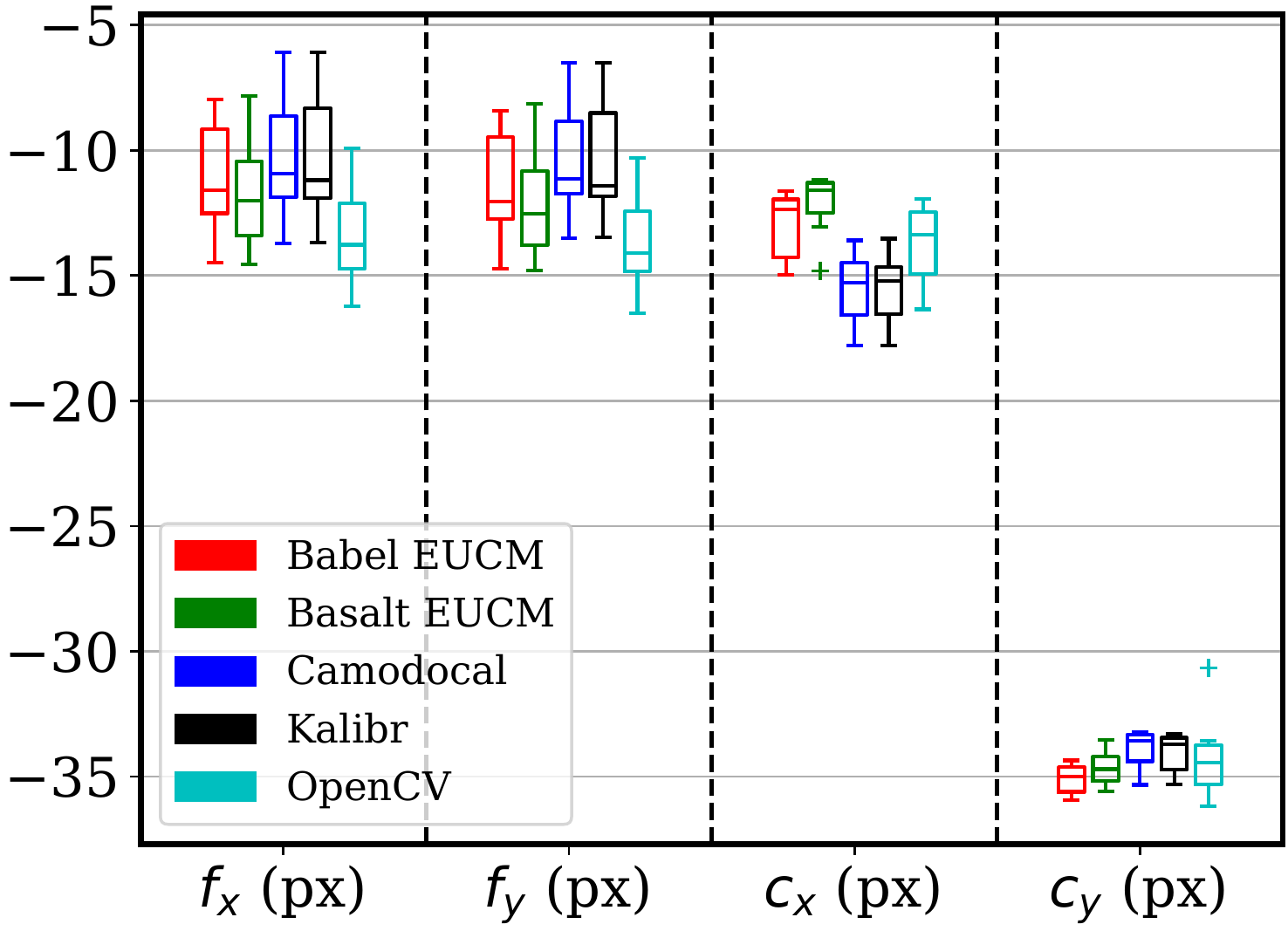}
	\includegraphics[width=0.48\columnwidth]{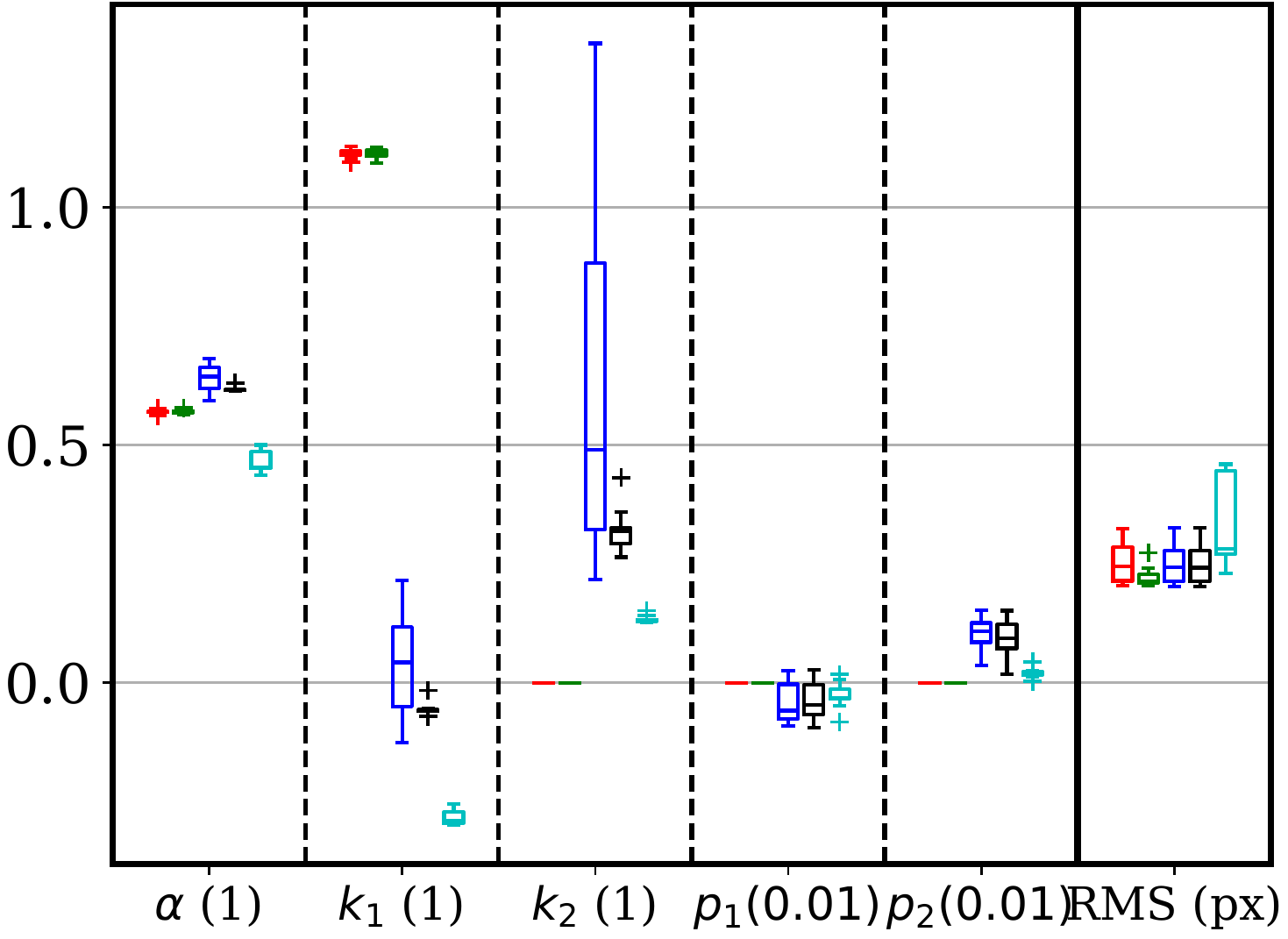} \\
	(BM4018) \\
	\includegraphics[width=0.48\columnwidth]{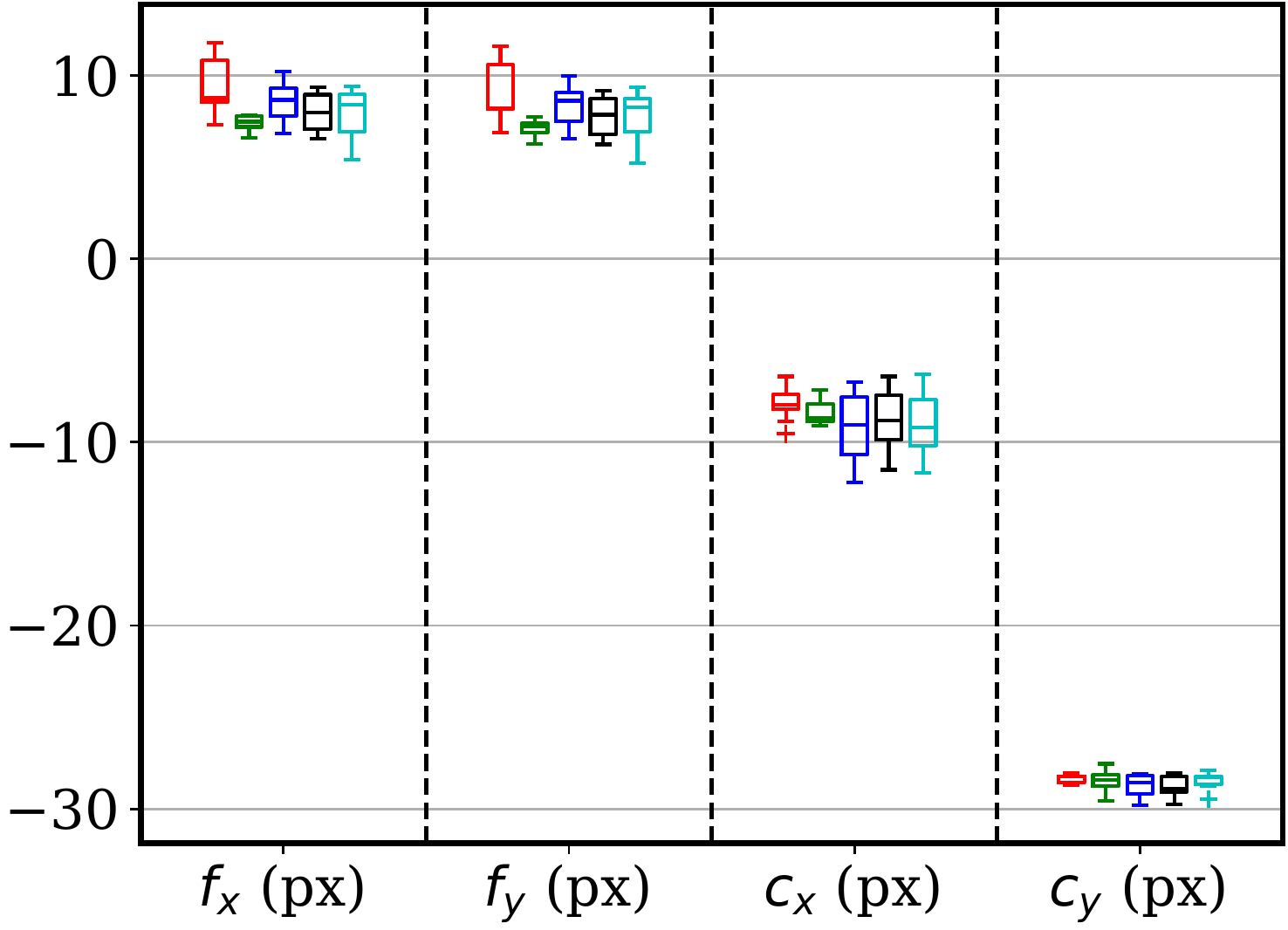}
	\includegraphics[width=0.48\columnwidth]{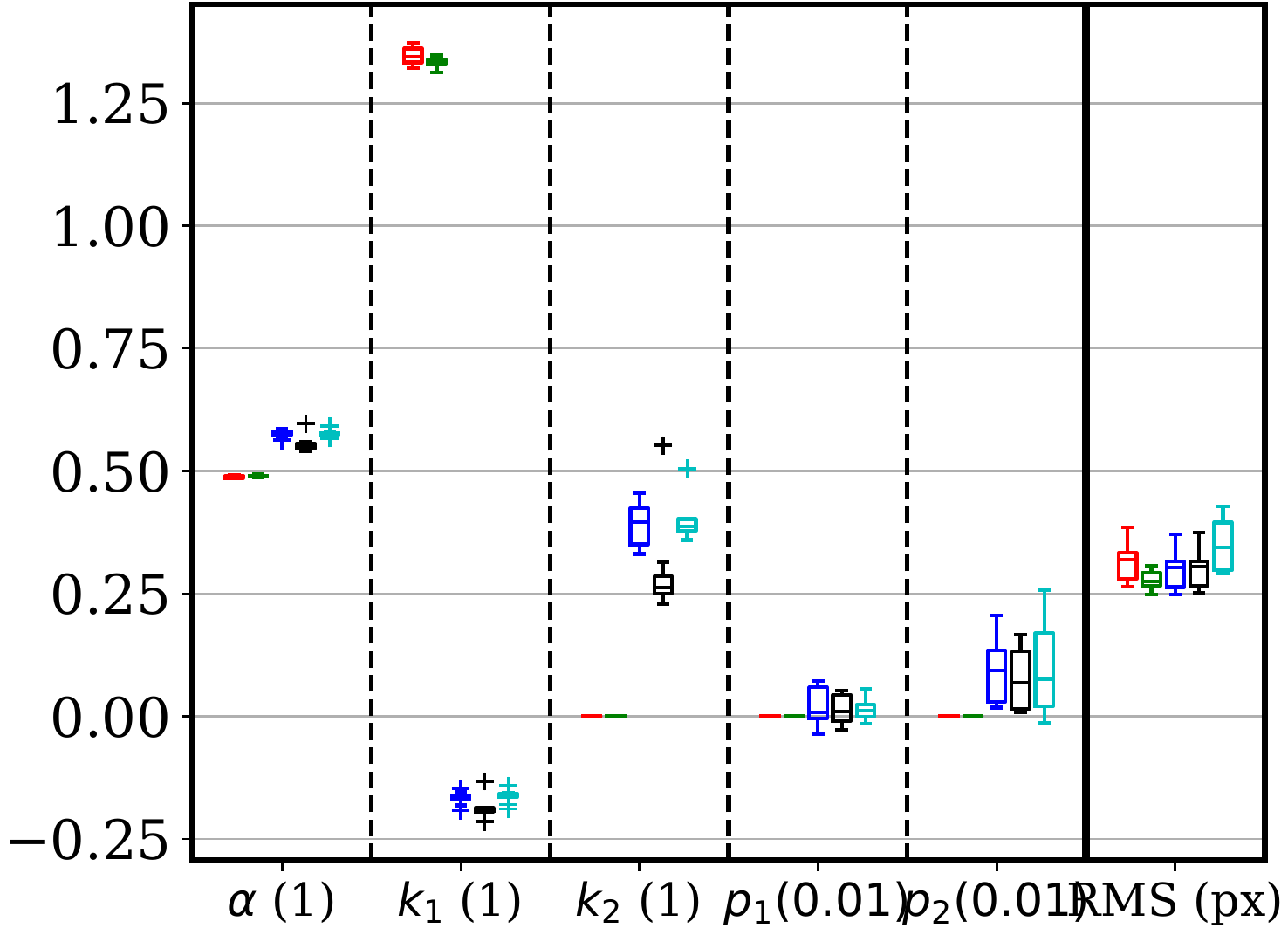} \\
	(BT2120) \\
	\includegraphics[width=0.48\columnwidth]{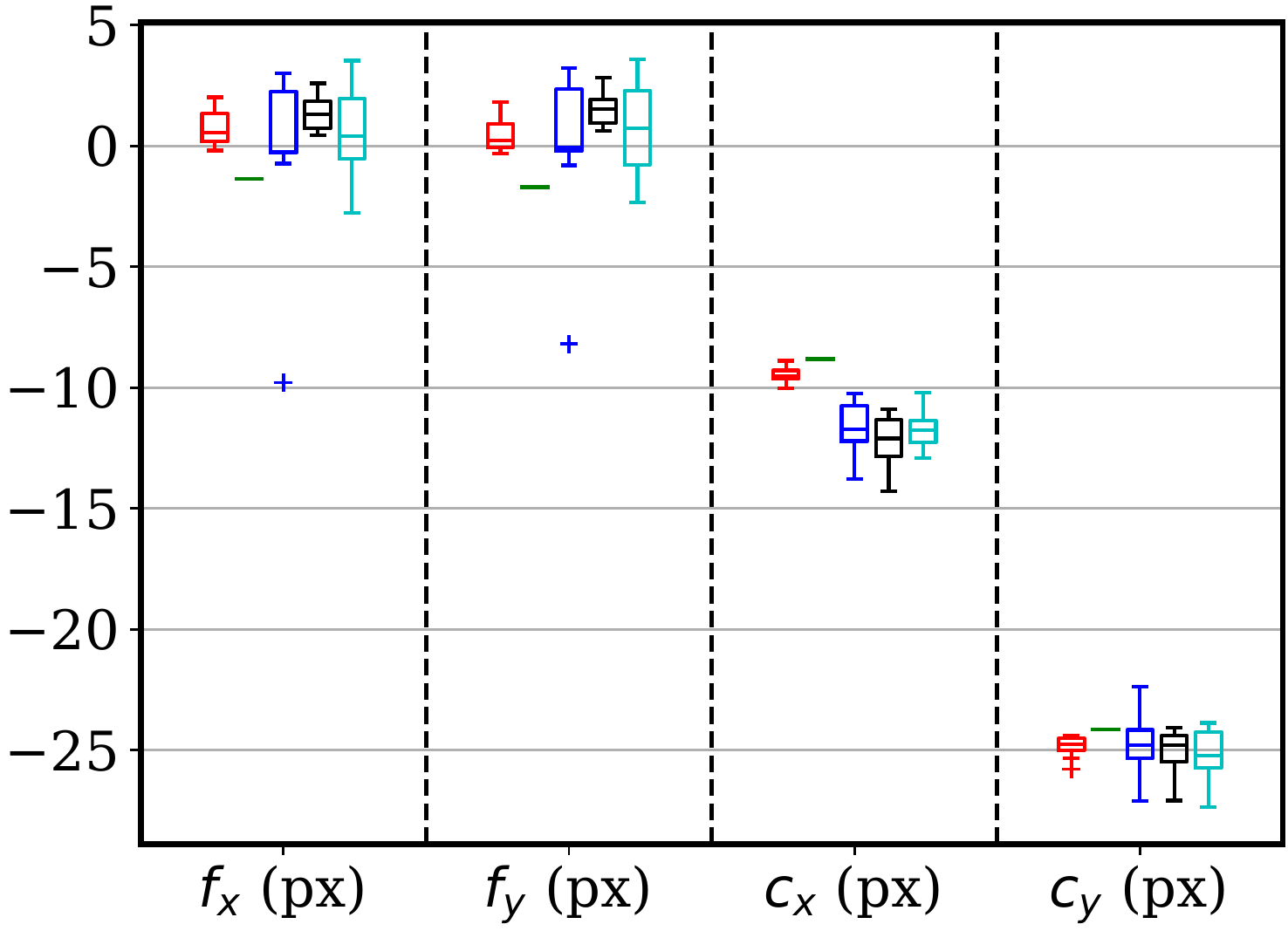}
	\includegraphics[width=0.48\columnwidth]{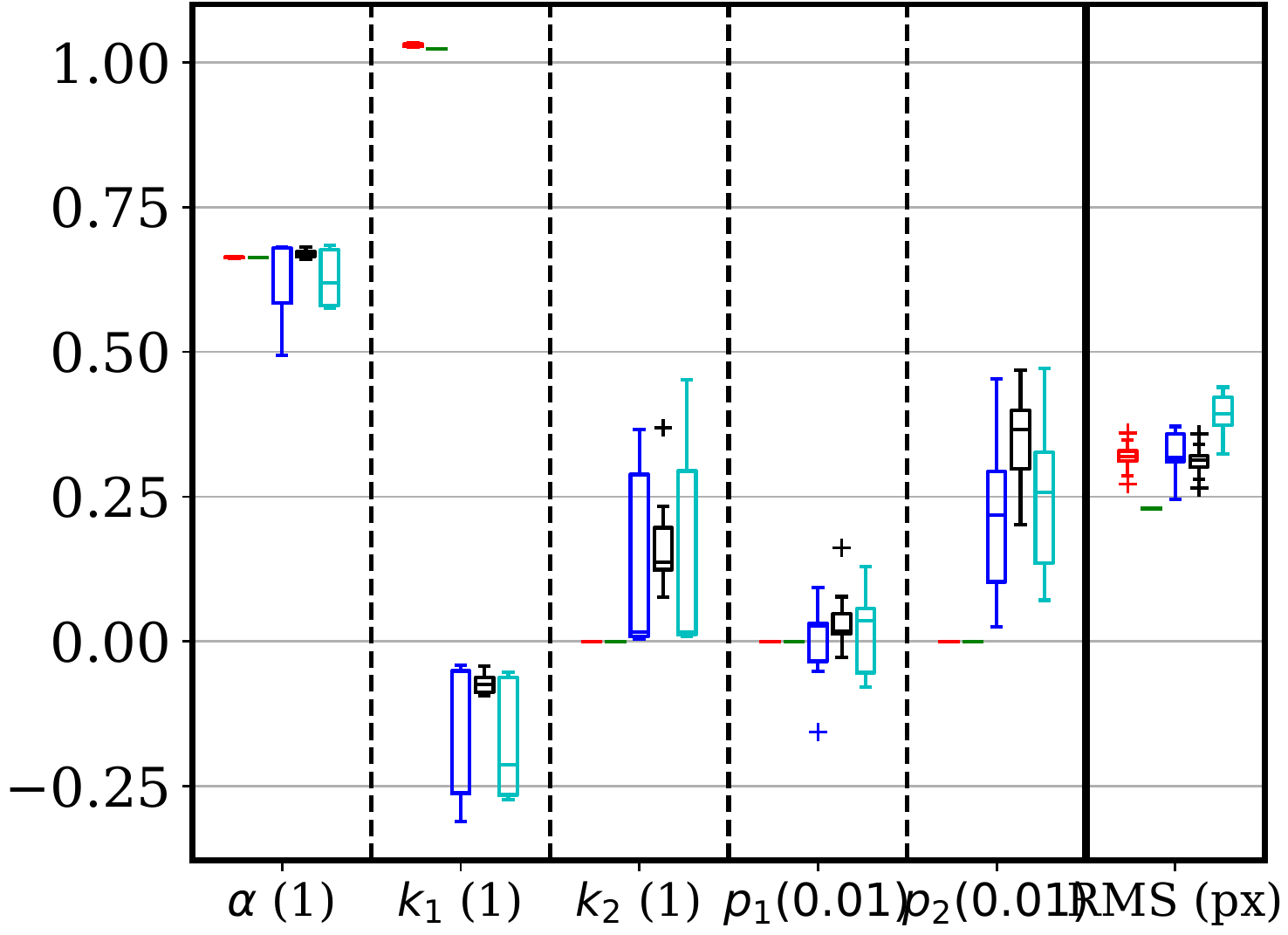} \\
	(MTV185)
	\caption{Mei parameters and the root mean square (RMS) reprojection errors by geometric camera calibration tools, BabelCalib with the extended unified camera model (EUCM), Basalt with the EUCM, 
	Camodocal, Kalibr / TartanCalib, and the ROS / OpenCV camera calibrator,
	on the real data captured with BM4018, BT2120, and MTV185 lenses, each of 9 sequences.
	Note that the EUCM model parameter $\beta$ is shown together with the Mei parameter $k_1$.
	The unavailable parameters, e.g., $k_2$ to the EUCM model, are zero by default.
	Basalt failed 8 out of 9 MTV185 sequences.
	}
	\label{fig:boxplot-mei}
\end{figure}

For the Mei model, we processed the BM4018, BT2120, and MTV185 sequences using tools including Kalibr, Camodocal, and the ROS calibrator.
For comparison to the EUCM model, these sequences were also processed by BabelCalib and Basalt.
The calibration parameters and the RMS reprojection errors are shown in Fig.~\ref{fig:boxplot-mei}.
With real data, these tools obtained similar values and dispersions for focal lengths and principal points, 
more consistent than the simulation shown in Fig.~\ref{fig:sim-mei} where the data were simulated with the Mei model.
The distortion parameters of the Mei model had large variance despite reasonable RMS reprojection errors, due to its parameter redundancy.
Otherwise, the EUCM model resulted in consistent values for $[\alpha, \beta]$.

\section{Conclusions and Research Trends}
\label{sec:conclusion}
For the prevalent wide-angle cameras, we survey the recent GCC tools from the perspectives of camera models, calibration targets, and algorithms,
providing an overview of the benefits and limitations of these tools.
We also evaluated six well-known calibration tools, including 
BabelCalib, Basalt, Camodocal, Kalibr, the MATLAB calibrator, and the OpenCV-based ROS calibrator, 
to study their consistency and repeatability on simulated and real data.

From the review and experiments, we summarize several findings.

(1) Outlier handling is crucial for optimization-based camera calibration tools.
These outliers are usually detected corners a few pixels away from their actual image locations, and often occur in somewhat blurry images.
Luckily, most GCC tools can deal with outliers.

(2) The GCC tools, Camodocal, Kalibr, and the MATLAB calibrator, support well the pinhole radial tangential model.
BabelCalib and Camodocal support well the KB-8 model, and TartanCalib supports well the KB-8 model for a camera with a DAOV $< 180^\circ$.
Camodocal, TartanCalib, and OpenCV support well the Mei model, but the model suffers from parameter instability and redundancy.

Moreover, the pinhole radial tangential model may become inadequate for cameras of a DAOV $>100^\circ$.
The KB-8 model is typically preferred for cameras of a large DAOV due to its wide support and good accuracy when a global camera model is to be obtained.

(3) The various failure cases revealed in our tests imply the intricacy in camera model initialization and optimization of a classic GCC tool.
Aside from these failures, these GCC tools in (2) agree well with each other on calibrating conventional, fisheye, and omnidirectional cameras with proper global camera models.

Based on this study, we point out several future research directions.

\textbf{Interactive Calibration}
It is well known that quality data and informative data are essential for GCC.
The opposite are two problems, image blur that may be caused by rapid motion or out of focus, and insufficient data. One way to ensure data quality and information is interactive calibration 
which provides quality check, selects the quality data, and gives next-move suggestions in real time, 
whether for target-based or target-free calibration.
AprilCal \cite{richardsonAprilCalAssistedRepeatable2013} and Calibration Wizard \cite{pengCalibration2019} are such interactive tools for target-based calibration.

\textbf{Static Calibration}
Target-based calibration often involves unrepeatable onerous movements 
which can be obviated in at least two ways, calibration with a programmed robot arm and static calibration.
Robot arm-based calibration has been studied in \cite{wernerEfficientPreciseConvenient2019}.
Static calibration usually relies on active targets.
Such methods have been developed in \cite{gaoScreenbasedMethodAutomated2019,tehraniPracticalMethodFully2017} with application-specific setups.
We think there is still much room in static calibration to explore.

\textbf{Reconstruction with Calibration}
The setup of the lab calibration is usually different from the in-situ setup, e.g., in focusing distance (depth of field), exposure, 
capture mode (snapshot or video), aperture,
and size of the objects of interest.
Some work has been done to mitigate the differences, e.g., out of focus, in \cite{bellMethodOutoffocusCamera2016,haAccurateCameraCalibration2015}.
An ultimate solution would be self-calibration or calibration based on prior maps.
These methods depend on a reconstruction engine that supports calibration.
Such an engine based on bundle adjustment is colmap \cite{schonbergerStructurefrommotionRevisited2016}.
New engines capable of calibration based on deep learning are on the surge, for instance,
\cite{jeongSelfcalibratingNeuralRadiance2021,lindenbergerPixelperfect2021}.

{\small
\bibliography{zoterox}
}

\end{document}